\documentclass[fleqn,usenatbib]{mnras}

\usepackage{newtxtext,newtxmath}

\usepackage[T1]{fontenc}
\usepackage{ae,aecompl}

\usepackage{graphicx}	\usepackage{amsmath}	\usepackage{amssymb}

\title[Sgr A* GRMHD parameter survey]{A parameter survey of Sgr A* radiative models from GRMHD simulations with self-consistent electron heating}

\author[Dexter, Jim\'{e}nez-Rosales, Ressler, Tchekhovskoy, et al.]{
J. Dexter,$^{1,2}$\thanks{jason.dexter@colorado.edu}
A. Jim\'{e}nez-Rosales,$^{1}$\thanks{ajimenez@mpe.mpg.de}
S. M. Ressler,$^{3}$
A. Tchekhovskoy,$^{4}$
M. Baub\"{o}ck,$^{1}$
\newauthor
P. T. de Zeeuw,$^{1}$ F. Eisenhauer,$^{1}$ S. von Fellenberg,$^{1}$ F. Gao,$^{1}$ R. Genzel,$^{1,5}$\newauthor
S. Gillessen,$^{1}$ M. Habibi,$^{1}$ T. Ott,$^{1}$ J. Stadler,$^{1}$ O. Straub,$^{1}$ F. Widmann$^{1}$
\\
$^{1}$Max Planck Institute for Extraterrestrial Physics (MPE), Giessenbachstr. 1, 85748 Garching, Germany\\
$^{2}$JILA and Department of Astrophysical and Planetary Sciences, University of Colorado, Boulder, CO 80309, USA\\
$^{3}$Kavli Institute for Theoretical Physics, University of California Santa Barbara, Kohn Hall, Santa Barbara, CA 93107, USA\\
$^{4}$Center for Interdisciplinary Exploration \& Research in Astrophysics (CIERA), Physics \& Astronomy, Northwestern University,\\ Evanston, IL 60202, USA\\
$^{5}$Departments of Physics and Astronomy, Le Conte Hall, University of California, Berkeley, CA 94720, USA
}

\date{Accepted XXX. Received YYY; in original form ZZZ}

\pubyear{2019}

\begin{document}
\label{firstpage}
\pagerange{\pageref{firstpage}--\pageref{lastpage}}
\maketitle

\begin{abstract}
The Galactic center black hole candidate Sgr A* is the best target for studies of low-luminosity accretion physics, including with near-infrared and submillimeter wavelength long baseline interferometry experiments. Here we compare images and spectra generated from a parameter survey of general relativistic MHD simulations  to a set of radio to near-infrared observations of Sgr A*. Our models span the limits of weak and strong magnetization and use a range of sub-grid prescriptions for electron heating. We find two classes of scenarios can explain the broad shape of the submillimeter spectral peak and the highly variable near-infrared flaring emission. Weakly magnetized ``disk-jet'' models where most of the emission is produced near the jet wall, consistent with past work, as well as strongly magnetized (magnetically arrested disk) models where hot electrons are present everywhere. Disk-jet models are strongly depolarized at submillimeter wavelengths as a result of strong Faraday rotation, inconsistent with observations of Sgr A*. We instead favor the strongly magnetized models, which provide a good description of the median and highly variable linear polarization signal. The same models can also explain the observed mean Faraday rotation measure and potentially the polarization signals seen recently in Sgr A* near-infrared flares.
\end{abstract}


\section{Introduction}

The most extensively studied low-luminosity accretion flow is that onto the Galactic centre massive black hole candidate Sagittarius A* (Sgr A*). Detected in the radio \citep{balick1974}, Sgr A* shows an inverted spectrum up to a peak at submillimeter (submm) wavelengths \citep{serabyn1997,falcke1998,stone2016,vonfellenberg2018,bower2019}. The bolometric luminosity is $\simeq 5\times10^{35} \,\rm erg\,\rm s^{-1}$ \citep{bower2019}, roughly 9 orders of magnitude sub-Eddington for the mass of $4 \times 10^6 M_\odot$ measured from the orbit of the star S2 \citep{ghez2008,gillessen2009,gillessen2017,gravityredshift,gravity2019,do2019}. Due to its proximity, Sgr A* is the largest black hole in angular size on the sky. It has long been a target of radio very long baseline interferometry  \citep[VLBI,][]{lo1975,backer1978,krichbaum1998,bower2006}. With $1.3$mm VLBI, the source size is as compact as $\simeq 8-10 r_g$ \citep{doeleman2008,fish2011,lu2018,johnson2018}, where $r_g = GM/c^2 \simeq 6\times10^{11}$ cm is the gravitational radius. 

Sgr A* shows large-amplitude ``flares'' simultaneously at near-infrared \citep[NIR,][]{genzel2003,ghez2004} and X-ray \citep{baganoff2001} wavelengths. They originate from transiently heated electrons, likely as a result of magnetic reconnection or shocks \citep{markoff2001} close to the black hole \citep{barriere2014,haggard2019}. The observed emission is due to synchrotron radiation at radio to NIR \citep[and likely also X-ray,][]{doddseden2009} wavelengths and is strongly polarized from the submm to NIR \citep{aitken2000,bower2003,marrone2006,eckart2006,trippe2007}. 

The flares can now be spatially resolved with the second generation VLT Interferometer beam combiner instrument GRAVITY. In 2018, 3 flares were shown to continuously rotate clockwise at relativistic speeds \citep{gravityflare}. The astrometric data are consistent with orbital motion around Sgr A* at a radius of $6-10 r_g$ \citep{bauboeck2020}. A matching polarization angle evolution suggests the presence of dynamically important magnetic fields in the emission region on event horizon scales. There is also evidence for ordered magnetic fields from $1.3$mm VLBI \citep{johnson2015}. 

Theoretical models of the near horizon regions of low-luminosity accretion flows \citep[][]{rees1982,narayan1995sgra,yuan2003} are now commonly realized using general relativistic MHD (GRMHD) simulations \citep{gammie2003,devilliers2003}. These calculations capture the self-consistent evolution of the magnetic field, which drives accretion via the magnetorotational instability \citep{mri} and extracts black hole spin energy to power relativistic jets \citep[][BZ]{blandfordznajek}. Past models of Sgr A* based on such calculations have found submm spectra \citep[e.g.,][]{noble2007,moscibrodzka2009,shcherbakov2012,moscibrodzka2014}, variability \citep{dexter2009,dexter2010,chan2015var}, source sizes \citep{moscibrodzka2009,dexter2010,shcherbakov2012}, and polarization \citep{shcherbakov2012,gold2017} consistent with observations. Similar to analytic models \citep{falcke2000shadow,bromley2001,broderick2006}, they generically find that the submm and shorter wavelength emission originates near the event horizon, resulting in  ``crescent'' shaped images.

Many of those calculations made two important, simplifying assumptions regarding the radiating electrons: i) that they are thermally distributed in energy, despite low densities implying a collisionless plasma \citep[e.g.,][]{mahadevan1997}; and ii) that the electron to proton temperature ratio is a constant value \citep{goldston2005}. Alternative, physically motivated prescriptions have instead put more of the available internal energy in electrons where the magnetic fields are stronger \citep{moscibrodzka2013,chan2015image} or according to kinetic prescriptions taking into account anisotropic viscosity \citep{sharma2007,shcherbakov2012}. \citet{anantua2020} recently studied a wide range of electron temperature prescriptions.

Those prescriptions were applied in post-processing to calculate the electron energy density (temperature) from single fluid MHD simulations. GRMHD algorithms can now evolve a separate electron fluid, which receive a fraction of the local dissipated energy according to a chosen sub-grid prescription \citep{ressler2015}. Such models self-consistently heat electrons, and can incorporate more directly results from kinetic calculations \citep[e.g.,][H10, R17, W18, K19]{howes2010,rowan2017,werner2018,kawazura2019,zhdankin2019}. \citet{ressler2017} showed that ``disk-jet'' models are a natural outcome of turbulent heating prescriptions based on gyrokinetic theory, where the electrons are heated preferentially in strongly magnetized regions (H10). \citet{chael2018} showed that electron heating is more uniform in alternative scenarios based on heating mediated by magnetic reconnection (R17).

Here we carry out a parameter survey, expanding the range of electron heating models and magnetic field strength compared to previous work (\autoref{sec:sims}). We constrain radiative models (\autoref{sec:models}) using the updated submm to NIR polarized spectrum, total intensity rms variability, and 86/230 GHz image size (\autoref{sec:observations}). We show (\autoref{sec:results}) that two combinations of heating prescriptions and magnetic field strengths can explain the broad submm peak and large amplitude NIR variability of Sgr A*. Those correspond to disk-jet models considered in previous work \citep{moscibrodzka2014,ressler2017}, as well as magnetically arrested accretion flow scenarios \citep[MADs,][]{shcherbakov2013,gold2017}. Our disk-jet models underproduce the observed submm linear polarization, and so are disfavored. We show that very long duration GRMHD simulations can produce the observed Faraday rotation measure of Sgr A*. Finally, we discuss the prospects of future observations and improvements to the theoretical models (\autoref{sec:discussion}). 

\begin{table*}
	\centering
	\caption{Parameters and convergence criteria of GRMHD simulations averaged over $8-10\times10^3 \,r_g/c$.}
	\label{tab:sim_table}
	\begin{tabular}{llccccccc} 		\hline
magnetic field & spin parameter & $\phi_{\rm BH}$ & $\langle b_r b^r \rangle / \langle b_\phi b^\phi \rangle$ & $\langle Q^{(\theta)} \rangle$ & $\langle Q^{(\phi)} \rangle$ & $\langle \beta \rangle$ & $r_{\rm eq}$ ($r_g$)\\
\hline
SANE & 0 & 7.3 & 0.14 & 16.3 & 23.7 & 19.9 & 19.4\\
SANE & 0.5 & 7.9 & 0.16 & 16.3 & 21.4 & 26.7 & 17.8\\
SANE & 0.9375 & 8.6 & 0.19 & 19.6 & 24.4 & 27.1 & 19.8\\
MAD & 0 & 49.7 & 0.21 & 55.6 & 56.4 & 7.9 & 31.1\\
MAD & 0.5 & 72.4 & 0.34 & 87.9 & 67.0 & 8.0 & 25.6\\
MAD & 0.9375 & 56.5 & 0.41 & 138.1 & 95.7 & 5.8 & 34.6\\
		\hline
	\end{tabular}
\end{table*}

\begin{table*}
\centering
	\caption{Time evolution of convergence criteria, magnetic field strength, and inflow equilibrium radius for our long duration SANE $a=0$ simulation.}
	\label{tab:sim_long_sane}
\begin{tabular}{lccccccc}
\hline
\hline

time ($10^3\,r_g/c$) & $\phi_{\rm BH}$ & $\langle b_r b^r \rangle / \langle b_\phi b^\phi \rangle$ & $\langle Q^{(\theta)} \rangle$ & $\langle Q^{(\phi)} \rangle$ & $\langle \beta \rangle$ & $r_{\rm eq}$ ($r_g$)\\
\hline
$5-10$ & 7.9 & 0.13 & 15.2 & 21.5 & 23.0 & 17.8 \\
$10-25$ & 6.0 & 0.16 & 18.7 & 24.8 & 21.1 & 28.8 \\
$25-50$ & 6.7 & 0.17 & 22.6 & 29.8 & 17.8 & 41.0 \\
$50-100$ & 8.0 & 0.18 & 26.8 & 35.3 & 14.5 & 61.3 \\
$100-200$ & 12.8 & 0.19 & 30.1 & 38.8 & 13.3 & 89.0 \\
\hline
\end{tabular}
\end{table*}

\begin{table*}
\centering
	\caption{Time evolution of convergence criteria, magnetic field strength, and inflow equilibrium radius for our long duration MAD $a=0.9375$ simulation.}
	\label{tab:sim_long_mad}
\begin{tabular}{lccccccc}
\hline
\hline

time ($10^3\,r_g/c$) & $\phi_{\rm BH}$ & $\langle b_r b^r \rangle / \langle b_\phi b^\phi \rangle$ & $\langle Q^{(\theta)} \rangle$ & $\langle Q^{(\phi)} \rangle$ & $\langle \beta \rangle$ & $r_{\rm eq}$ ($r_g$)\\
\hline
$5-10$ & 61.4 & 0.37 & 112.5 & 91.3 & 6.4 & 13.4 \\
$10-20$ & 59.2 & 0.37 & 111.4 & 92.6 & 6.5 & 30.0 \\
$20-30$ & 62.7 & 0.38 & 125.9 & 100.2 & 7.8 & 52.5 \\
$30-40$ & 62.7 & 0.34 & 112.7 & 96.1 & 8.2 & 66.7 \\
$40-60$ & 62.6 & 0.40 & 141.8 & 103.9 & 7.5 & 65.9 \\
\hline
\end{tabular}
\end{table*}

\begin{figure*}
\begin{tabular}{cc}
\includegraphics[width=0.48\textwidth]{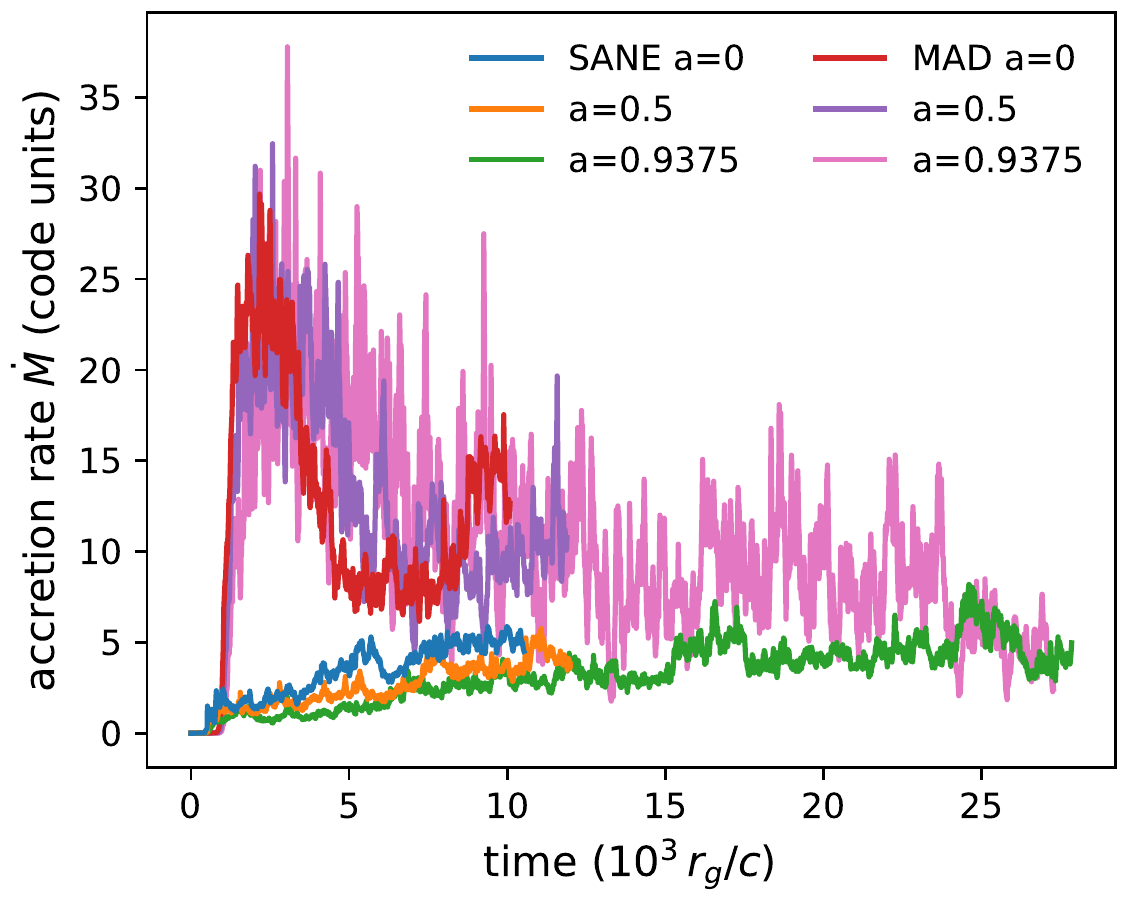} &
\includegraphics[width=0.48\textwidth]{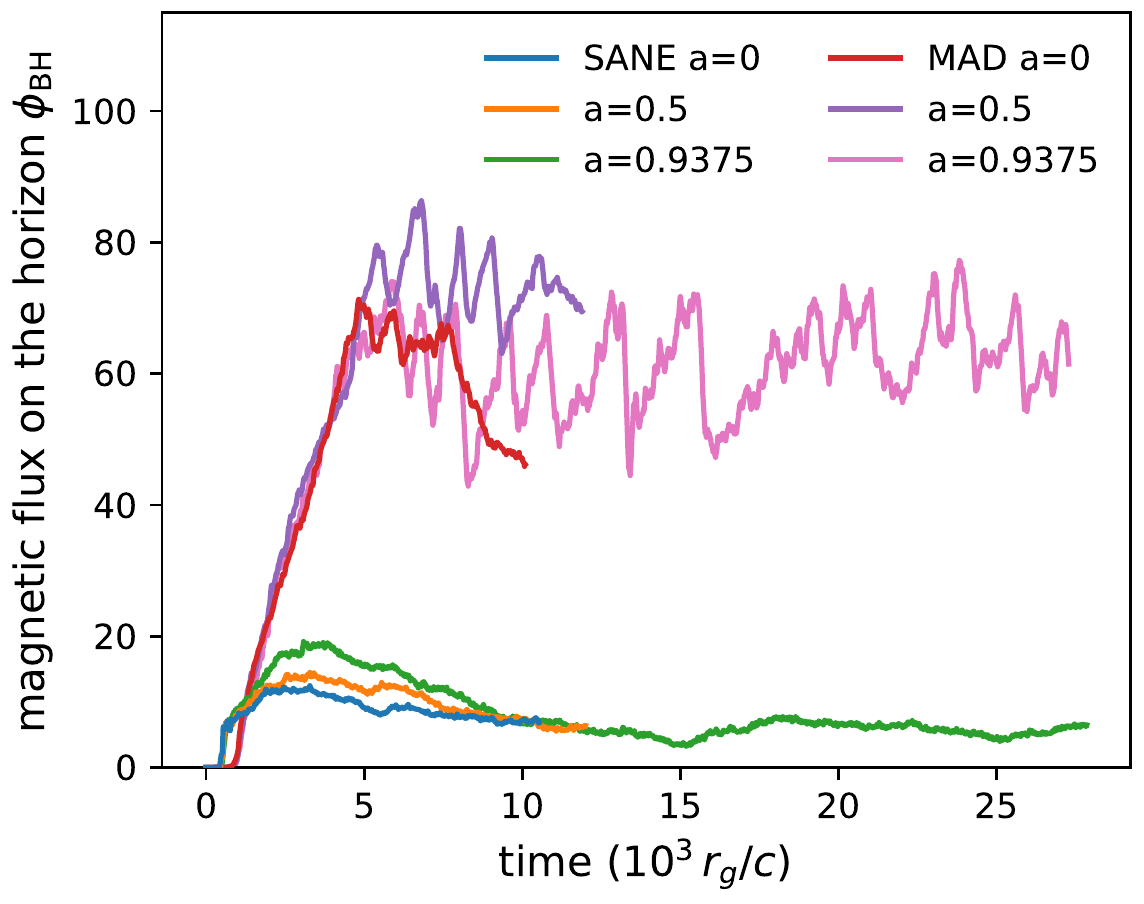}
\end{tabular}
\caption{\label{fig:sim_plots_t}Accretion rate and dimensionless magnetic flux on the black hole event horizon $\phi_{\rm BH}$ as a function of time for the simulations used here. The SANE and MAD limits are reached as intended, with low and saturated dimensionless magnetic flux accumulated on the black hole. In the SANE case, the accretion rate remains steady or rises as the large initial torus drains. In the MAD case, strong magnetic fields lead to rapid accretion of the torus.}
\end{figure*}

\begin{figure*}
\begin{tabular}{cc}
\includegraphics[width=0.48\textwidth]{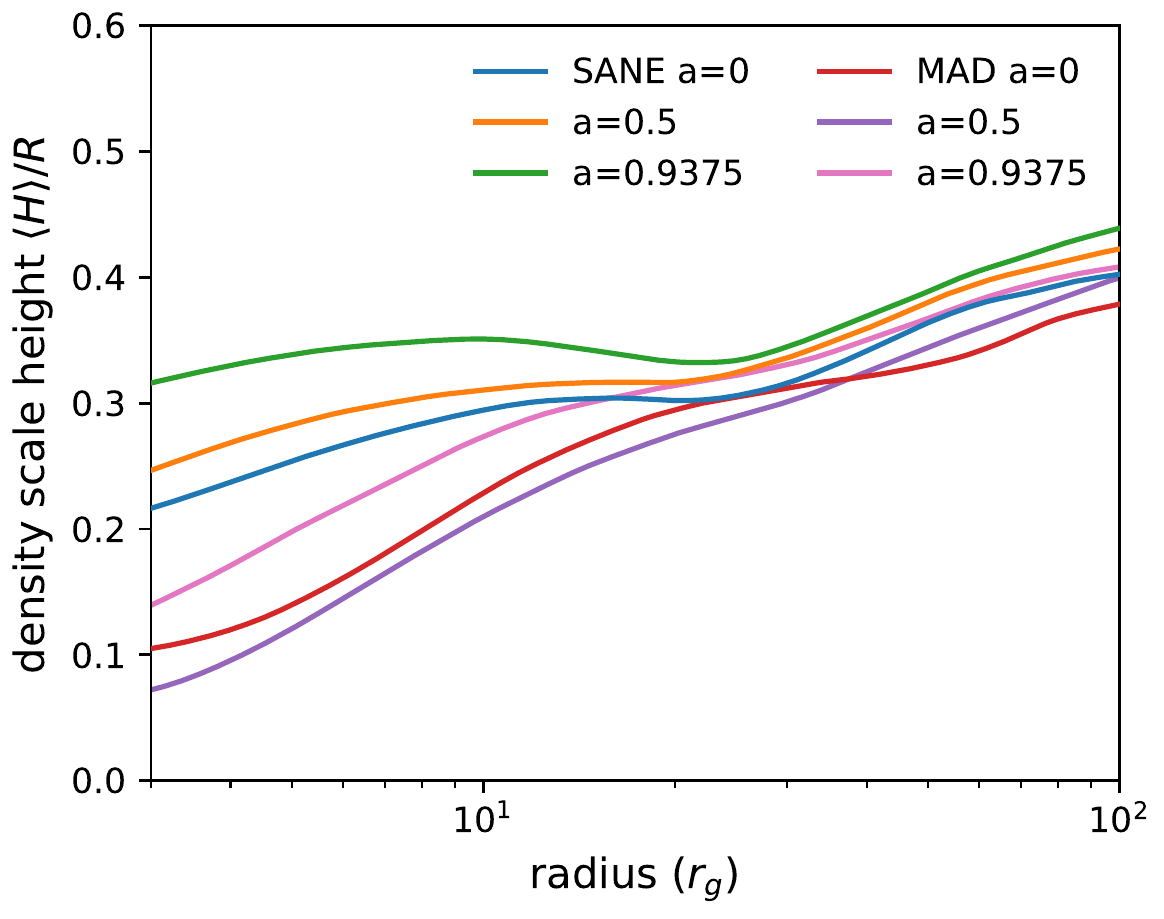} &
\includegraphics[width=0.48\textwidth]{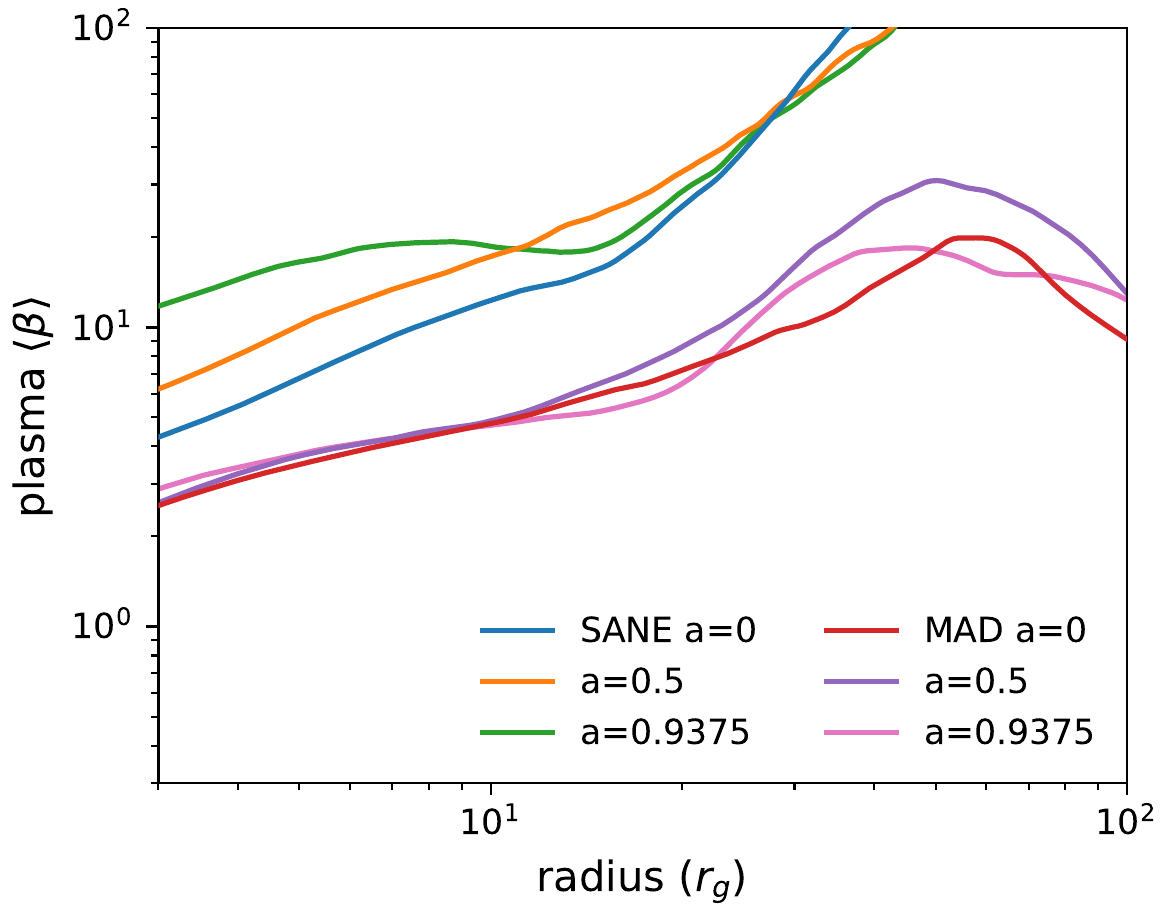}
\end{tabular}
\caption{\label{fig:sim_plots_r}Radial profiles of density scale height and plasma $\beta$ for the simulations used here. SANE simulations show roughly constant scale height in the region of inflow equilibrium ($r \lesssim 20 r_g$) with plasma $\beta \simeq 10-30$. MADs are geometrically compressed at small radii by the wide jet base \citep{mckinney2012} and are more strongly magnetized with lower values of $\beta < 10$.}
\end{figure*}

\begin{figure}
    \centering
    \includegraphics[width=0.48\textwidth]{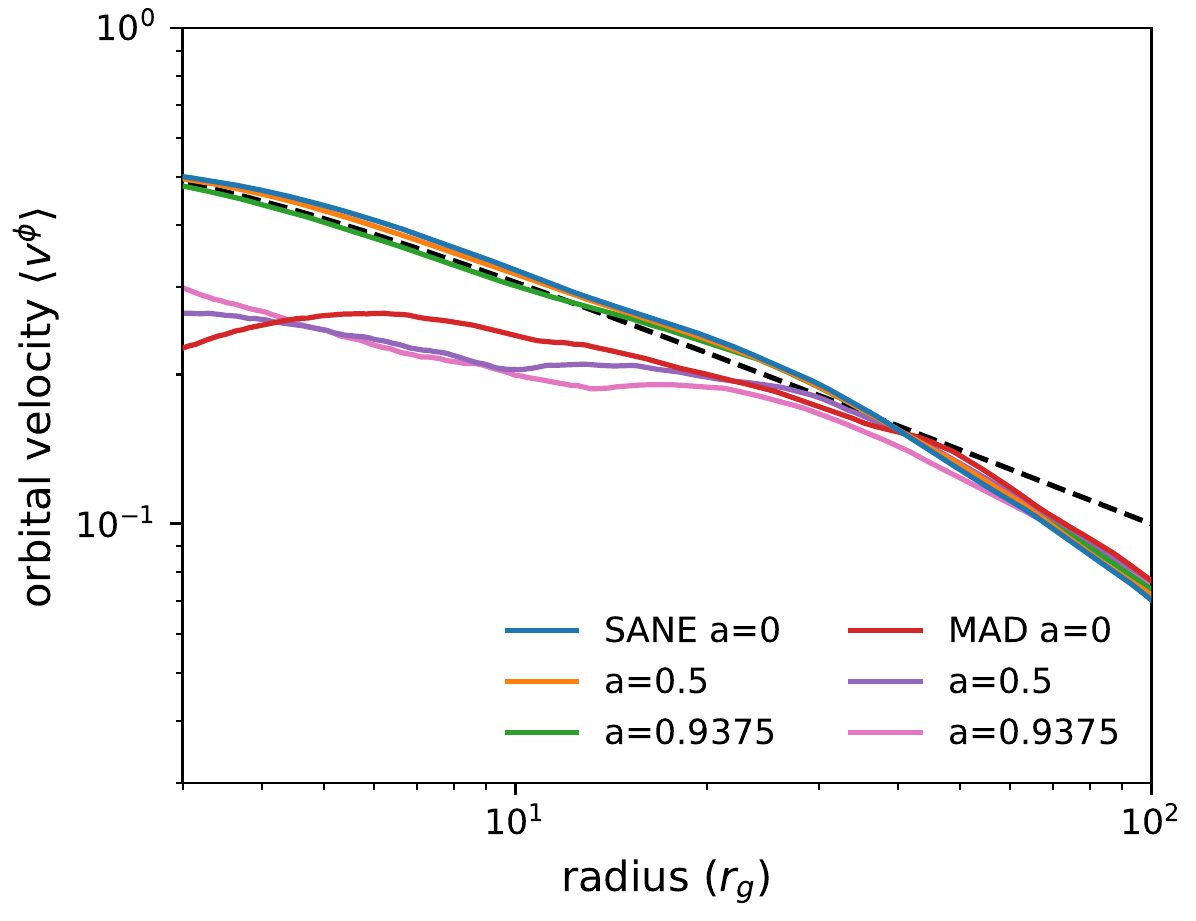}
    \caption{Radial profile of azimuthal velocity for our simulations. The black line shows the Keplerian coordinate velocity for a spin of $a=0.9375$. SANE simulations show Keplerian rotation, while MADs are significantly sub-Keplerian in the inner regions as a result of magnetic pressure support.}
    \label{fig:sim_plots_vphi}
\end{figure}

\begin{figure*}
\begin{tabular}{cc}
\includegraphics[width=0.48\textwidth]{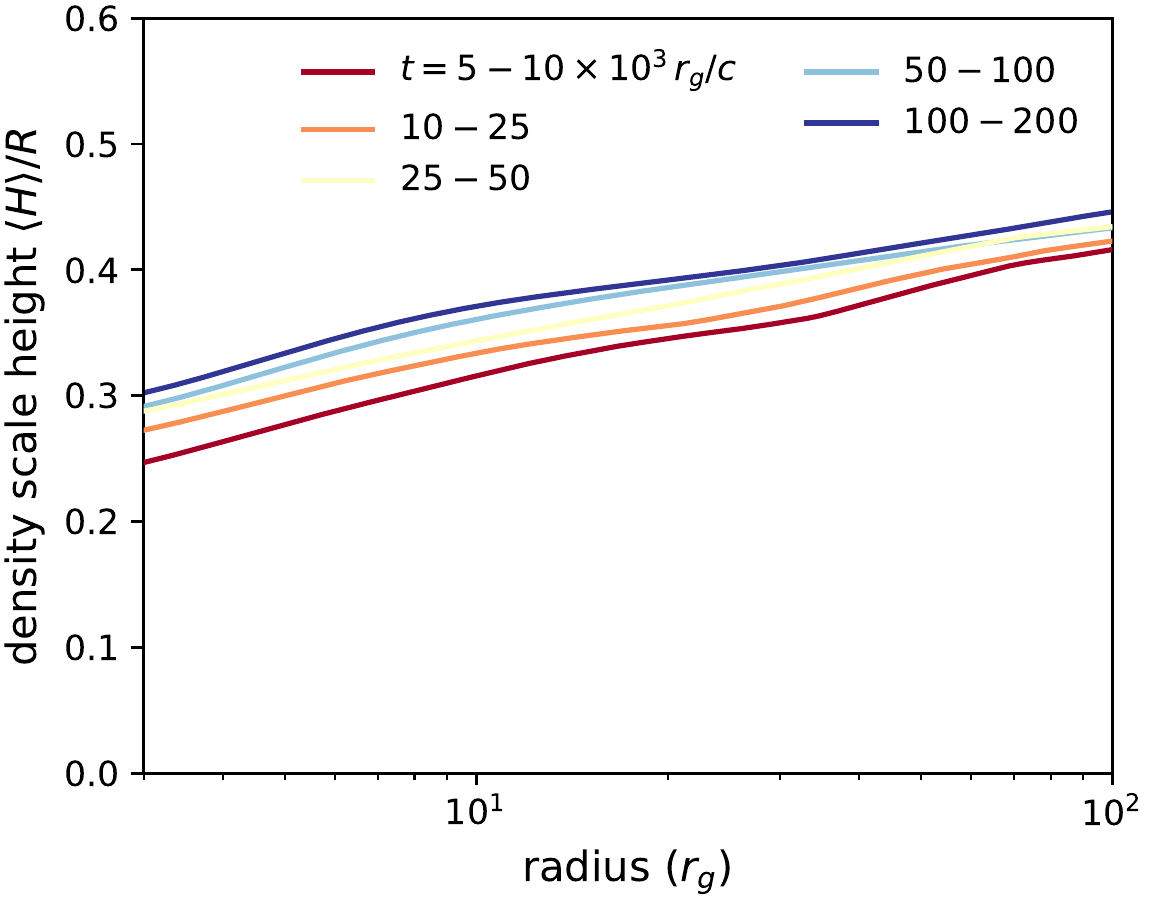} &
\includegraphics[width=0.48\textwidth]{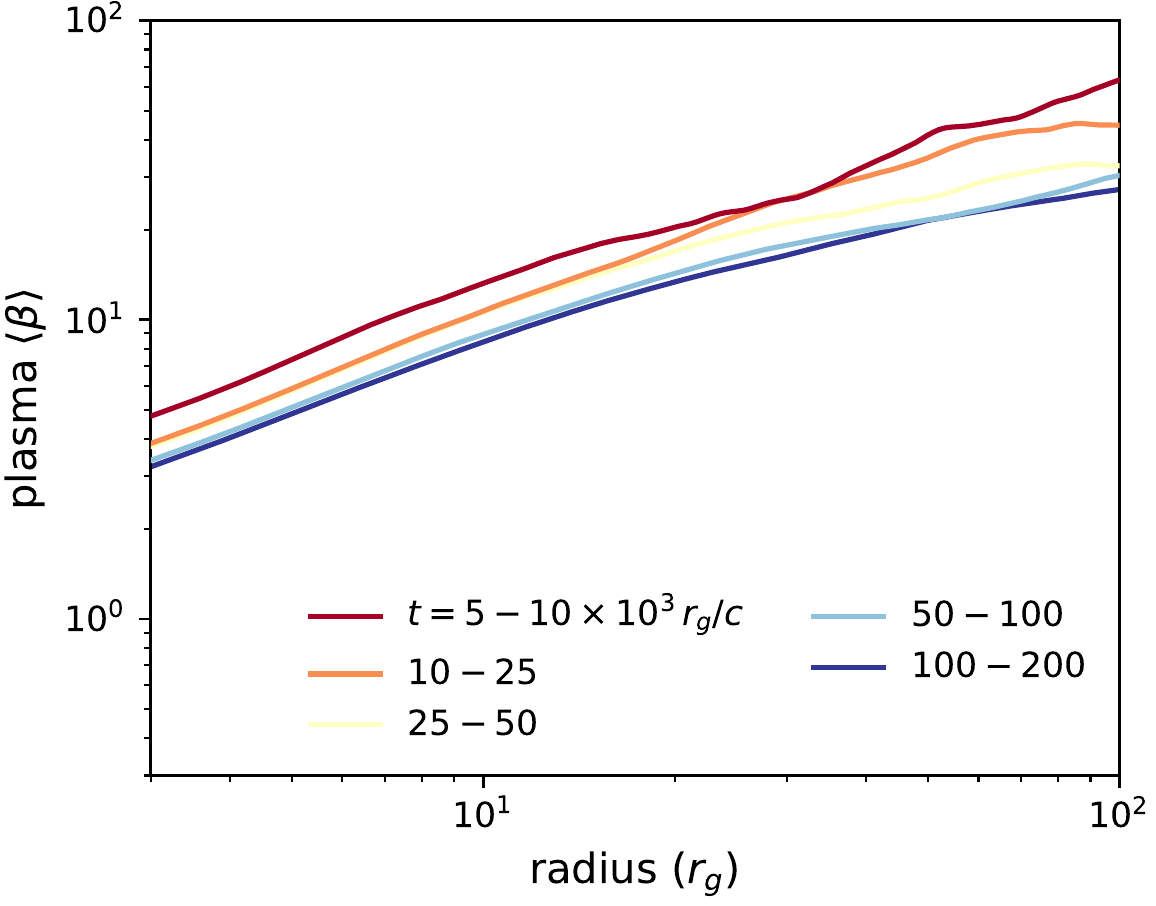}\\
\includegraphics[width=0.48\textwidth]{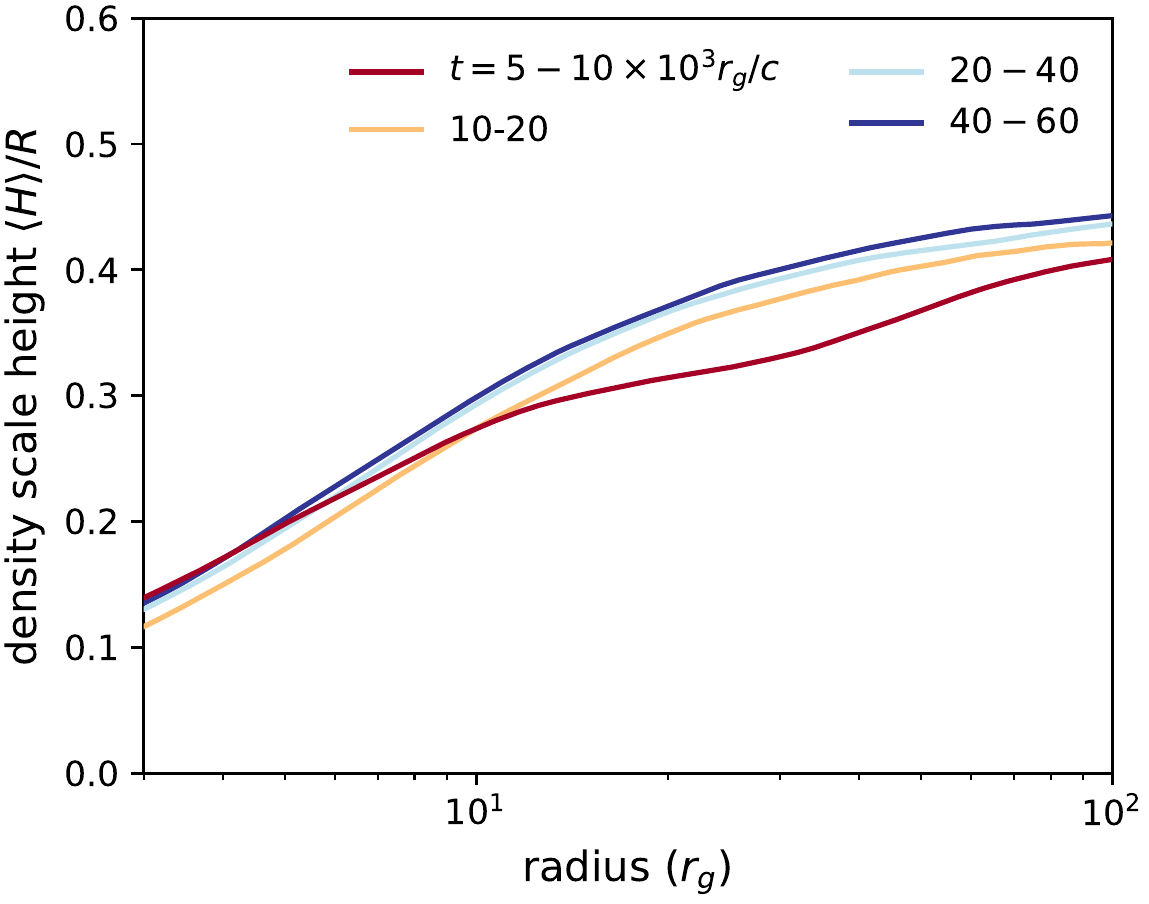} &
\includegraphics[width=0.48\textwidth]{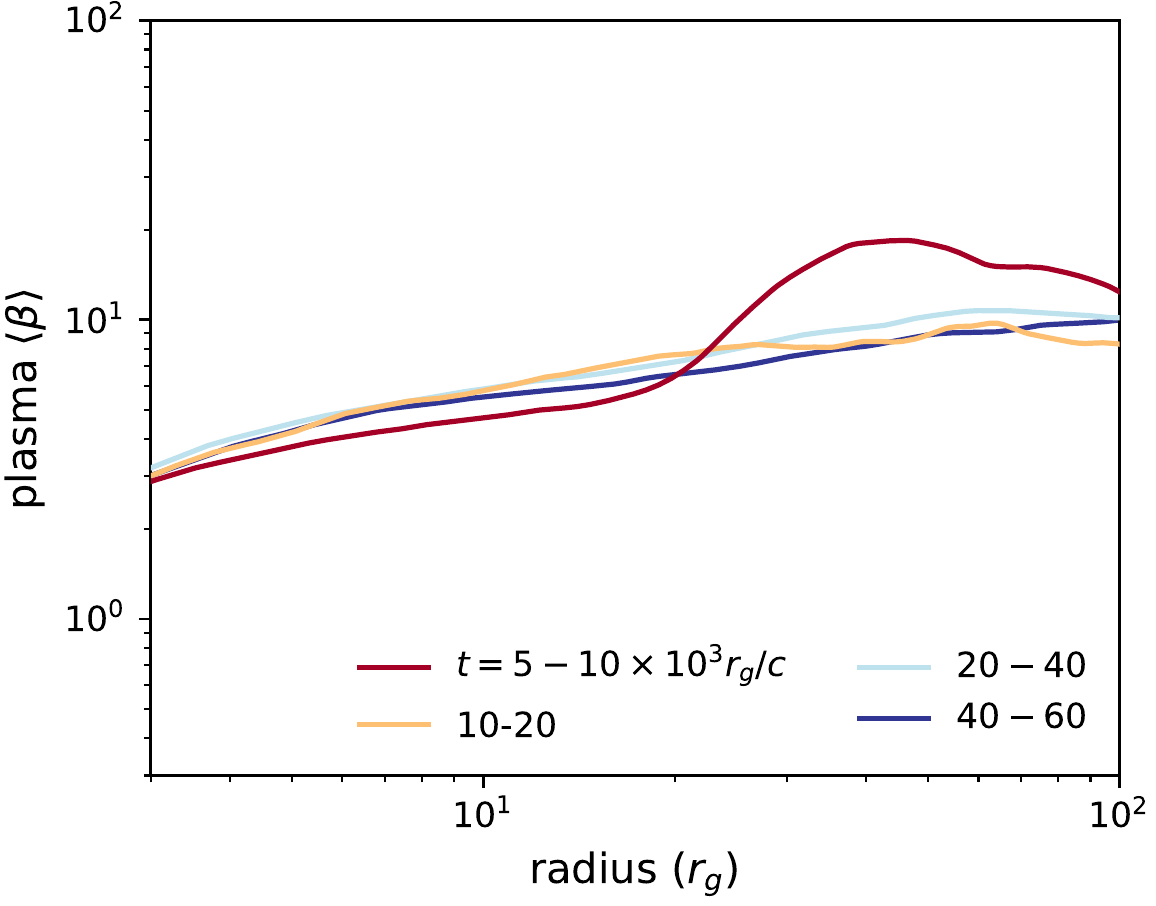}
\end{tabular}
\caption{\label{fig:sim_plots_r_long}Radial profiles of density scale height and plasma $\beta$ averaged over early (red) to late (blue) time intervals for the long duration SANE (top) and MAD (bottom) simulations used here. Over time the average scale height and magnetization increase, at all radii for SANE simulations and at larger radii for MAD simulations.}
\end{figure*}

\begin{figure*}
	\includegraphics[width=\textwidth]{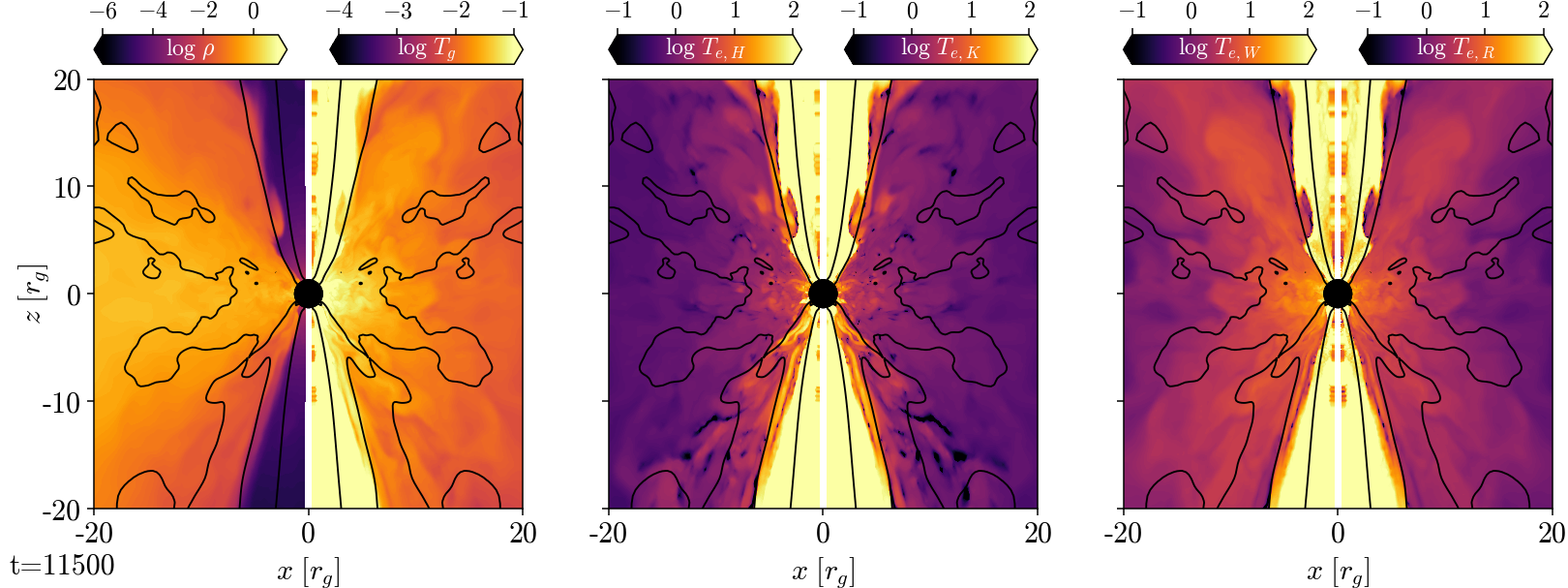}
    \caption{Snapshot azimuthal averages from the SANE a=0.9375 simulation of $\rho$, $T_{\rm gas}$ (left, in units of $m_p c^2 / k$), and $T_e$ for the four electron heating schemes (middle and right panels, H10, K19, W18, R17, in units of $m_e c^2 / k$). The H10/K19 (``turbulence'') and W18/R17 (``reconnection'') pairs show similar behavior. The turbulence models heat electrons significantly only in polar jet regions where the magnetization is high, while the reconnection models also substantially heat the dense accretion flow near the midplane.}
    \label{fig:3panelsane}
\end{figure*}

\begin{figure*}
	\includegraphics[width=\textwidth]{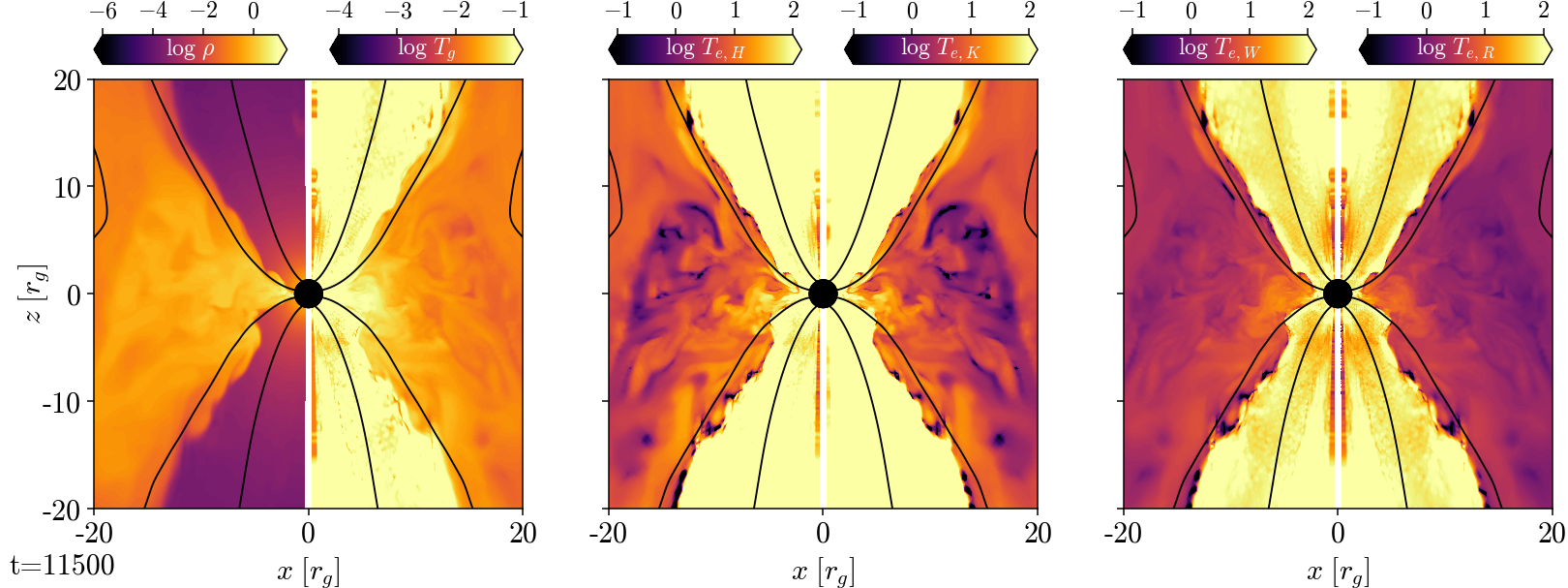}
    \caption{Snapshot azimuthal averages from the MAD a=0.9375 simulation of $\rho$, $T_{\rm gas}$, and $T_e$ for the 4 electron heating schemes (H10, K19, W18, R17). Again the turbulence and reconnection pairs show very similar behavior. Here, there is strong electron heating near the midplane in both scenarios due to regions of high magnetization in the MAD accretion flow.}
    \label{fig:3panelmad}
\end{figure*}

\section{GRMHD simulations with multiple electron heating prescriptions}
\label{sec:sims}

We have carried out a set of 3D GRMHD black hole accretion simulations using the \texttt{harmpi}\footnote{\url{https://github.com/atchekho/harmpi}} code \citep{harmpiascl}. \texttt{harmpi} is a 3D implementation of the \texttt{HARM} algorithm \citep{gammie2003,noble2006} for conservative MHD in a fixed spacetime. Simulations were initialized from a Fishbone-Moncrief steady state hydrodynamic equilibrium torus in a Kerr spacetime with inner radius $r_{\rm in} = 12 r_g$ and pressure maximum at $r_{\rm max} = 25 r_g$. The torus was seeded with a single loop of poloidal magnetic field, whose radial profile is designed to supply either relatively modest (SANE) or maximal (MAD) magnetic flux. For each case, black hole spin values of $a = 0$, $0.5$, $0.9375$ were used. The simulations were carried out in modified spherical polar Kerr-Schild coordinates, with resolution concentrated towards the equatorial plane to resolve the accretion flow at small radius and towards the pole to resolve the jet at larger radius. The outer radial boundary was extended to $10^5 r_g$ using a super-exponential radial coordinate. The grid was chosen to provide a cylindrical innermost cell in polar angle \citep{tchekhovskoy2011,ressler2017}. This significantly increases the time step allowed by the Courant condition. The resolution used was $320\times256\times160$ cells, including the full $2\pi$ azimuthal domain. All simulations are evolved for at least $10^4 \, r_g/c$, and we analyze the period from $5\times10^3-10^4 \, r_g/c$, once the emission region has established inflow equilibrium. We also evolve one SANE $a=0$ and one MAD $a=0.9375$ simulation for a much longer time ($2\times10^5 \,r_g/c$ and $6\times10^4 \,r_g/c$, respectively), in order to allow larger radii $\simeq 100 \,r_g$ to reach inflow equilibrium. That location is thought to produce the bulk of the observed submm Faraday rotation from Sgr A* \citep{marrone2007,ressler2018}.

We use the version of \texttt{harmpi} with a separate electron fluid as implemented by \citet{ressler2015}. Magnetic and kinetic energy dissipated at the grid scale is recaptured as internal energy. A fraction is also assigned to a separate electron internal energy. This electron internal energy is evolved independently of the fluid dynamics, which allows us to incorporate multiple electron heating prescriptions within a single simulation. Some error is introduced in this approximation, since it effectively assumes that the electron and proton adiabatic indices are the same \citep[e.g., see][]{sadowski2017,ressler2017}.

The electron heating mechanism is uncertain, as is its sub-grid implementation. We explore a total of 4 prescriptions based on 2 physical scenarios. We use fitting formulae derived from gyrokinetic linear theory (H10) and numerical simulations (K19) of heating in a turbulent cascade, as well as fitting formulae derived from particle-in-cell simulations of electron heating in magnetic reconnection (R17, W18). 

We assess the convergence of our simulations using criteria from the literature \citep{hawley2011,shiokawa2012,hawley2013}. We define the 1D radial, density-weighted profile of a quantity $X$ as:

\begin{equation}
    \langle X \rangle = \frac{\int d\theta d\phi \sqrt{-g} \rho X}{\int  d\theta d\phi \sqrt{-g} \rho},
\end{equation}

\noindent where $\rho$ is the fluid mass density and $\sqrt{-g}$ is the Jacobian. We define the shell-averaged plasma $\beta = p_{\rm gas}/p_B$ as,

\begin{equation}
\langle \beta \rangle = 8\pi \frac{\langle p_{\rm gas} \rangle}{\langle b^2 \rangle}.
\end{equation}

\noindent We evaluate simulation resolution quality using $Q$ values \citep{hawley2011} calculated in the locally non-rotating frame (LNRF), $Q^{(i)} = \lambda_{\rm MRI}^{(i)} / \Delta x^{(i)}$ \citep{porth2019}, where $\lambda_{\rm MRI}$ is the fastest growing MRI wavelength,

\begin{equation}
    \lambda_{\rm MRI}^{(i)} = \frac{2\pi \, b^\mu e_\mu^{(i)}}{\sqrt{\rho+\gamma u+b^2}},
\end{equation}

\noindent $u$ is the fluid internal energy and $\gamma$ is the adiabatic index. The tetrad vectors $e_\mu^{(i)}$ describe the transformation to the LNRF \citep[e.g.,][]{takahashi2008}, and $\Delta x^{(i)} = \Delta x^\mu e_\mu^{(i)}$ is the LNRF grid cell spacing. Finally we report on the relative strength of the radial and azimuthal field, here taken to be Kerr-Schild coordinate values of $b_r b^r / b_\phi b^\phi$.

\section{Radiative models of Sgr A*}
\label{sec:models}

From each simulation and electron heating prescription, we compute radiative models of Sgr A*. The electron temperature is taken directly from the GRMHD electron internal energy density. We then scale the mass density until the median observed flux density at $230$ GHz is $\simeq 3$ Jy \citep[e.g.,][]{dexter2014,bower2015}. We exclude emission from regions where $\sigma = b^2/\rho > 1$.

Observables are calculated using a ray tracing method with the public code \texttt{grtrans} \citep{dexteragol2009,dexter2016}. We follow Kerr photon geodesics corresponding to uniformly sampled camera pixels of a distant observer at a viewing orientation of $i=25$, $45$, and $60$ degrees. The image resolution is $192$ pixels over a $42 \, r_g$ ($210 \, \mu$as) field of view. Rays are sampled evenly in $1/r$ from an outer boundary $r_{\rm out}$ until they either reach the black hole event horizon or return to $r_{\rm out}$. Near radial turning points, we switch to sampling evenly in $\cos{\theta}$ to avoid taking large steps. Along each ray, we solve the full polarized radiative transfer equations for synchrotron emission and absorption and Faraday rotation and conversion, assuming a purely thermal electron distribution function. Coefficients are taken from \citet{dexter2016}, with the Faraday rotation coefficient $\rho_V$ modified to correctly reproduce the non-relativistic limit (see appendix \ref{app:rhov_appendix}). This is important for calculations of the Faraday rotation measure through the extended torus where the dimensionless electron temperature $\theta_e = kT_e/mc^2 \lesssim 1$. We do not include inverse Compton scattering, which allows comparisons to the observed X-ray luminosity \citep[e.g.,][]{dolence2009,moscibrodzka2009}. We calculate images at radio to NIR frequencies for all snapshots from $(5-10)\times10^3 \, r_g/c$ spaced by $10 \, r_g/c$. Here we focus on results for time-averaged observables and their rms variability. 

\section{Observational constraints}
\label{sec:observations}

We compare our models to observational constraints derived from millimeter to NIR observations of Sgr A*.

\subsection{Spectrum and variability}
\label{sec:spectra_data}

First, we consider the shape of the submm to NIR total intensity spectrum  \citep[e.g.,][]{falcke1998,bower2015,stone2016,vonfellenberg2018,bower2019,schoedel2011,doddseden2011,witzel2018} and the rms variability fraction as a function of wavelength \citep{doddseden2011,witzel2018,dexter2014,bower2015}.

Quantitatively, we require a submm spectral index between 230 and 690 GHz of $-0.35 - 0.25$ \citep{marronephd,bower2019} and an upper limit to the median NIR flux density of $< 1.4$ mJy \citep[][GRAVITY collaboration 2020, submitted]{doddseden2011,witzel2018}, and a $230$ GHz total flux density rms of $20-40\%$ for a model to be considered viable.

\subsection{86 and 230 GHz source sizes and NIR centroid motion}

We also consider constraints on the image size at 86 and 230 GHz \citep{krichbaum1998,doeleman2008,fish2011,johnson2018,issaoun2019}. We adopt semi-major axis size constraints of $a_{\rm 86} = 86-154\,\mu$as \citep{issaoun2019} and $a_{\rm 230} = 51-63\,\mu$as \citep{johnson2018}. In the latter case, we have for simplicity assumed that the intrinsic source semi-major axis aligns with that of the interstellar scattering kernel. 
We further compare our results with the evolution seen in the NIR centroid \citep{gravityflare}.

\subsection{Polarization}

We also compare to median submm observed linear and circular polarization fractions \citep{aitken2000,eckart2006,trippe2007,munoz2012,liu2016,bower2018,gravityflare}. Specifically, we enforce constraints on the median polarization fractions of $LP = 2-8\%$ \citep{bower2018} and $|CP| = 0.5-2\%$ \citep{munoz2012}. 

Sgr A* also shows a dependence of electric vector position angle EVPA $\propto \lambda^2$ as expected for Faraday rotation ``external'' to the emission region (so that the polarized source has its EVPA coherently rotated, e.g. appears point-like on the Faraday screen). The rotation measure is RM $\simeq -6\times10^5 \, \rm rad / \rm m^{-2}$ \citep{marrone2007}, with a consistent sign in measurements over many years \citep[e.g.,][]{bower2003,marrone2006,marrone2007,bower2018}. The RM is thought to result from the extended accretion flow, and has been used to constrain the accretion rate onto the black hole \citep{marrone2006}. We do not use the RM to select models, since it originates outside of the region of inflow equilibrium in standard GRMHD models. We show that that an RM signature can be generated with approximately the right magnitude in radiative models from our long duration simulations.

\begin{figure*}
	\includegraphics[width=0.7\textwidth]{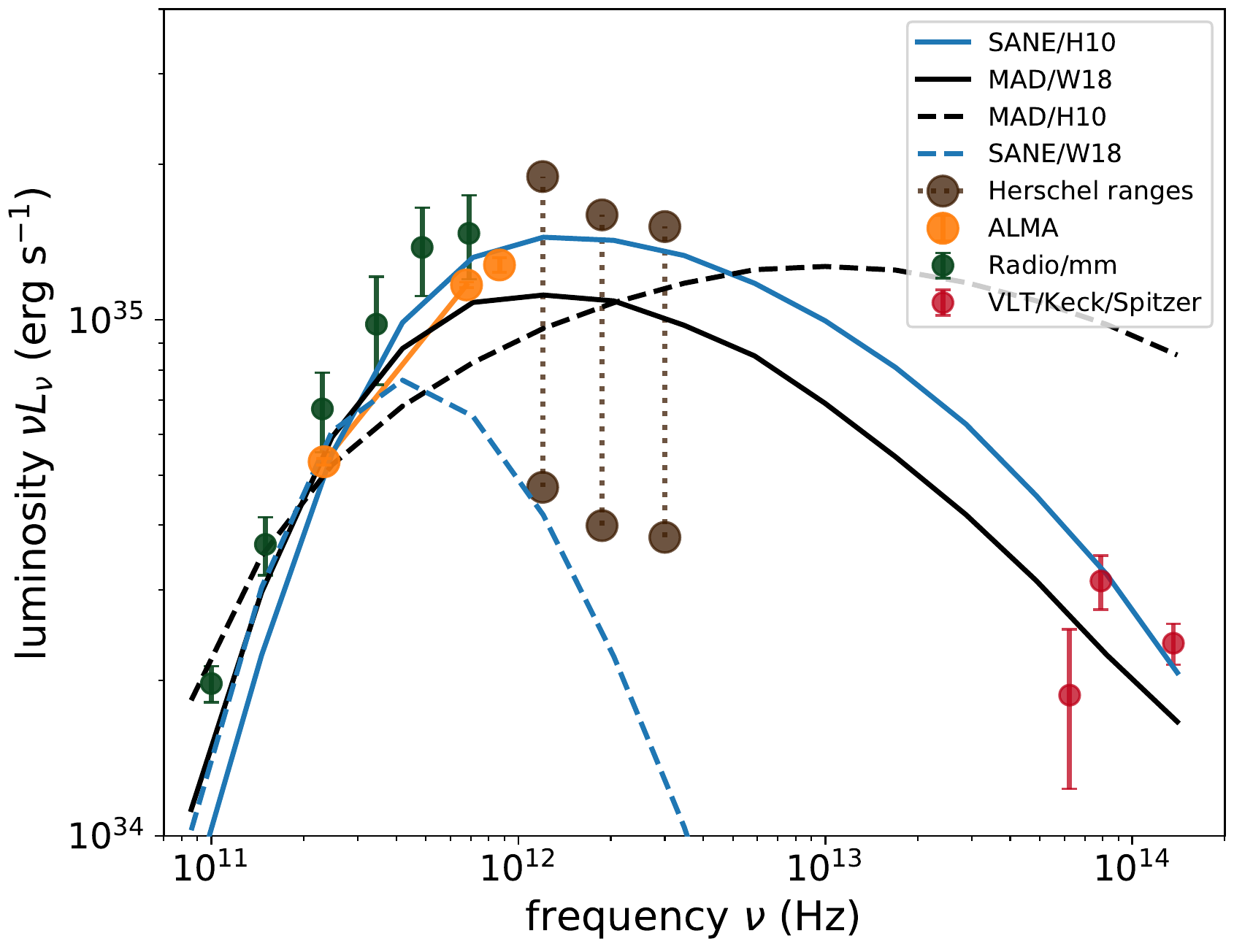}
    \caption{Median spectra from sample SANE (blue) and MAD (black) models with $a=0.5$ and $i=45^\circ$ and the H10 and W18 electron heating models compared to Sgr A* mm to NIR data. The SANE/H10 and MAD/W18 models are consistent with the observed spectral shape. The SANE/W18 model does not produce sufficiently hot electrons and can't explain the broad submm peak in Sgr A*. The MAD/H10 model produces too many hot electrons. It fails to match the submm peak and overproduces the observed NIR emission.}
    \label{fig:spectra_sample}
\end{figure*}

\begin{figure*}
\begin{tabular}{cc}
	\includegraphics[width=0.48\textwidth]{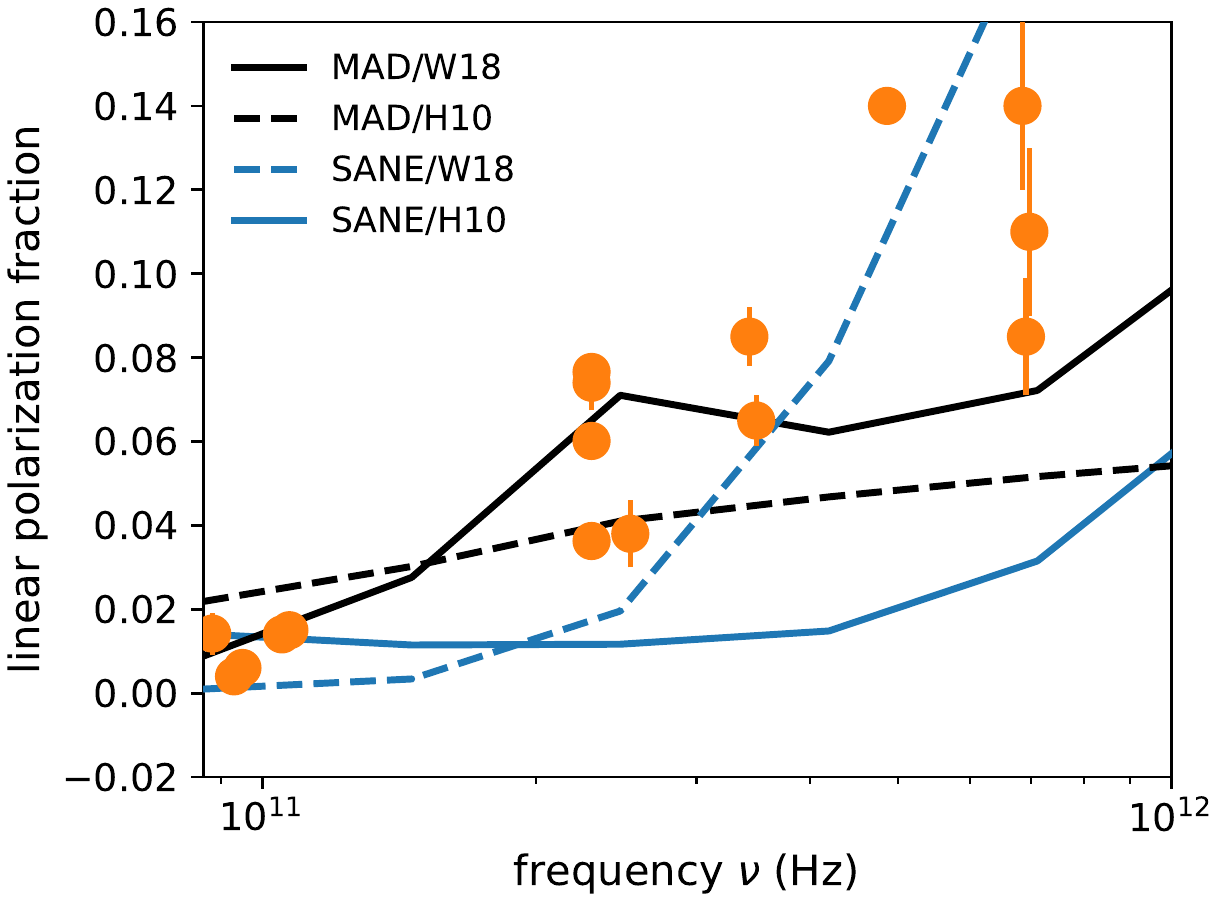} & 
	\includegraphics[width=0.46\textwidth]{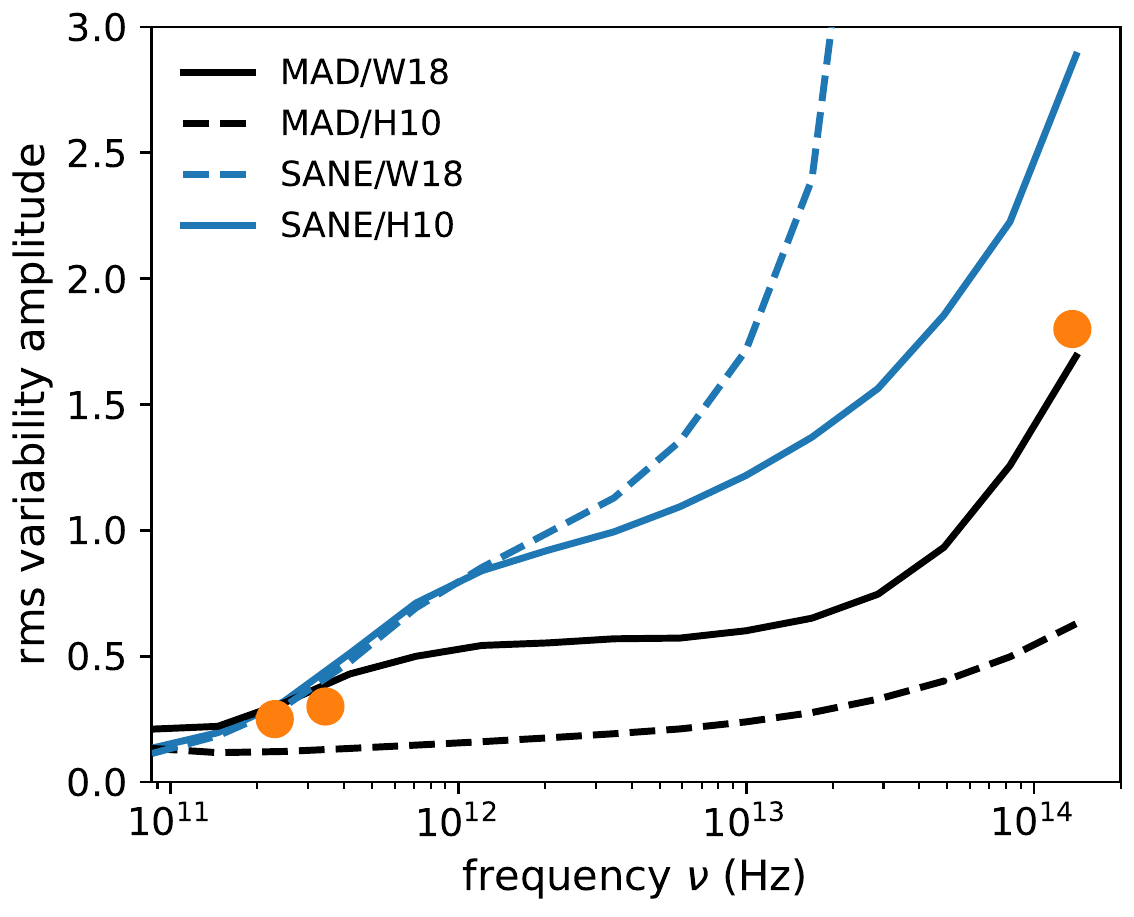}
	\end{tabular}
    \caption{Linear polarization fraction (left) and rms variability amplitude (right) for sample SANE (blue) and MAD (black) simulations with $a=0.5$ and $i=45^\circ$ for the H10 and W18 electron heating models compared to Sgr A* mm to NIR data. The SANE/H10 (disk-jet) model is heavily depolarized as a result of Faraday rotation in the emission region. The MAD/W18 model is consistent with both constraints.}
    \label{fig:lprms_sample}
\end{figure*}

\begin{figure*}
\begin{tabular}{cc}
	\includegraphics[width=0.48\textwidth]{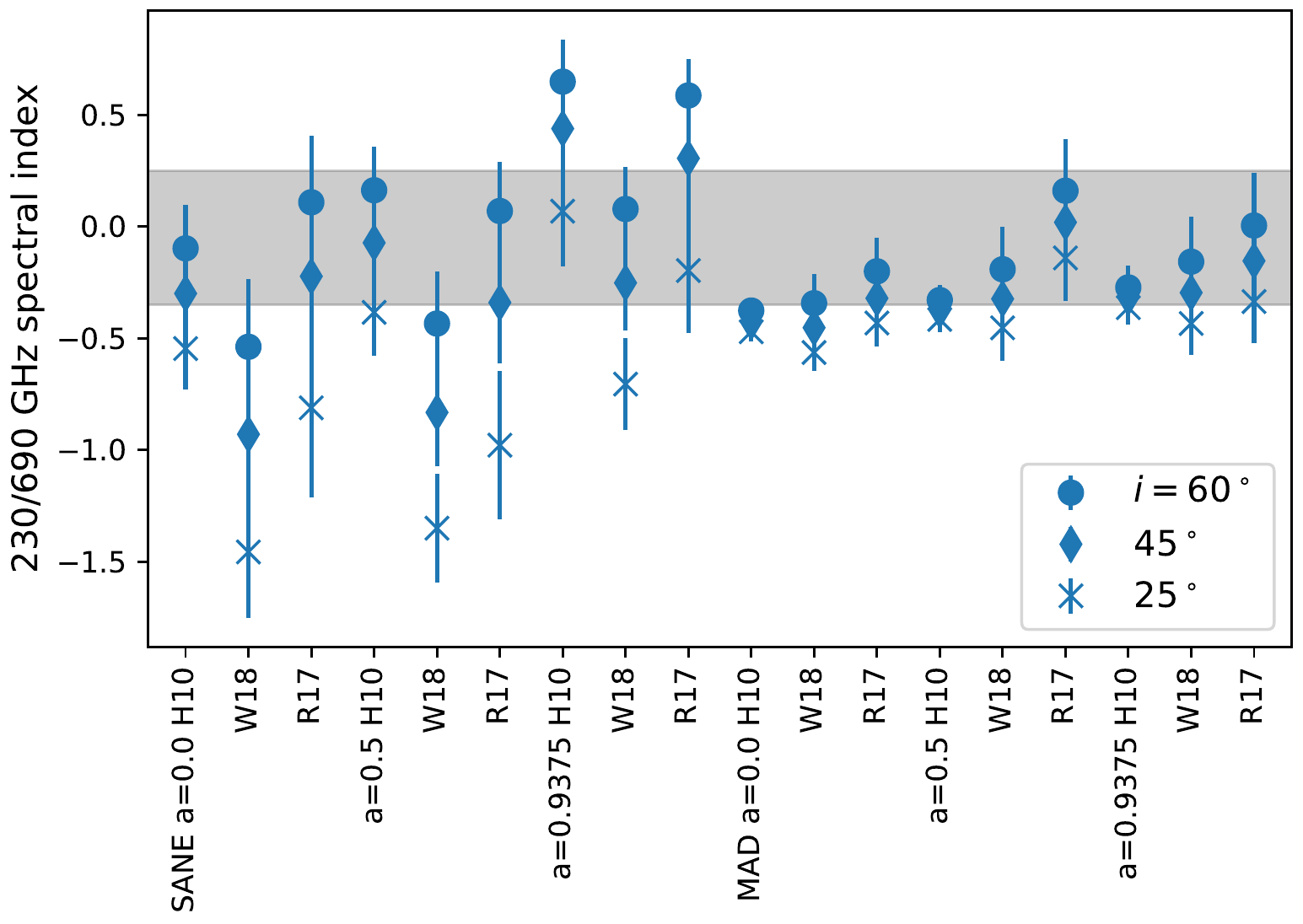} & 
	\includegraphics[width=0.48\textwidth]{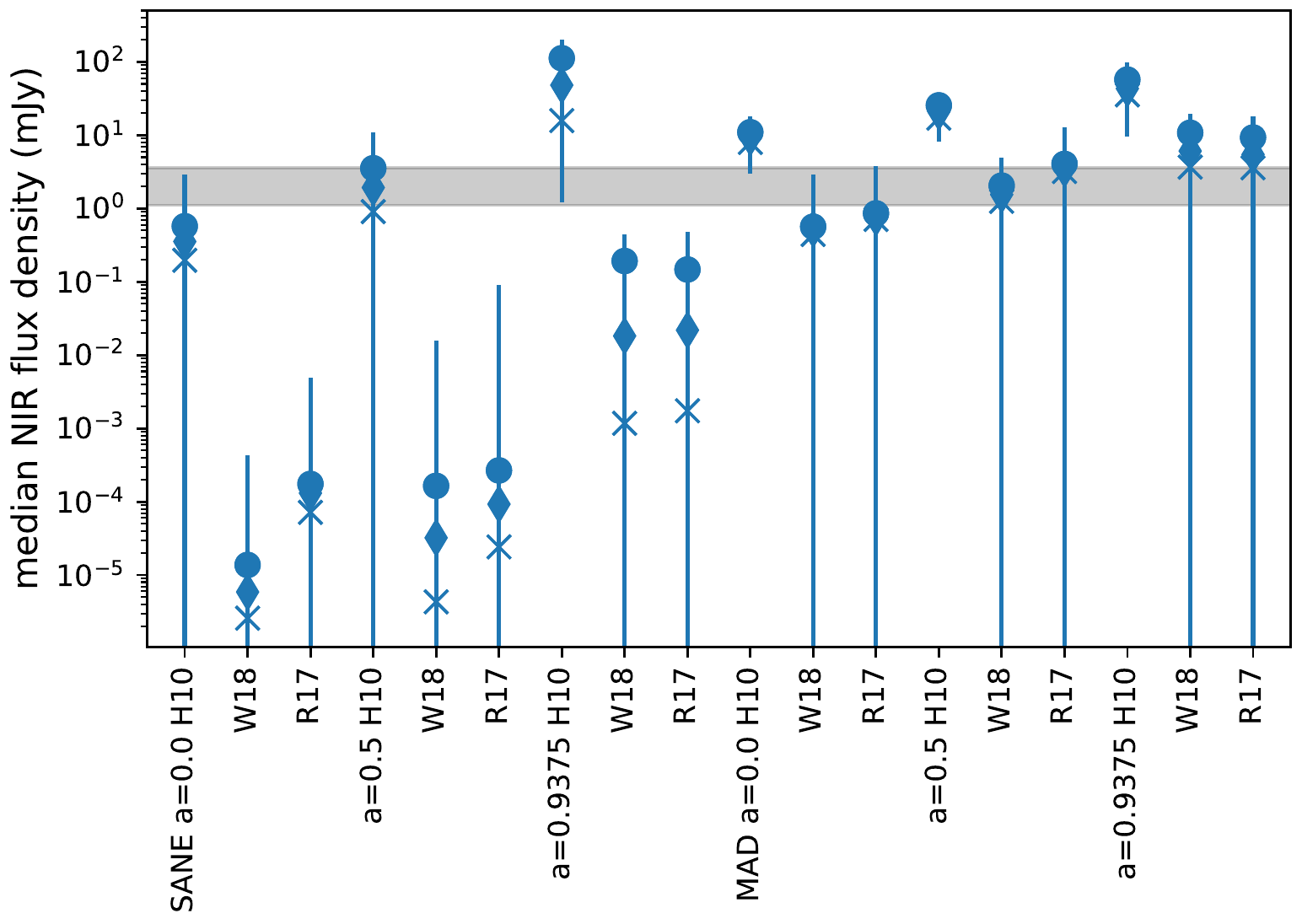}
	\end{tabular}
    \caption{Model median (points) and rms (error bars) submm spectral indices (left) and NIR flux densities (right). The different symbols for each model are 3 values of the observer inclination angle. The shaded regions correspond to the observed ranges, although we use the NIR flux density as an upper limit when constraining models. Higher inclinations correspond to larger submm spectral indices and NIR flux densities as a result of increased Doppler beaming. The SANE/reconnection models produce negligible NIR emission for the thermal distribution function assumed here.}
    \label{fig:alphafnir}
\end{figure*}

\begin{figure*}
\begin{tabular}{ll}
	\includegraphics[width=0.31\textwidth]{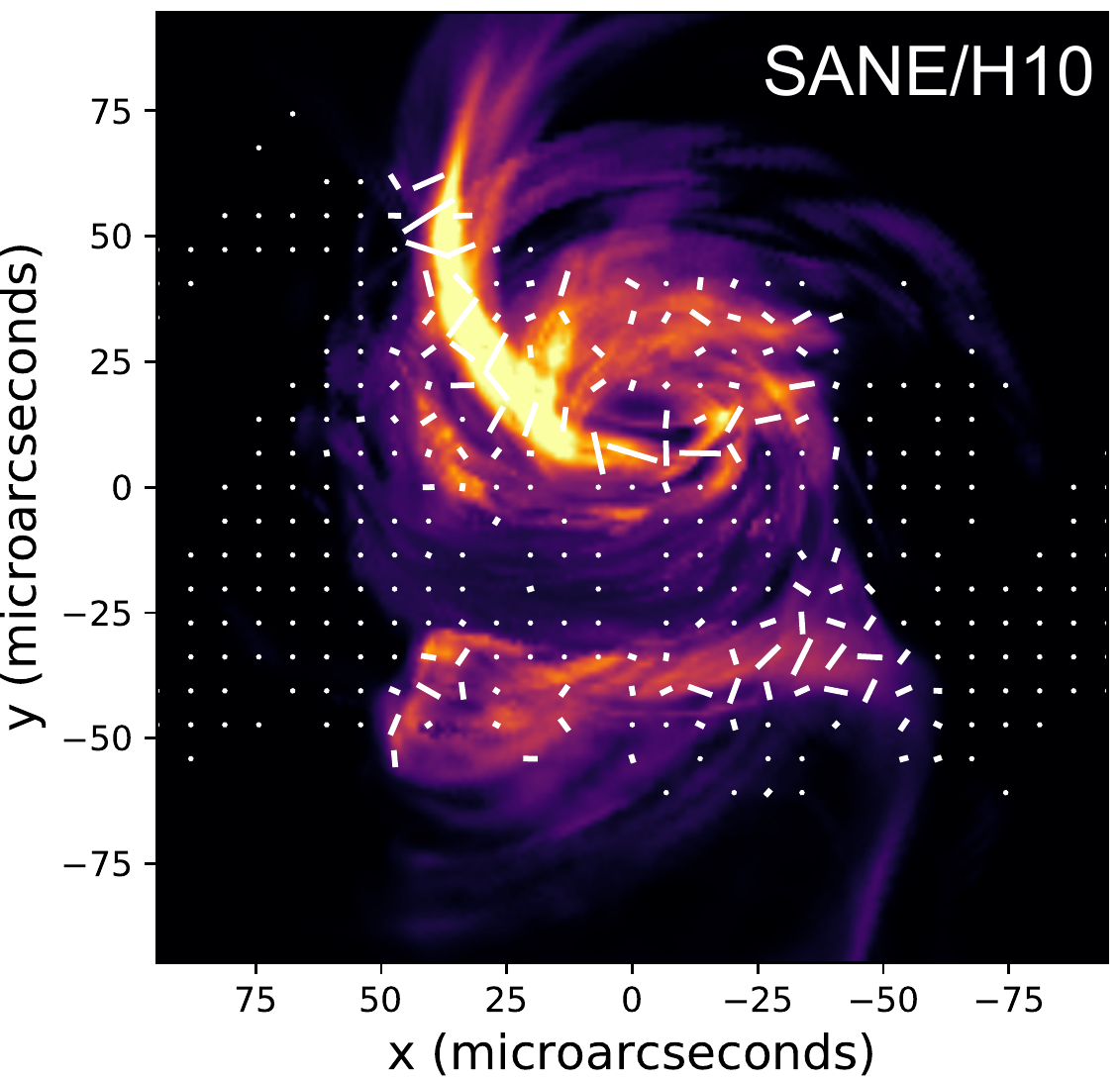} & 
			\includegraphics[width=0.31\textwidth]{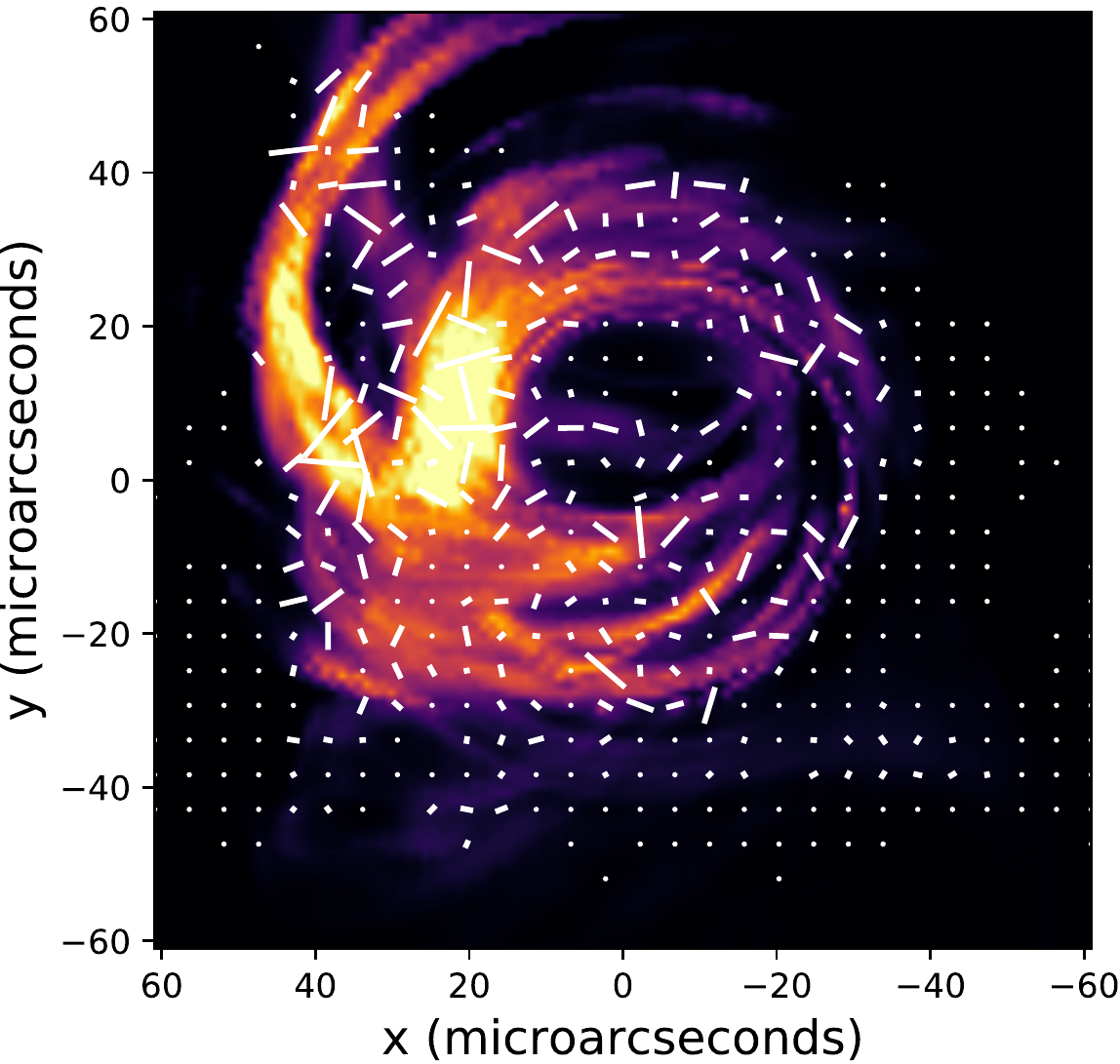}\\
	\includegraphics[width=0.31\textwidth]{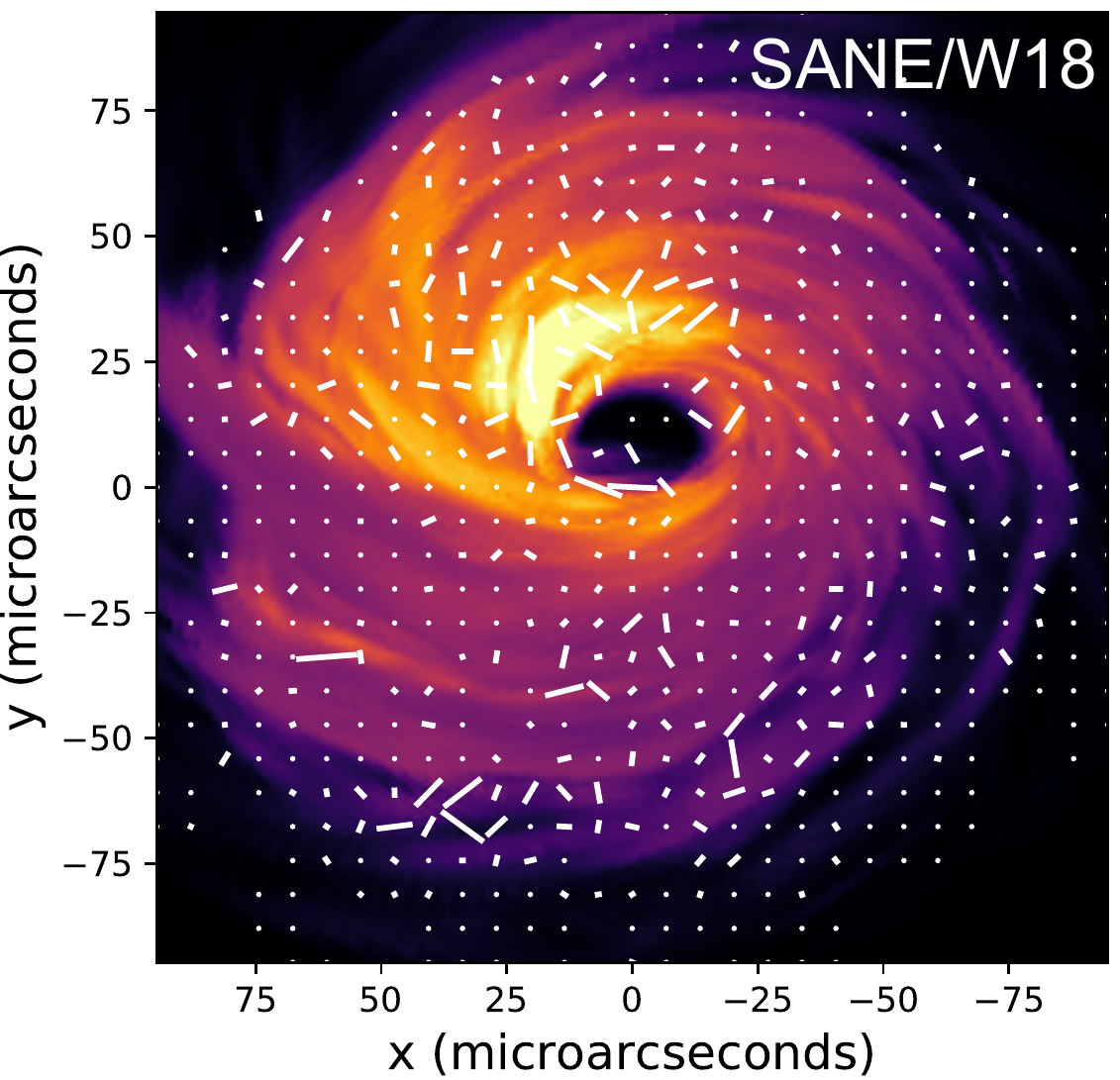} &
		\includegraphics[width=0.31\textwidth]{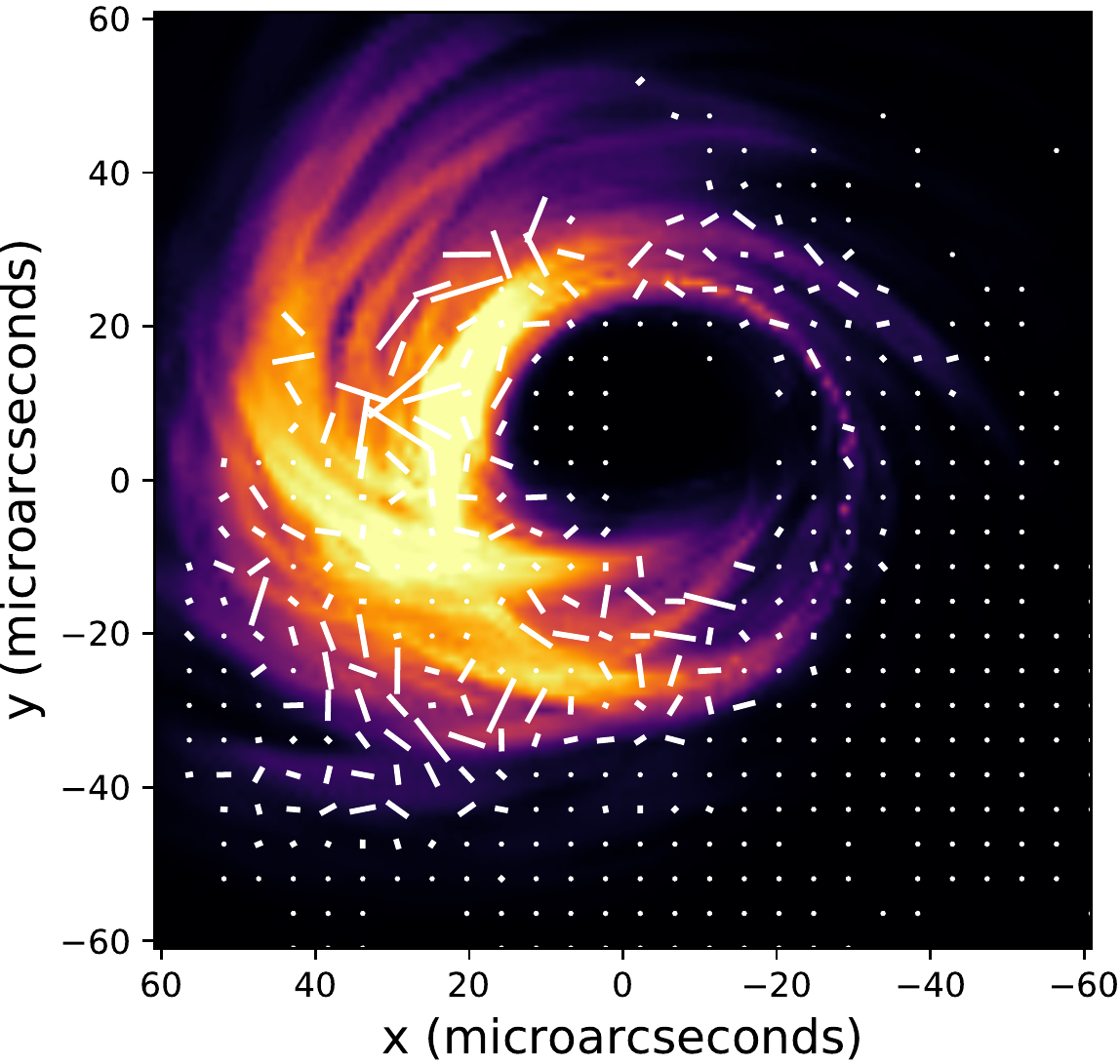}\\
	\includegraphics[width=0.31\textwidth]{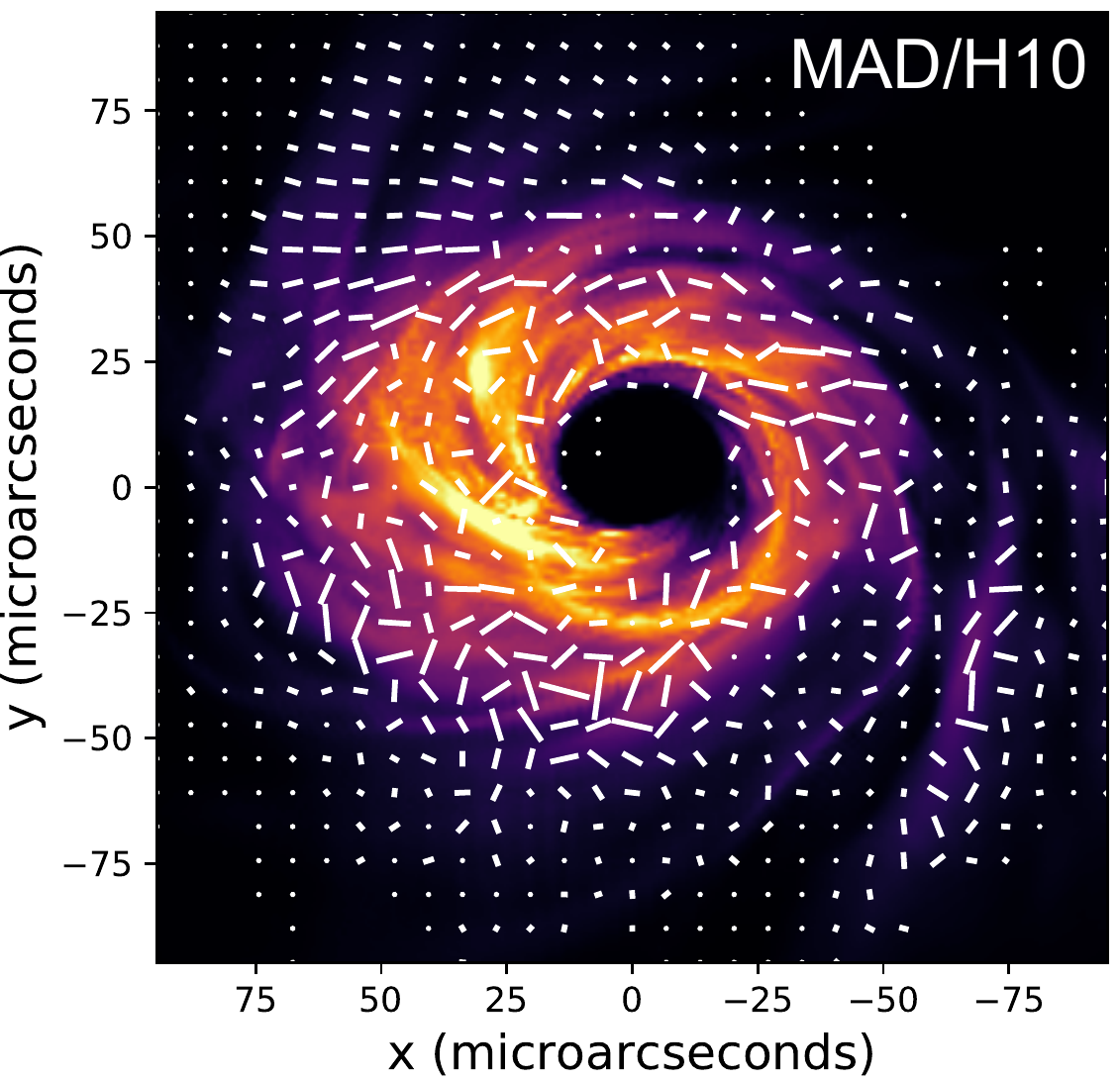} & 
		\includegraphics[width=0.31\textwidth]{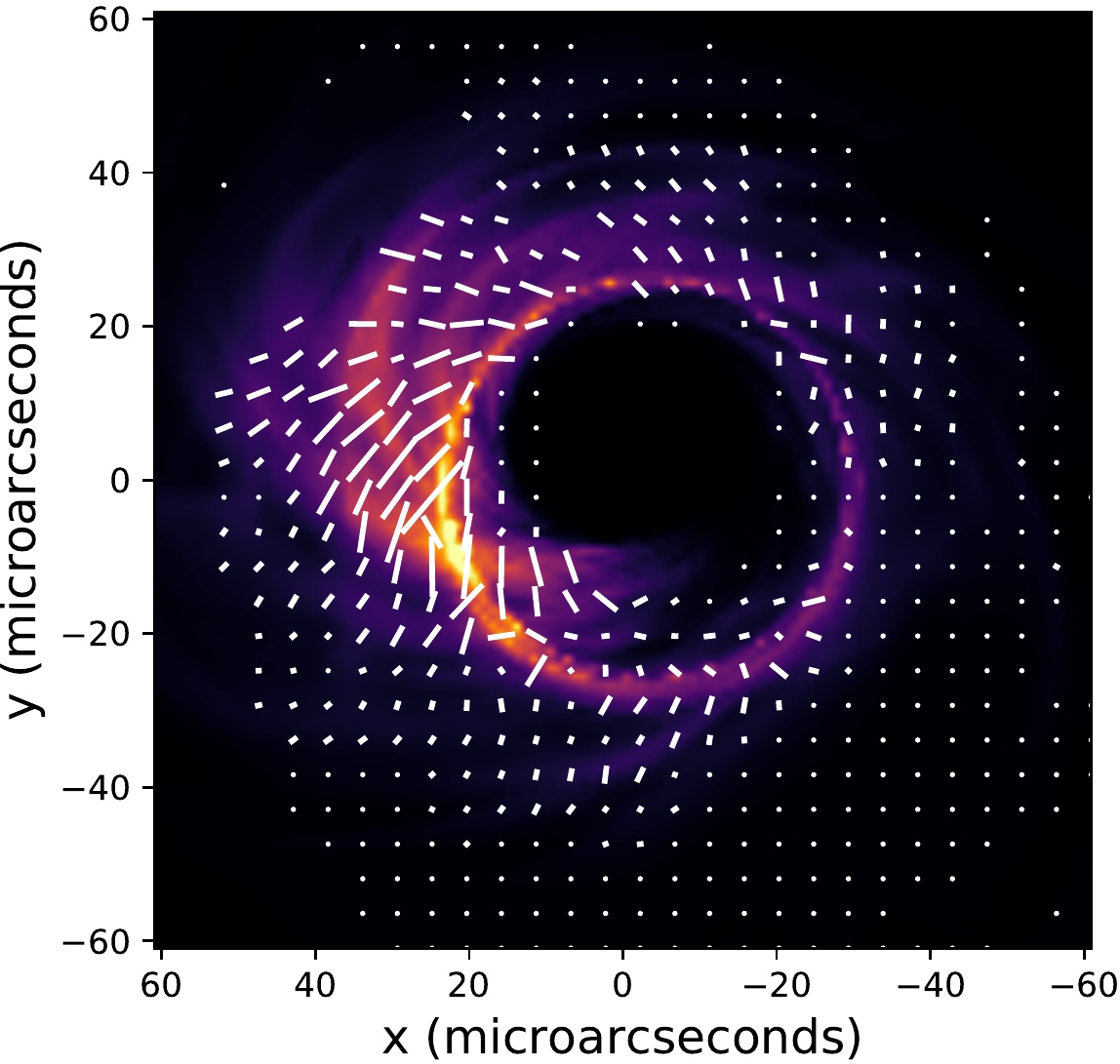}\\
	\includegraphics[width=0.31\textwidth]{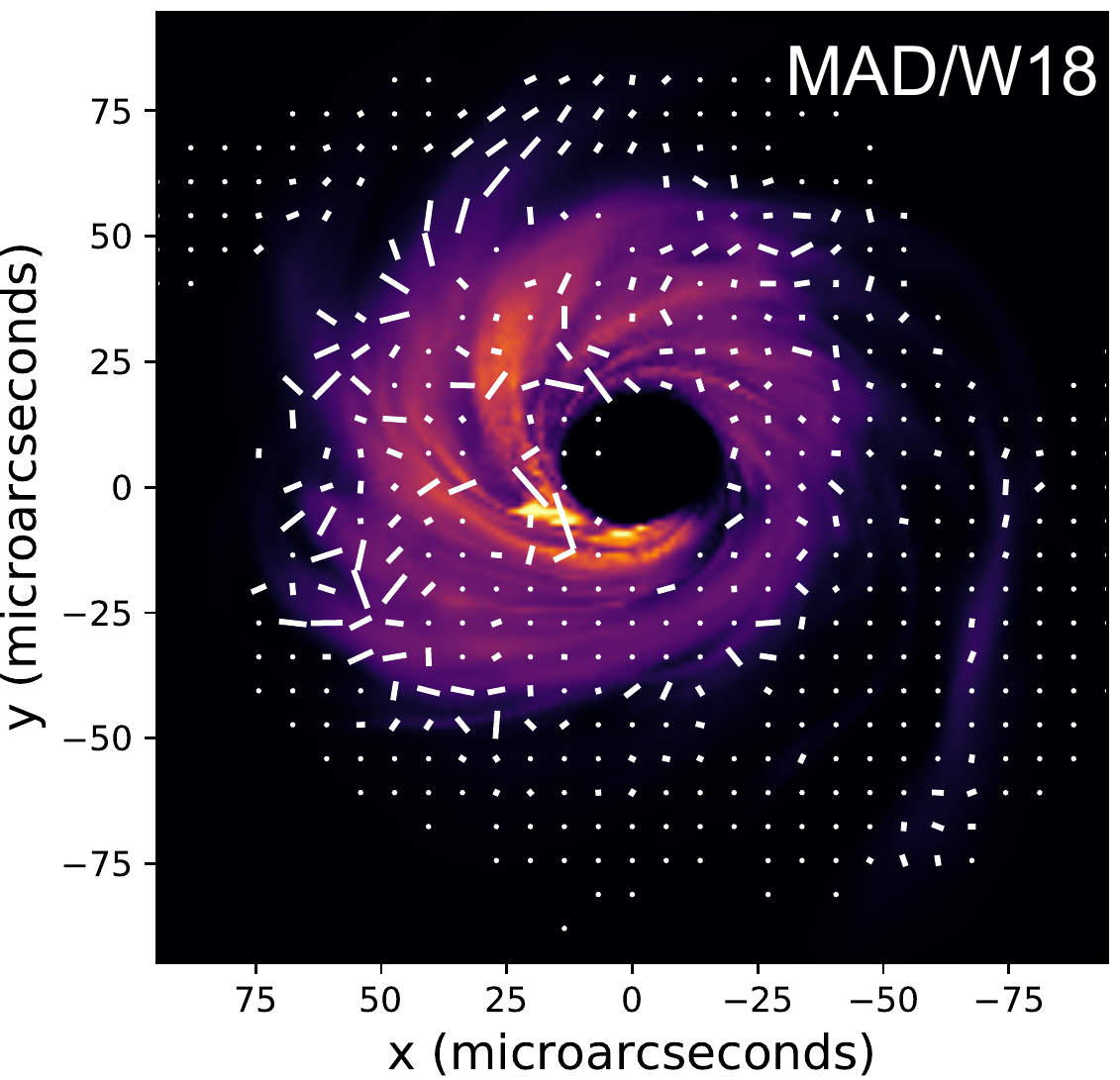} & 
		\includegraphics[width=0.31\textwidth]{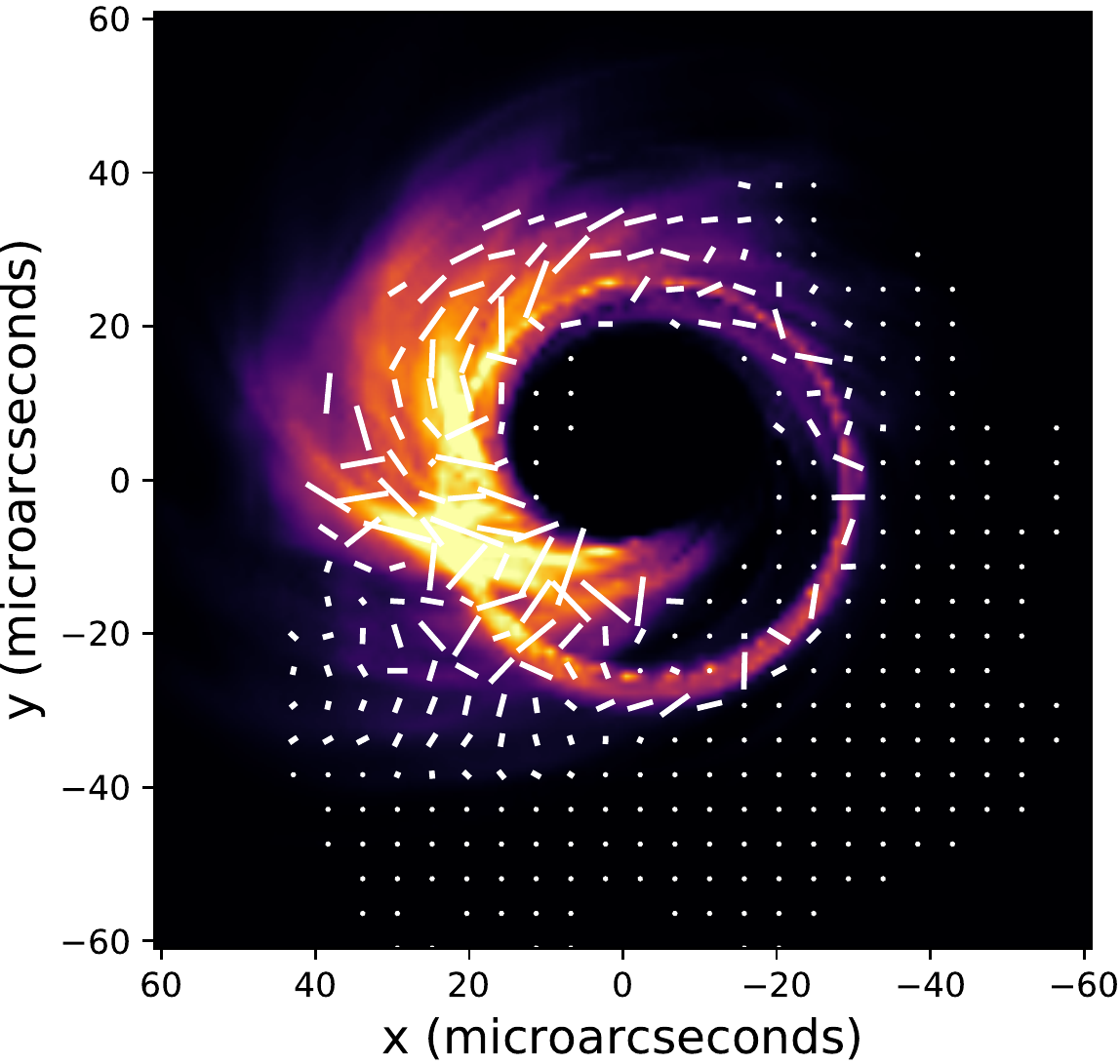}\\
	\end{tabular}
    \caption{Sample snapshot 86 (left) and 230 (right) GHz linearly scaled false color images and polarization maps for $a=0.5$ and $i=45^\circ$. Polarization tick length is proportional to polarized flux. The models are ordered from top to bottom as SANE/H10, SANE/W18, MAD/H10, MAD/W18. All models at 230 GHz show a characteristic crescent morphology from the combination of Doppler beaming and light bending. The SANE/H10 model shows a ``disk-jet'' structure with prominent polar emission from the jet wall at 86 GHz. In the other cases, the emission is predominantly from close to the midplane. All models are substantially depolarized from Faraday rotation at 86 GHz, and the SANE models are also depolarized at 230 GHz. Images of the K19 and R17 models are similar to those of the H10 and W18 models, respectively.}
    \label{fig:pol_maps}
\end{figure*}

\begin{figure*}
\begin{tabular}{cc}
	\includegraphics[width=0.48\textwidth]{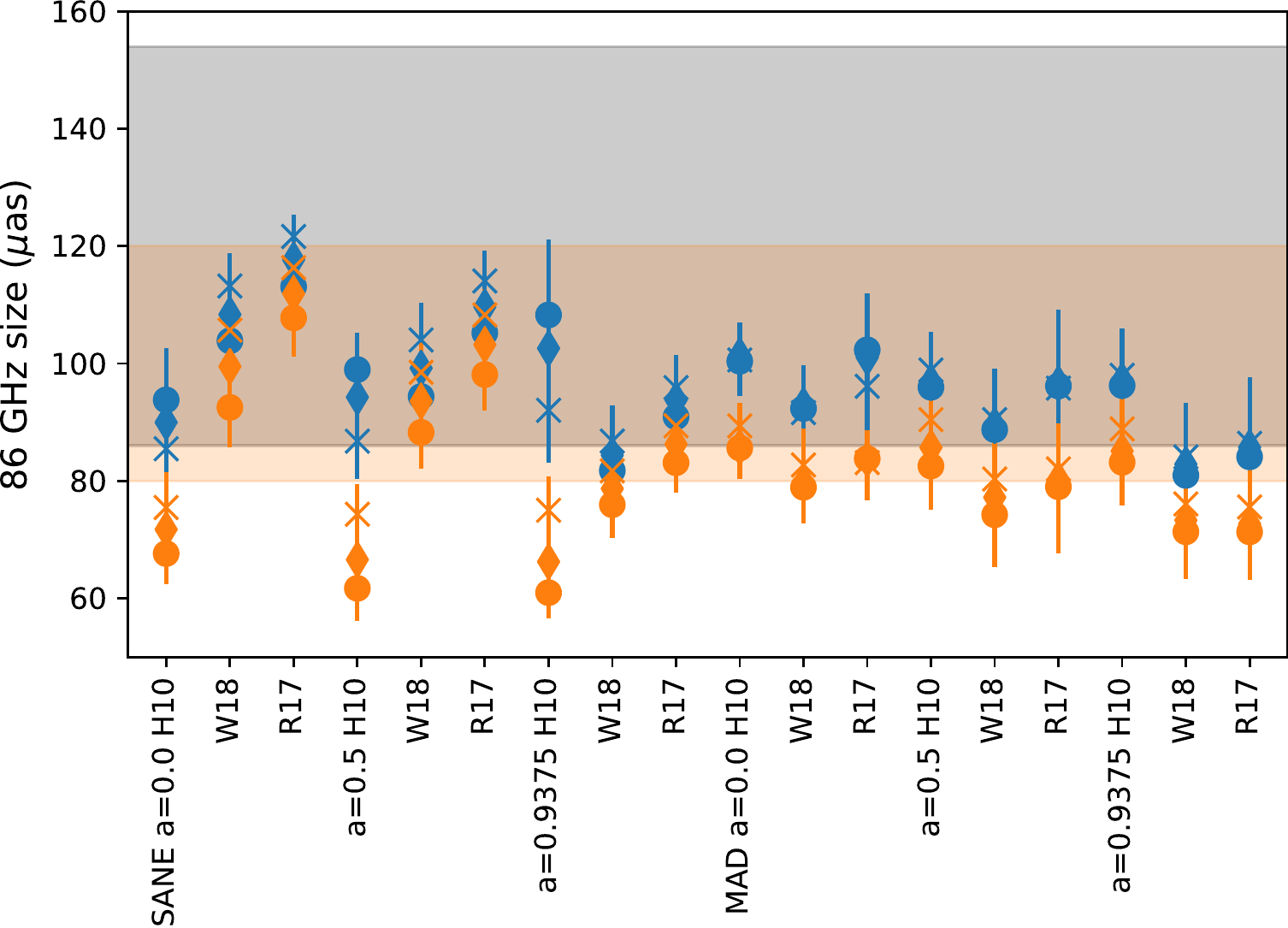} & 
	\includegraphics[width=0.48\textwidth]{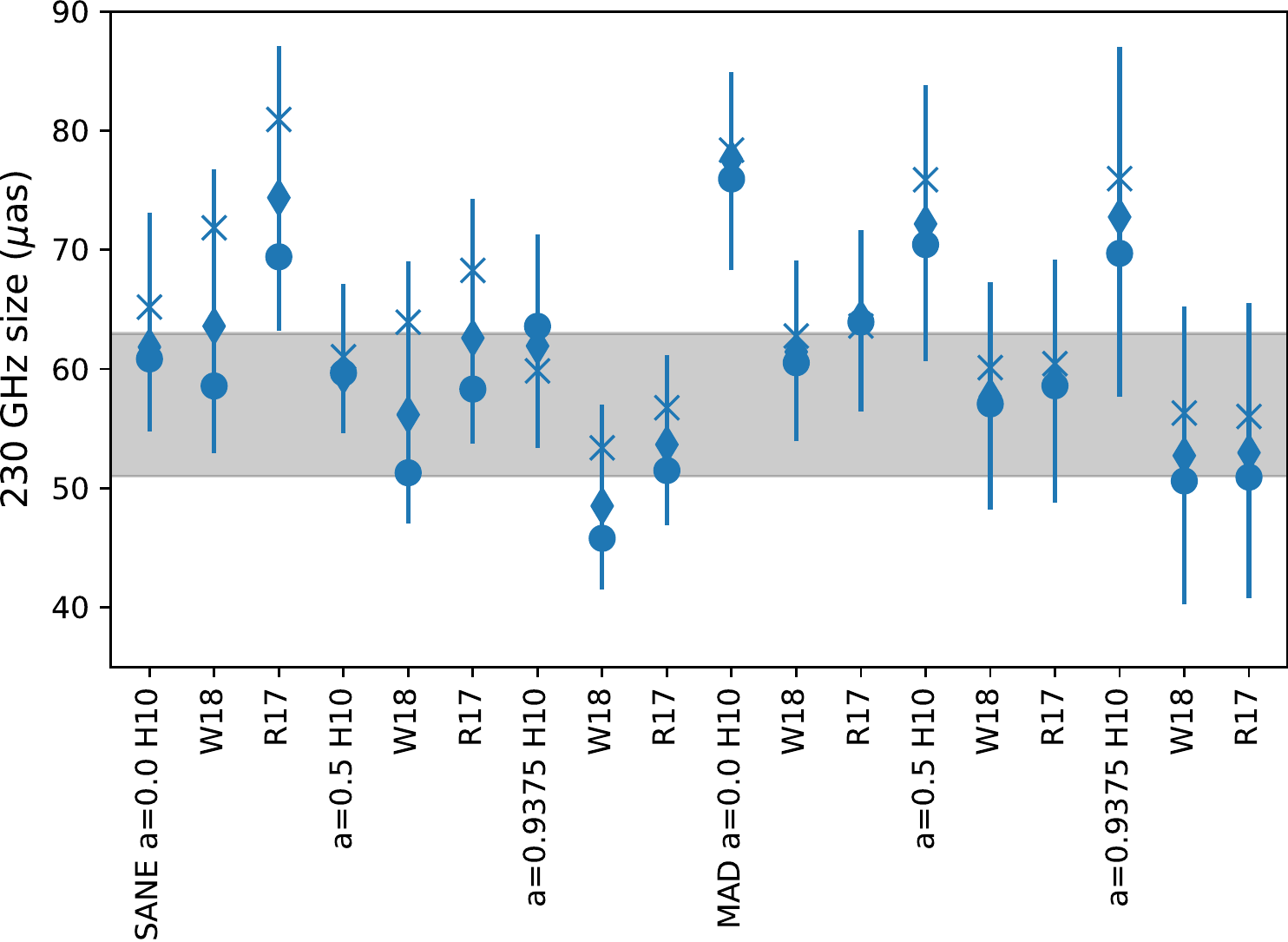}
	\end{tabular}
    \caption{Model image sizes at 86 (left) and 230 (right) GHz. The image second moments are shown along semi-major (blue) and semi-minor (orange, 86 GHz only) axes, multiplied by a factor of $2.35$ for comparison with the Gaussian FWHM sizes reported in the literature (gray bands). In each model column, the three points correspond to three viewing inclinations and the error bars correspond to the rms scatter over time. Many models can satisfy both constraints, although they are generally small compared with the $86$ GHz size. SANE/H10 models are too elliptical at 86 GHz for high viewing inclinations, and MAD/H10 models are too large at 230 GHz.}
    \label{fig:sizes}
\end{figure*}

\begin{figure*}
\begin{tabular}{cc}
	\includegraphics[width=0.48\textwidth]{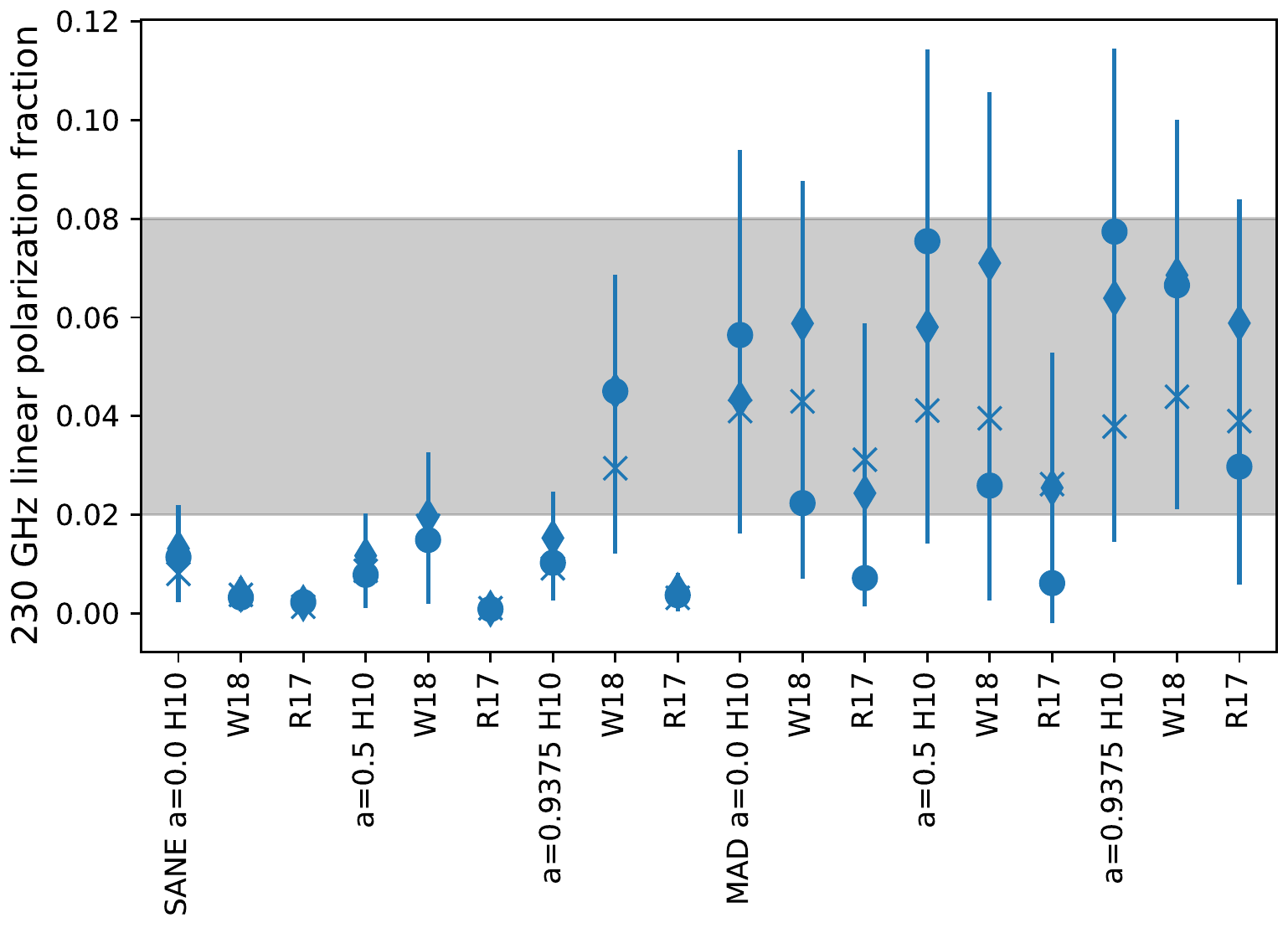} & 
	\includegraphics[width=0.48\textwidth]{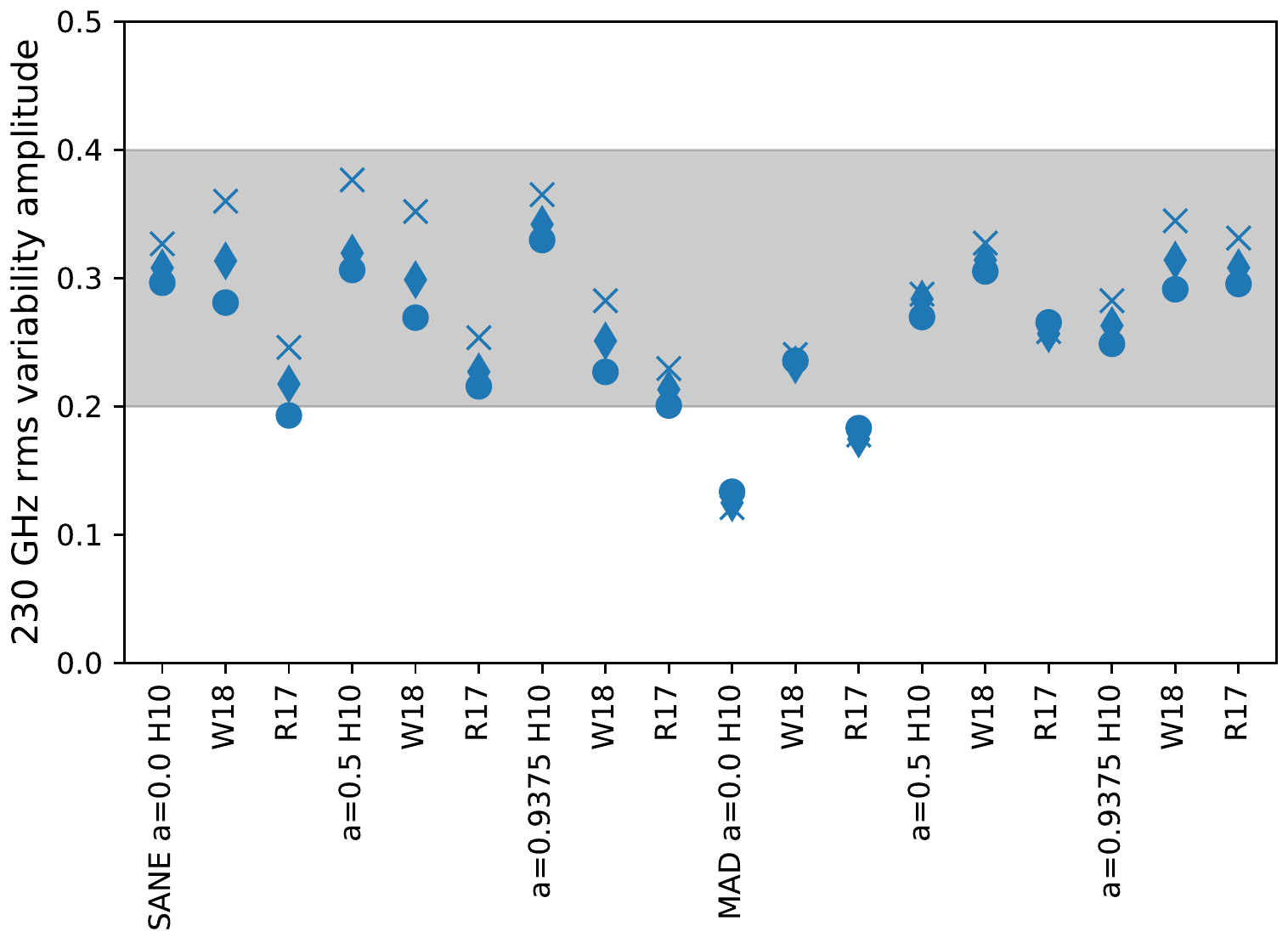}
	\end{tabular}
    \caption{Model 230 GHz linear polarization fractions (left) and rms variability amplitudes (right). All SANE models except $a=0.9375$ W18 show low linear polarization fractions as a result of Faraday rotation internal to the emission region. MAD models by contrast are frequently consistent with the range of median 230 GHz linear polarization seen from Sgr A*. All scenarios considered here can show 230 GHz variability with an rms amplitude $\simeq 20-40\%$, consistent with that observed. The variability is the result of turbulence driven by the MRI.}
    \label{fig:lprms}
\end{figure*}

\begin{figure}
    \centering
    \includegraphics[width=0.48\textwidth]{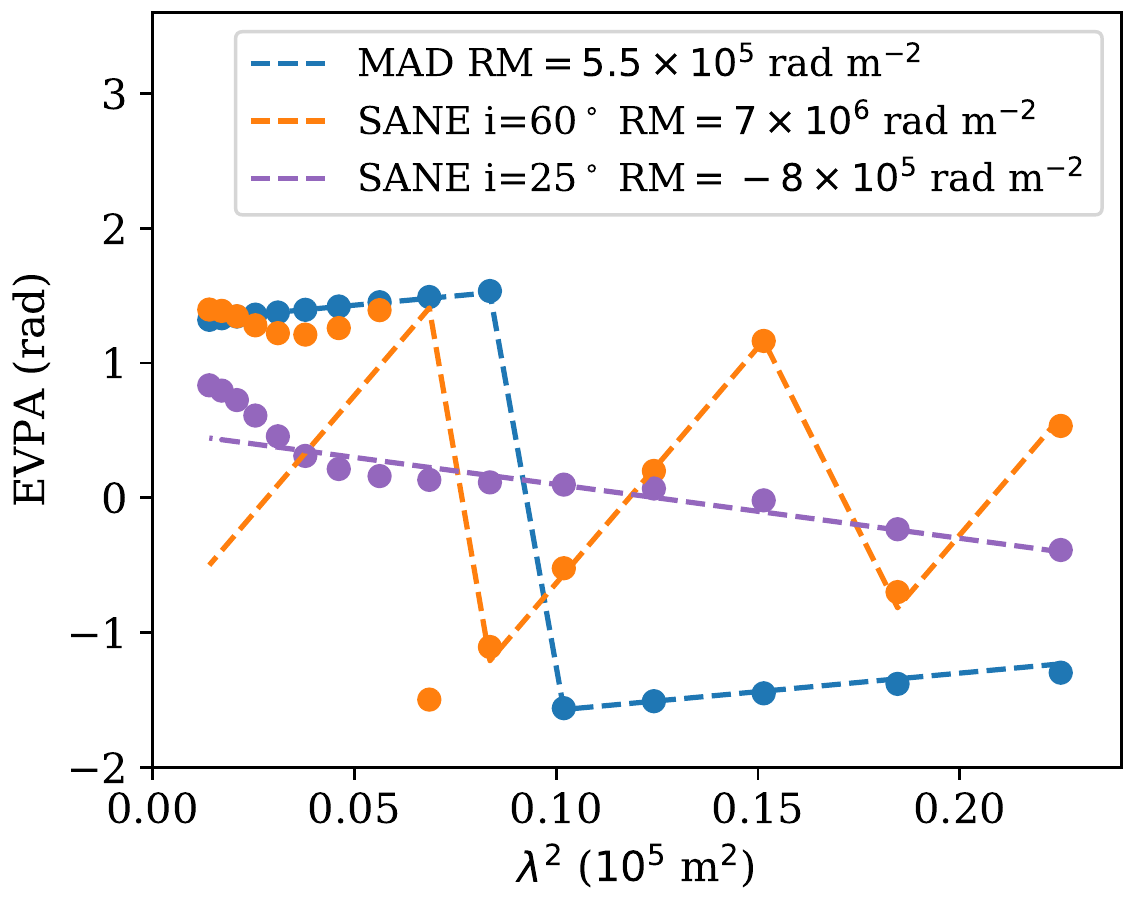}
    \caption{Image-integrated EVPA as a function of squared wavelength for late time snapshots at $i=60^\circ$ of sample MAD and SANE models, as well as the same SANE model at $i=25^\circ$. The EVPA behavior is consistent with external Faraday rotation. The inferred RM values for the MAD and low inclination SANE cases are consistent with that found in submm observations of Sgr A*. Many models (e.g. the SANE models shown here) show significant departures from a $\lambda^2$ dependence at short wavelengths $\lesssim 1$mm.}
    \label{fig:external_rm}
\end{figure}

\section{Results}
\label{sec:results}

We discuss accretion flow and convergence properties for our GRMHD simulations, compare their properties in radiative models of Sgr A* to the above observational constraints, and then study the properties of their multi-wavelength images and polarization maps.

\subsection{Convergence and accretion flow properties}

Our GRMHD simulations reach the MAD and SANE limits as expected. \autoref{fig:sim_plots_t} show the accretion rate histories and accumulated dimensionless magnetic flux on the horizon. SANE simulations have a relatively low net magnetic flux, $\phi_{\rm BH} \simeq 5-10$, while the flux saturates in the MAD case at a maximum value of roughly $\phi_{\rm BH} \simeq 50-70$ \citep{tchekhovskoy2011}. 

Table \ref{tab:sim_table} lists average values of those quantities for all of our simulations. The magnetic flux $\phi_{\rm BH}$ is measured on the horizon, while the $Q$ values and magnetic field tilt angle are averaged over the region $r=6-15 r_g$. MAD models are well resolved according to these criteria, while SANE models are more difficult to resolve due to their lower net vertical magnetic flux. Still, we find satisfactory convergence in all cases, generally defined as $\langle Q^{(\phi)} \rangle \langle Q^{(\theta)} \rangle \gtrsim 200$, tilt $\simeq 0.15-0.20$, and $\langle \beta \rangle \simeq 10-20$ (radial profiles of $\beta$ are shown in \autoref{fig:sim_plots_r}). We define inflow equilibrium following \citet{narayan2012} as the outermost radius where the elapsed time exceeds a viscous time, $t \langle v^r \rangle (r_{\rm eq}) / r_{\rm eq} \gtrsim 1$. Inflow equilibrium at these early times only reaches out to $r_{\rm eq} \simeq 20 r_g$. This is sufficient to capture the submm to NIR emission \citep[e.g.,][]{moscibrodzka2009}, but complicates our studies of linear polarization and Faraday rotation. All convergence criteria are readily satisfied for MAD simulations. The MAD models show lower equilibrium values of plasma $\beta \simeq 5-10$ and inflow equilibrium reaches larger radius $r_{\rm eq} \simeq 25-30 r_g$.

In SANE models the scale height is roughly constant in the region of inflow equilibrium ($H/R \simeq 0.3-0.4$), similar to past GRMHD \citep{shiokawa2012,narayan2012} and recent pseudo-Newtonian \citep{dhang2019} results.  In the MAD case, we find a strong decrease in the inner radii. This is due to magnetic pressure from the strong surrounding magnetosphere \citep{mckinney2012}. The azimuthal velocity profile is nearly Keplerian for SANE models, while MADs are sub-Keplerian in the region of inflow equilibrium due to magnetic pressure support \citep[e.g.,][]{mckinney2012}.

\subsection{Long duration runs and inflow equilibrium}

We have also run sample MAD and SANE simulations to long durations, similar to what was done by \citet{narayan2012} and \citet{white2020riaf}. Averaged properties of those simulations over various time intervals are listed in \autoref{tab:sim_long_sane} and \autoref{tab:sim_long_mad}. Our SANE model evolves to a slightly larger scale height, lower median $\beta$, higher magnetic tilt angle, and larger $Q^{(\phi)}$ and $Q^{(\theta)}$ convergence values. Radial profiles averaged over different time intervals are shown in \autoref{fig:sim_plots_r_long}. The inflow equilibrium radius $r_{\rm eq} \simeq 90 r_g$ reached is similar to that from \citet{narayan2012} in the same elapsed coordinate time of $2 \times 10^5 r_g/c$. Our MAD model is run for much shorter ($\simeq 6 \times 10^4 r_g/c$) but nonetheless reaches inflow equilibrium out to $r_{\rm eq} \simeq 70 r_g$.

\subsection{Electron heating}

Azimuthally averaged snapshots of electron temperature $T_e$ are shown in \autoref{fig:3panelsane} and \autoref{fig:3panelmad}. We find similar behavior for the various electron heating models as in previous work for SANE  \citep{ressler2015,ressler2017,chael2018,ryan2018} and MAD \citep{chael2019} simulations. In the SANE case, the dense accretion flow near the midplane has relatively large plasma $\beta \simeq 10-20$. The electrons there are only significantly heated for the reconnection models. In the turbulent case, electrons in the disk body remains cold. For MAD solutions, the average plasma $\beta$ is lower and electrons are heated efficiently everywhere in both scenarios. We find mild spin dependence, in the sense that for higher black hole spin the fluid and in turn the electrons have higher energies. More details on the time evolution and convergence of the electron heating solutions are provided in appendix \ref{app:te_appendix}.

\subsection{Spectrum and variability}

Sample spectra are shown in \autoref{fig:spectra_sample} for the four general classes of model considered. The spectra show median values over time at each frequency. The black hole spin is fixed at $a=0.5$ and the inclination at $i=45^\circ$. We compare SANE (blue) with MAD (black), and turbulence (H10) and reconnection (W18) cases. Data are taken from the references listed in \autoref{sec:spectra_data}. The combinations of SANE/reconnection (blue dashed) generically underproduce the observed THz emission from Sgr A*. They do not produce sufficiently hot electrons, and therefore show steep spectral breaks which are ruled out by recent Herschel and ALMA data. The combinations of MAD/turbulence by contrast strongly heat electrons to the degree that the spectral energy distribution (SED) peaks in the infrared. These models strongly overproduce the median NIR while underproducing the submm emission. The other two combinations, SANE/turbulence and MAD/reconnection, can match the Sgr A* SED shape at least for some combinations of inclination angle and black hole spin. The same general results hold for our full parameter survey. In particular, we have not found  combinations of parameters for SANE/reconnection or MAD/turbulence models which reproduce the submm spectral index and NIR flux density upper limit.

Most models we consider produce submm variability consistent with that observed from Sgr A*. The SANE/turbulence and MAD/reconnection models show highly variable NIR emission, which can in principle account for the flaring emission seen from Sgr A*. In those cases, the frequency-dependent rms (\autoref{fig:lprms_sample}) is often in fairly good agreement with that observed, rising from the radio through the submm and NIR. A more detailed comparison to the NIR flux distribution is forthcoming (GRAVITY collaboration 2020, in prep).

The detailed results for submm spectral index and median NIR flux density are further separated into our 6 simulations (MAD/SANE at 3 values of black hole spin) and 3 electron heating models (H10/W18/R17) in \autoref{fig:alphafnir}. The 3 points at each x-axis location correspond to 3 values of observer inclination ($i=25^\circ$, $45^\circ$, $60^\circ$). The gray bands show our allowed ranges for Sgr A*. The general trends from our chosen $a=0.5$ models can be found there, but also with systematic trends of higher submm spectral index and median NIR flux density with increasing black hole spin and observer inclination angle. Both effects result in part from increased Doppler beaming at higher inclination.

The SANE/turbulence and MAD/reconnection models studied produce interesting NIR flaring events (large rms variability in \autoref{fig:alphafnir}). MAD models show associated time-variable polarization and image photocenter (centroid) motion. Typical linear polarization fractions are $\simeq 10-30\%$, consistent with observations of NIR flares from Sgr A* \citep{eckart2006,trippe2007,gravityflare}. The rms centroid motion during flares is only $\simeq 10 \mu$as, a factor of $2-3$ smaller than seen from Sgr A*. Given the uncertainty in electron heating and distribution function, we nonetheless consider MAD accretion flows as promising for explaining the NIR (and X-ray) flares from Sgr A*.

\subsection{86 and 230 GHz image sizes}

Sample snapshot images and polarization maps corresponding to those same 4 model combinations are shown in \autoref{fig:pol_maps}. The SANE/turbulence model shows a jet-like, elongated 86 GHz image, while the others are dominated by emission from the dense inflow in the midplane. The SANE/reconnection images have much higher optical depth and a larger source size, particularly at 86 GHz.

\autoref{fig:sizes} shows semi-major (blue) and semi-minor (orange, 86 GHz only) axis sizes for each model considered. Except when viewed at low inclination, SANE/turbulence models are disfavored due to their small semi-minor axis size (large axis ratio). Most other model combinations can satisfy both the 86 and 230 GHz semi-major axis size constraints.

We generally find increasing 86 GHz size with decreasing emission region electron temperature in the order H10, W18, R17. The trend is stronger in SANE than in MAD models. At 230 GHz, the emission region size systematically decreases with black hole spin for the reconnection heating models (W18, R17) although all spins considered can produce median values falling within the allowed range. 

We also note that all models produce 86 GHz sizes on the smaller end of the allowed range. In particular, all sizes found are too small when compared to the model-fitting results of \citet{johnson2018} at 86 GHz. To be viable, many or all models may require an additional non-thermal component to the electron distribution function which can produce mm-wavelength emission further from the black hole.

\subsection{Polarization}

The snapshot polarization maps show that all SANE models explored are highly depolarized as a result of strong Faraday rotation internal to the emission region. We also find that the outer torus ($r \gtrsim 20 r_g$) can significantly alter the linear polarization degree and polarization map structure for SANE models. That region is not in inflow equilibrium, and we omit it from the radiative transfer calculations presented here. Even excluding that material altogether, few SANE models can match the median observed linear polarization fraction of Sgr A*. Our MAD polarization maps frequently show signatures of strong poloidal field, leading to azimuthal EVPA structure \citep[e.g.,][]{gravityflare}.

\autoref{fig:lprms_sample} shows the median image-integrated linear polarization fraction for our 4 sample models. The SANE models, and particularly SANE/turbulence, are too depolarized compared to observations of Sgr A*. MAD/reconnection models capture the frequency-dependent polarization fraction fairly well. In detail (\autoref{fig:lprms}), MAD models show highly variable LP within the observed range of Sgr A*. The polarization fraction is lowest for the R17 model, and slightly higher at higher black hole spin values. Our SANE models are too depolarized to explain the relatively large 230 GHz LP from Sgr A*, except the W18 model at high spin. The depolarization is the result of Faraday rotation near the dense accretion flow midplane. 

In all cases, the polarized emission is  more time variable than the total intensity. This is due to a variable EVPA pattern over the images resulting from turbulence and/or Faraday rotation. Time-variable beam depolarization then drives large variability of the integrated Stokes parameters. One way to see this is to note that the integrated polarized flux $\sqrt{Q^2+U^2}$ from each image pixel shows similar variability as in Stokes I, while both are much less variable than the net polarization integrated over images. This finding agrees with Sgr A* submm polarization observations \citep[e.g.,][]{marrone2006,bower2018}. By contrast, the NIR polarization degree seems much less variable \citep{eckart2008pol,shahzamanian2015}, likely as a result of a more compact emission region and negligible Faraday rotation at high frequency ($\nu/\nu_c \gtrsim 10^3$ with $\nu_c \sim T_e^2 B$ the critical synchrotron frequency).

\subsection{Summary of comparison with observational constraints}

\autoref{tab:summary} provides median observed and model values of the quantitative constraints used here: the submm spectral index, NIR flux density, 230 GHz LP and CP, image semi-major axis size at 230 and 86 GHz, the 230 GHz rms variability fraction. Comparing to the set of observed ranges, we produce a final pass/fail score for each model. Italicized entries indicate where models fail to match Sgr A* data. 

Several MAD models can match all constraints considered, and several others fail in only one category. They also produce highly variable NIR emission, similar to that observed. While we have not exhaustively explored the parameter space of either the simulations or the radiative transfer models, we expect the results to hold for other viewing angles and black hole spin values. All SANE models are ruled out by multiple constraints. Matching the spectral shape with only thermal electrons requires disk-jet models where the jet wall electrons are heated and the accretion flow is cold. Those models are too strongly depolarized to explain the measured submm linear and circular polarization of Sgr A*. 

\subsection{Faraday rotation in long duration runs}
\label{sec:faraday}

Using our long duration SANE $a=0$ model, we have also explored Faraday rotation and depolarization once inflow equilibrium has reached large radius $r \simeq 100 r_g$. For the turbulent electron heating, there is little difference. The accretion flow remains cold, and the emission from close to the black hole is depolarized to a maximum of $1-2\%$. For the reconnection heating models, at late times the large-scale accretion flow can substantially heat. This increases the observed polarization fraction to values consistent with Sgr A* data. 

For both long duration SANE and MAD models, we also find behavior consistent with external Faraday rotation, where the EVPA $\propto \lambda^2$ over a range of frequencies $\simeq 200-300$ GHz. \autoref{fig:external_rm} shows example fits to a MAD $a=0.9375$ model with an inflow equilibrium radius of $r_{\rm eq} \simeq 50 r_g$ ($t \simeq 2.7 \times 10^4 r_g/c$) and a SANE $a=0$ model with $r_{\rm eq} \simeq 100 r_g$ ($t \simeq 2 \times 10^5 r_g/c$). Both models are viewed at an inclination $i = 60^\circ$. The EVPA shows a clear linear trend with $\lambda^2$, particularly at longer wavelength. We infer Faraday rotation measures of $\simeq 6 \times 10^5$ rad m$^{-2}$ (MAD) and $7 \times 10^6$ rad m$^{-2}$ (SANE). The typical MAD values are in good agreement with the observed Faraday rotation measure of Sgr A* \citep[e.g.,][]{bower2003,marrone2006,marrone2007,bower2018}. The external RM decreases to $\lesssim \times 10^6$ rad m$^{-2}$ for $i=25^\circ$. We also find rapid changes in the sign of the RM at high inclination, while low inclinations can show a persistent sign (although our longest simulation only spans a few weeks for Sgr A*). Studies of the RM time variability, departures from $\lambda^2$, and spatially resolved maps are left to future work. Small RM values for the SANE long duration models are possible even when there is strong depolarization due to Faraday rotation internal to the emission region. For more details see appendix  \ref{app:rm_appendix}.

\begin{table*}
	\centering
	\caption{Comparison of the allowed ranges of various observational constraints considered with median values calculated for each model and a final pass/fail score.}
	\label{tab:summary}
	\begin{tabular}{lccccccccc} 		\hline
model & $i$ ($^\circ$) & $\alpha_{\rm submm}$ & $\log F_{\rm NIR}$ (mJy) & LP & |CP| & $a_{\rm 230}$ ($\mu$as) & $a_{\rm 86}$ ($\mu$as) & rms & summary\\
 & & $-0.35-0.25$ & $< 0.3$ & $0.02-0.08$ & $0.005-0.02$ & $51-63$ & $86-154$ & $0.2-0.4$ & \\
 \hline
SANE a=0.0 H10 & 25 & \emph{-0.55} & -0.7 & \emph{0.008} & \emph{0.002} & \emph{65.2} & \emph{85.5} & 0.33 & \emph{fail}\\
 & 45 & -0.30 & -0.5 & \emph{0.004} & \emph{0.003} & 61.8 & 90.0 & 0.31 & \emph{fail}\\
 & 60 & -0.10 & -0.2 & \emph{0.004} & \emph{0.003} & 60.8 & 93.8 & 0.30 & \emph{fail}\\
SANE a=0.0 W18 & 25 & \emph{-1.46} & -5.6 & \emph{0.003} & \emph{-0.004} & \emph{71.8} & 113.2 & 0.36 & \emph{fail}\\
 & 45 & \emph{-0.93} & -5.2 & \emph{0.002} & \emph{-0.002} & \emph{63.6} & 108.3 & 0.31 & \emph{fail}\\
 & 60 & \emph{-0.54} & -4.9 & \emph{0.002} & \emph{-0.001} & 58.6 & 103.9 & 0.28 & \emph{fail}\\
SANE a=0.0 R17 & 25 & \emph{-0.81} & -4.1 & \emph{0.001} & \emph{-0.001} & \emph{80.9} & 121.6 & 0.25 & \emph{fail}\\
 & 45 & -0.22 & -3.9 & \emph{0.001} & \emph{0.001} & \emph{74.4} & 117.7 & 0.22 & \emph{fail}\\
 & 60 & 0.11 & -3.8 & \emph{0.001} & \emph{0.001} & \emph{69.4} & 113.1 & \emph{0.19} & \emph{fail}\\
SANE a=0.5 H10 & 25 & \emph{-0.38} & -0.0 & \emph{0.008} & \emph{0.002} & 61.0 & 86.8 & 0.38 & \emph{fail}\\
 & 45 & -0.07 & 0.3 & \emph{0.004} & \emph{0.001} & 59.5 & 94.3 & 0.32 & \emph{fail}\\
 & 60 & 0.16 & 0.5 & \emph{0.004} & \emph{0.001} & 59.7 & 99.0 & 0.31 & \emph{fail}\\
SANE a=0.5 W18 & 25 & \emph{-1.35} & -5.4 & \emph{0.017} & -0.010 & \emph{63.9} & 104.0 & 0.35 & \emph{fail}\\
 & 45 & \emph{-0.83} & -4.5 & \emph{0.004} & -0.007 & 56.2 & 99.2 & 0.30 & \emph{fail}\\
 & 60 & \emph{-0.43} & -3.8 & \emph{0.003} & \emph{-0.005} & 51.3 & 94.4 & 0.27 & \emph{fail}\\
SANE a=0.5 R17 & 25 & \emph{-0.98} & -4.6 & \emph{0.001} & \emph{-0.002} & \emph{68.3} & 114.1 & 0.25 & \emph{fail}\\
 & 45 & -0.34 & -4.0 & \emph{0.001} & \emph{0.001} & 62.6 & 109.6 & 0.23 & \emph{fail}\\
 & 60 & 0.07 & -3.6 & \emph{0.001} & \emph{0.001} & 58.3 & 105.2 & 0.22 & \emph{fail}\\
SANE a=0.9375 H10 & 25 & 0.07 & \emph{1.2} & \emph{0.011} & \emph{0.000} & 59.8 & 92.0 & 0.37 & \emph{fail}\\
 & 45 & \emph{0.44} & \emph{1.7} & \emph{0.003} & \emph{-0.000} & 61.9 & 102.6 & 0.34 & \emph{fail}\\
 & 60 & \emph{0.65} & \emph{2.0} & \emph{0.003} & \emph{-0.001} & \emph{63.6} & 108.3 & 0.33 & \emph{fail}\\
SANE a=0.9375 W18 & 25 & \emph{-0.71} & -2.9 & 0.038 & \emph{-0.029} & 53.4 & 86.8 & 0.28 & \emph{fail}\\
 & 45 & -0.25 & -1.7 & \emph{0.015} & \emph{-0.028} & \emph{48.5} & \emph{84.4} & 0.25 & \emph{fail}\\
 & 60 & 0.08 & -0.7 & \emph{0.005} & -0.019 & \emph{45.8} & \emph{81.8} & 0.23 & \emph{fail}\\
SANE a=0.9375 R17 & 25 & -0.20 & -2.8 & \emph{0.003} & \emph{-0.003} & 56.7 & 95.9 & 0.23 & \emph{fail}\\
 & 45 & \emph{0.31} & -1.7 & \emph{0.002} & \emph{-0.002} & 53.7 & 93.8 & 0.21 & \emph{fail}\\
 & 60 & \emph{0.59} & -0.8 & \emph{0.001} & \emph{-0.001} & 51.5 & 90.9 & 0.20 & \emph{fail}\\
MAD a=0.0 H10 & 25 & \emph{-0.47} & \emph{0.9} & 0.041 & 0.007 & \emph{78.4} & 100.6 & \emph{0.12} & \emph{fail}\\
 & 45 & \emph{-0.43} & \emph{1.0} & 0.043 & 0.006 & \emph{77.6} & 101.4 & \emph{0.13} & \emph{fail}\\
 & 60 & \emph{-0.38} & \emph{1.0} & 0.056 & \emph{0.005} & \emph{75.9} & 100.4 & \emph{0.13} & \emph{fail}\\
MAD a=0.0 W18 & 25 & \emph{-0.56} & -0.4 & 0.043 & 0.011 & 62.8 & 91.6 & 0.24 & \emph{fail}\\
 & 45 & \emph{-0.45} & -0.3 & 0.059 & 0.010 & 61.4 & 92.9 & 0.23 & \emph{fail}\\
 & 60 & -0.34 & -0.2 & 0.022 & 0.008 & 60.5 & 92.3 & 0.24 & \textbf{pass}\\
MAD a=0.0 R17 & 25 & \emph{-0.43} & -0.1 & 0.031 & 0.018 & \emph{63.6} & 96.1 & \emph{0.18} & \emph{fail}\\
 & 45 & -0.32 & -0.1 & 0.024 & 0.013 & \emph{64.2} & 101.0 & \emph{0.17} & \emph{fail}\\
 & 60 & -0.20 & -0.1 & \emph{0.007} & 0.009 & \emph{63.9} & 102.3 & \emph{0.18} & \emph{fail}\\
MAD a=0.5 H10 & 25 & \emph{-0.42} & \emph{1.2} & 0.041 & \emph{0.003} & \emph{75.9} & 98.9 & 0.29 & \emph{fail}\\
 & 45 & \emph{-0.37} & \emph{1.3} & 0.058 & \emph{0.003} & \emph{72.2} & 96.8 & 0.28 & \emph{fail}\\
 & 60 & -0.33 & \emph{1.4} & 0.075 & \emph{0.002} & \emph{70.4} & 96.0 & 0.27 & \emph{fail}\\
MAD a=0.5 W18 & 25 & \emph{-0.45} & 0.1 & 0.040 & 0.008 & 60.1 & 90.4 & 0.33 & \emph{fail}\\
 & 45 & -0.32 & 0.2 & 0.071 & 0.008 & 57.7 & 89.4 & 0.31 & \textbf{pass}\\
 & 60 & -0.19 & 0.3 & 0.026 & \emph{0.003} & 57.1 & 88.7 & 0.31 & \emph{fail}\\
MAD a=0.5 R17 & 25 & -0.14 & 0.5 & 0.026 & 0.013 & 60.4 & 95.8 & 0.26 & \textbf{pass}\\
 & 45 & 0.02 & 0.5 & 0.025 & 0.007 & 58.9 & 96.5 & 0.26 & \textbf{pass}\\
 & 60 & 0.16 & \emph{0.6} & \emph{0.006} & \emph{0.001} & 58.6 & 96.2 & 0.27 & \emph{fail}\\
MAD a=0.9375 H10 & 25 & \emph{-0.36} & \emph{1.5} & 0.038 & \emph{0.004} & \emph{76.0} & 98.0 & 0.28 & \emph{fail}\\
 & 45 & -0.32 & \emph{1.6} & 0.064 & \emph{0.002} & \emph{72.8} & 97.1 & 0.26 & \emph{fail}\\
 & 60 & -0.27 & \emph{1.8} & 0.077 & \emph{0.000} & \emph{69.7} & 96.2 & 0.25 & \emph{fail}\\
MAD a=0.9375 W18 & 25 & \emph{-0.43} & \emph{0.6} & 0.044 & 0.008 & 56.3 & \emph{84.1} & 0.34 & \emph{fail}\\
 & 45 & -0.30 & \emph{0.8} & 0.069 & \emph{0.002} & 52.7 & \emph{82.8} & 0.31 & \emph{fail}\\
 & 60 & -0.16 & \emph{1.0} & 0.067 & \emph{-0.002} & \emph{50.6} & \emph{80.9} & 0.29 & \emph{fail}\\
MAD a=0.9375 R17 & 25 & -0.34 & 0.6 & 0.039 & 0.012 & 56.0 & 86.4 & 0.33 & \textbf{pass}\\
 & 45 & -0.15 & \emph{0.7} & 0.059 & \emph{0.004} & 53.0 & \emph{85.7} & 0.31 & \emph{fail}\\
 & 60 & 0.00 & \emph{1.0} & 0.030 & \emph{-0.002} & \emph{50.9} & \emph{84.1} & 0.30 & \emph{fail}\\
		\hline
	\end{tabular}
\end{table*}

\begin{table*}
	\centering
	\caption{Average physical parameters of our radiative models at 230 GHz.}
	\label{tab:phys_props}
	\begin{tabular}{lcccccccc}
		\hline
		\hline
model & $\dot{M}$ ($10^{-8}\,M_{\odot}\,\rm yr^{-1}$) & $\tau_I$ & $\tau_{\rho_V}$ & $r$ ($r_g$) & $\theta$ & $n$ ($10^6\,\rm cm^{-3}$) & $B$ (G) & $\theta_e$\\
\hline
SANE a=0.0 H10 & 9.0 & $1.3$ & $394.7$ & $6.7$ & $1.7$ & $7.7$ & $63.3$ & $10.9$\\
SANE a=0.0 W18 & 8.1 & $2.1$ & $124.0$ & $7.8$ & $1.6$ & $15.4$ & $59.4$ & $4.0$\\
SANE a=0.0 R17 & 30.5 & $7.9$ & $1586.0$ & $8.8$ & $1.5$ & $53.0$ & $100.4$ & $2.4$\\
SANE a=0.5 H10 & 6.0 & $1.7$ & $431.8$ & $6.2$ & $1.7$ & $7.9$ & $66.4$ & $12.2$\\
SANE a=0.5 W18 & 3.3 & $1.7$ & $52.4$ & $7.1$ & $1.5$ & $9.6$ & $48.3$ & $5.1$\\
SANE a=0.5 R17 & 10.5 & $5.4$ & $544.1$ & $7.6$ & $1.5$ & $29.8$ & $81.5$ & $3.0$\\
SANE a=0.9375 H10 & 3.4 & $3.2$ & $332.8$ & $6.1$ & $1.6$ & $10.0$ & $77.8$ & $13.4$\\
SANE a=0.9375 W18 & 1.0 & $1.5$ & $13.6$ & $5.6$ & $1.6$ & $5.1$ & $43.4$ & $8.9$\\
SANE a=0.9375 R17 & 3.1 & $5.3$ & $129.5$ & $6.1$ & $1.5$ & $14.1$ & $68.7$ & $4.9$\\
MAD a=0.0 H10 & 3.4 & $0.2$ & $6.6$ & $8.1$ & $1.6$ & $1.1$ & $40.2$ & $24.3$\\
MAD a=0.0 W18 & 1.0 & $0.6$ & $18.5$ & $6.3$ & $1.6$ & $3.2$ & $71.9$ & $11.9$\\
MAD a=0.0 R17 & 3.1 & $1.1$ & $50.9$ & $6.4$ & $1.6$ & $4.8$ & $100.5$ & $10.0$\\
MAD a=0.5 H10 & 0.6 & $0.2$ & $4.4$ & $7.9$ & $1.6$ & $1.3$ & $40.3$ & $24.1$\\
MAD a=0.5 W18 & 1.2 & $0.9$ & $16.2$ & $6.2$ & $1.6$ & $4.2$ & $78.0$ & $10.6$\\
MAD a=0.5 R17 & 2.1 & $2.2$ & $60.8$ & $6.2$ & $1.5$ & $7.6$ & $115.5$ & $8.2$\\
MAD a=0.9375 H10 & 0.4 & $0.2$ & $3.1$ & $7.8$ & $1.5$ & $0.9$ & $38.8$ & $30.7$\\
MAD a=0.9375 W18 & 1.0 & $0.7$ & $8.0$ & $5.7$ & $1.6$ & $2.3$ & $72.7$ & $13.6$\\
MAD a=0.9375 R17 & 2.0 & $1.1$ & $18.4$ & $5.8$ & $1.6$ & $3.2$ & $91.9$ & $11.0$\\
		\hline
	\end{tabular}
\end{table*}

\section{Discussion}
\label{sec:discussion}

We have carried out a parameter space survey of Sgr A* models using ray tracing radiative transfer calculations based on GRMHD simulation data output. We consider both low (SANE) and saturated (MAD) magnetic flux limits, and sub-grid prescriptions for dividing dissipated energy at the grid scale between electrons and protons. 

Both prescriptions assign more dissipated heat to electrons in strongly magnetized regions, resulting in higher jet wall (outflow) than accretion flow (inflow) electron temperatures. The electron temperature contrast between jet wall and disk body is larger in the SANE case and for the turbulent heating prescription. This results in a relatively cold accretion flow ($\theta_e \lesssim 1$ for $r \gtrsim 30 r_g$). For strongly magnetized MAD models, there are hot electrons everywhere. Our results for electron heating are consistent with those from recent work \citep{ressler2015,ressler2017,chael2018,chael2019}.

We find two general paradigms for successfully reproducing the observed Sgr A* submm to NIR time-variable spectrum (including NIR flares), as well as 86 and 230 GHz resolved image sizes. Those combinations are SANE simulations with turbulent electron heating, and MAD simulations with reconnection electron heating. Other combinations underproduce or overproduce hot electrons for the black hole spin and inclination angles tried. The SANE/turbulence models have been proposed previously and result in a ``disk-jet'' morphology where jet emission becomes prominent at longer 3mm and 7mm wavelengths \citep{moscibrodzka2014,ressler2017,chael2018}. 

We have also calculated submm to NIR polarized images and spectra from our models. We find that all SANE/turbulence models are strongly depolarized due to Faraday rotation in the cold disk body, and cannot explain the net linear polarization observed from Sgr A* at $1.3$mm. The SANE/turbulence models also show elongated morphologies at 86 GHz due to extended jet structure, inconsistent with recent measurements except at low inclination \citep[as found previously by ][]{issaoun2019}. 

MAD/reconnection models can explain the highly time-variable net linear polarization of Sgr A* in the submm, and are only mildly depolarized. As a result, we favor MAD/reconnection models of Sgr A*. Within those models, the moderate black hole spin of $a=0.5$ is more successful at matching the observed properties than the high spin of $a=0.9375$. At high black hole spin, the models tend to overproduce hot electrons, resulting in too much NIR emission relative to that in the submm. They are also too compact at 86 GHz.

\subsection{Comparison to past work}

Many groups have done similar studies in the last several years. \citet{ressler2015} introduced the method used here for evolving multiple electron temperatures with sub-grid heating prescriptions. This method has been used in recent work on Sgr A* \citep{ressler2017,chael2018} and radiation GRMHD models of M87 \citep{ryan2018,chael2019}. Compared to those studies, we have considered a larger parameter space with multiple electron heating models, varying black hole spin, and considering both MAD and SANE solutions.

Our SANE/reconnection models are similar to early GRMHD models \citep{noble2007,moscibrodzka2009,dexter2009,dexter2010,drappeau2013} based on the assumption of constant proton-electron temperature ratio $T_p/T_e$ everywhere. This behavior is also seen in R17 models at low and high spin from \citet{chael2018}. Our SANE/turbulence models are similar to ``disk-jet'' models realized by assuming highly magnetized regions receive more heat in post-processing \citep{moscibrodzka2013,moscibrodzka2014,chan2015image} or using self-consistent electron heating with the H10 model \citep{ressler2017}. We have shown that such models, while otherwise promising, are highly depolarized as the result of Faraday rotation internal to the emission region (all inclination angles). Additionally they show very high Faraday rotation measure unless viewed at low inclination (here $i = 25^\circ$).

Previous MAD models of Sgr A* \citep{shcherbakov2013,gold2017} have used different post-processing electron heating prescriptions. Some of those models are at least broadly consistent with the spatially resolved submm polarization of Sgr A* \citep{johnson2015}. We seem to find more coherent submm polarization maps than in those studies.

Average properties of our radiative models are given in \autoref{tab:phys_props}. The accretion rate $\dot{M}$ is measured at the event horizon, while the others are intensity-weighted averages taken along each ray and over each pixel of the 230 GHz images. Viable models we identify have $\dot{M} = (1.0-2.1) \times 10^{-8} \, M_\odot \, \rm yr^{-1}$, resulting in low radiative efficiencies $\lesssim 0.1\%$. The plasma parameters generally agree with one zone estimates \citep[e.g.,][]{vonfellenberg2018,bower2019} and physical conditions in previous analytic \citep[e.g.,][]{falcke2000jet,ozel2000,yuan2003} and GRMHD \citep[e.g.,][]{moscibrodzka2009,dexter2010} models.

\begin{figure*}
\begin{tabular}{ll}
	\includegraphics[width=0.33\textwidth]{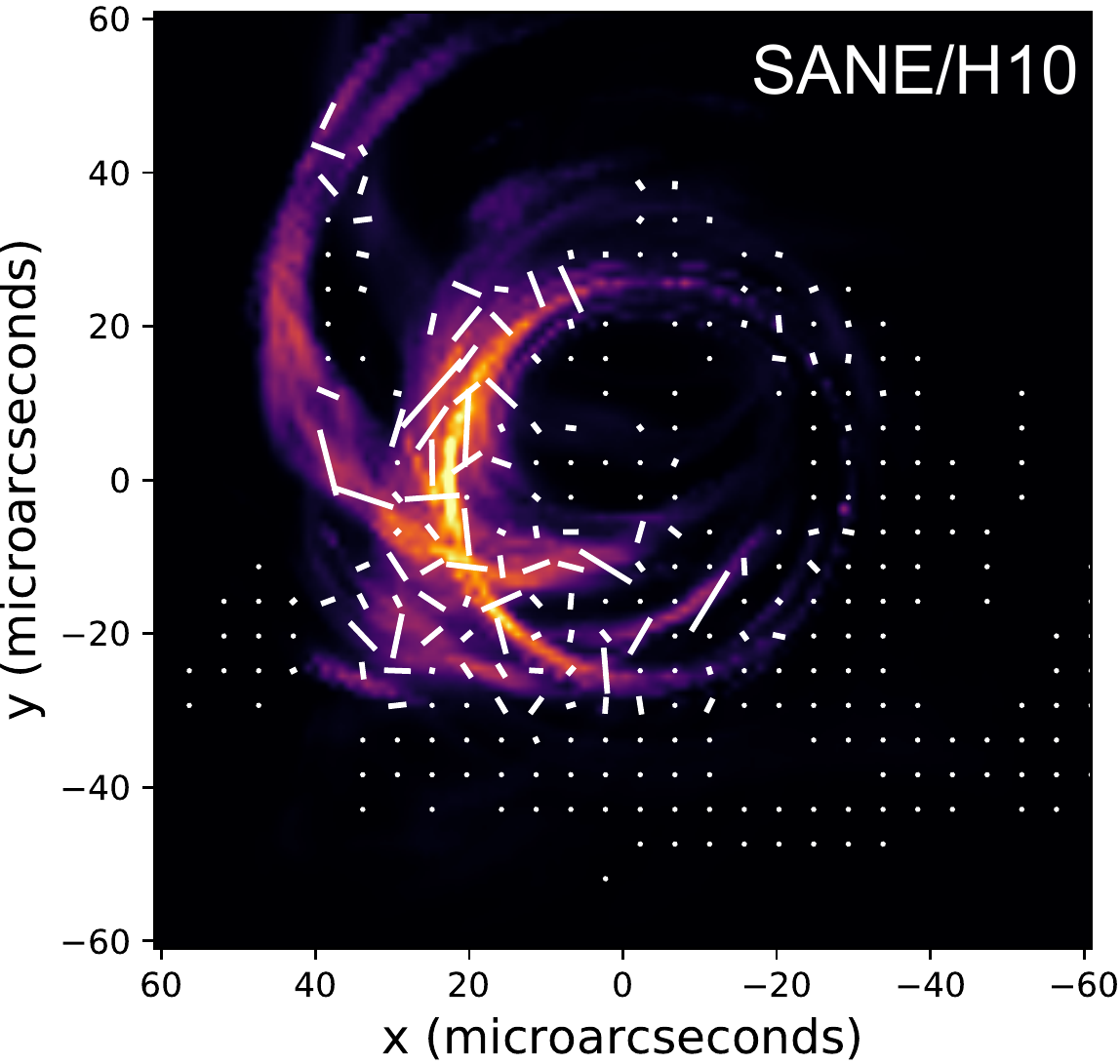} & 
	\includegraphics[width=0.33\textwidth]{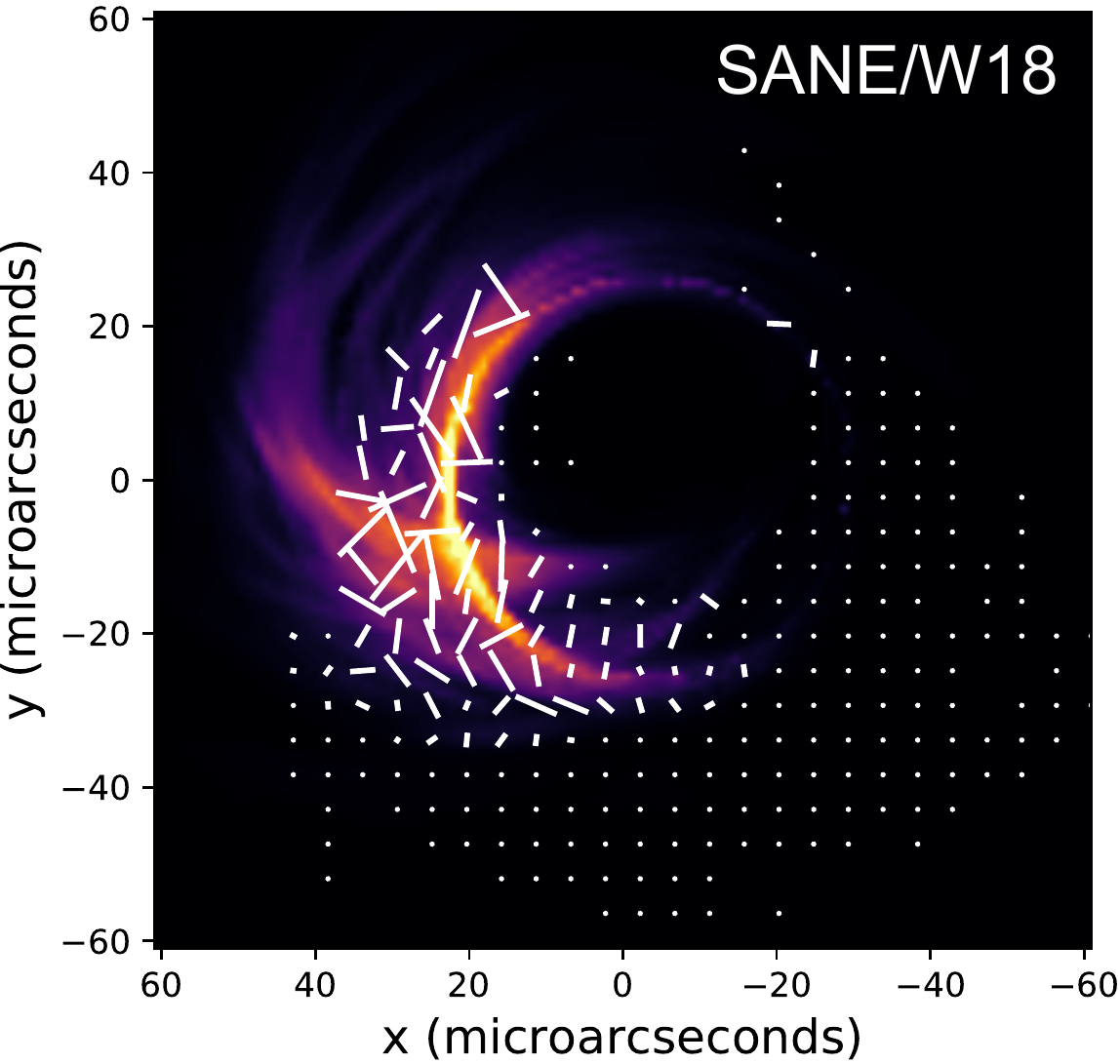}\\
	\includegraphics[width=0.33\textwidth]{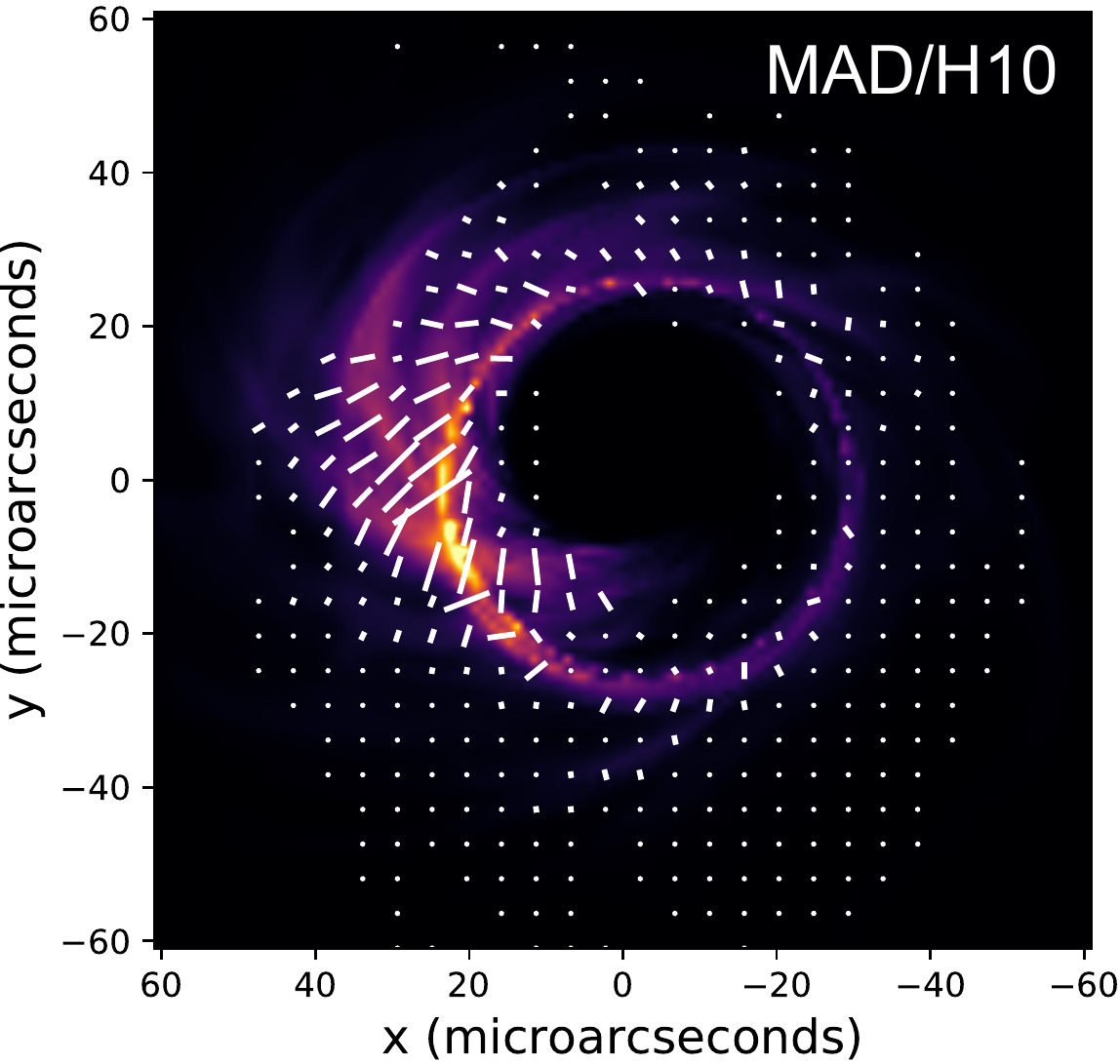} & 
	\includegraphics[width=0.33\textwidth]{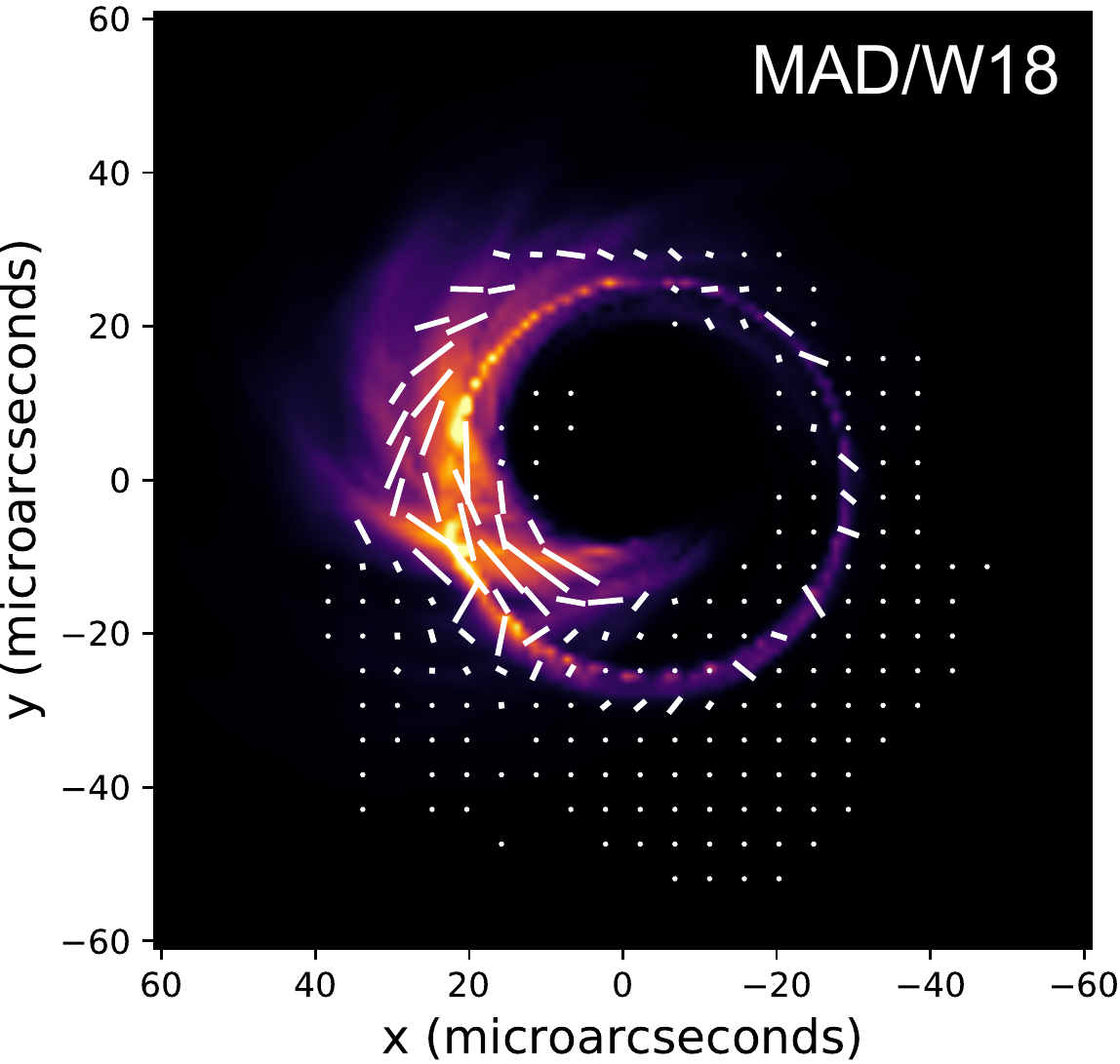}
	\end{tabular}
    \caption{Sample snapshot 345 GHz linearly scaled false color images and linear polarization maps for $a=0.5$ and $i=45^\circ$. Polarization tick length is proportional to polarized flux. At 345 GHz, the MAD polarization maps show negligible scrambling from Faraday rotation. Instead, the polarization maps trace the underlying magnetic field configuration.}
    \label{fig:345ghz}
\end{figure*}

\subsection{Future measurements}

The 230 GHz image morphology is primarily the result of relativistic effects of Doppler beaming and light bending, and does not depend strongly on the details of the magnetic field, black hole spin, or electron model \citep[e.g.,][]{kamruddin2013} when the emission region is optically thin. 

Polarization encodes plasma and magnetic field properties \citep[e.g.,][]{shcherbakov2012,dexter2016,moscibrodzka2017,jimenezrosales2018}. For the viable MAD models studied here, we find fairly coherent, azimuthal (``twisty'') EVPA maps (\autoref{fig:pol_maps}). The structure is due to significant poloidal magnetic field near the event horizon. Radial magnetic field and light bending both produce azimuthal EVPA structure, which is balanced by vertical magnetic field which results in a preferred horizontal EVPA direction \citep[e.g.,][]{gravityflare}. The azimuthal EVPA maps are far more evident at 345 GHz (or potentially in NIR flares) than at 230 GHz where external Faraday rotation can coherently rotate the EVPA by $\simeq 20-40$ degrees. Faraday rotation internal to the emission region produces additional disorder, but does not strongly depolarize the source or scramble the EVPA map.

We have also studied the image-integrated polarization angle as a function of wavelength for the models run to late times. The EVPA is defined as EVPA $= 1/2 \tan^{-1} U/Q$, where $U$ and $Q$ are image-integrated Stokes parameters. The EVPA $\sim \lambda^2$ behavior matches the expectation of Faraday rotation produced in the extended accretion flow \citep[\autoref{fig:external_rm}, ][]{marrone2006}. The favored MAD models show rotation measure values consistent with those observed from Sgr A* \citep{marrone2007,bower2018} for all inclination angles considered. The SANE models show Faraday rotation measures a factor $\simeq 10$ too large when viewed at high inclination. At low inclination ($i=25^\circ$), the rotation measure is much lower and can be consistent with Sgr A* data.

\autoref{fig:345ghz} shows 345 GHz images and polarization maps for our four sample models. At 345 GHz, the emission region is more compact than at 230 GHz. The morphology is similar in all cases. Faraday effects are weaker, resulting in higher net linear polarization and a more coherent polarization map. The emission is concentrated to still smaller radius, producing a bright photon ring feature. At moderate to high inclination, however, the strong Doppler beaming makes this feature highly asymmetric.

\subsection{Limitations}

The parameter survey presented here is both sparsely sampled and incomplete. We have used simplistic analytic models for sub-grid electron heating, based on a small number of recent kinetics calculations. As those calculations improve, so will the predictive power of our radiative models. In general, we expect that successful models of Sgr A* will need relatively high electron temperatures in the disk body, to avoid significant depolarization of the submm radiation. Within the prescriptions tried, that disfavors SANE/turbulence models. Successful models of the submm spectrum also require a range of electron temperatures, which disfavors the SANE/reconnection scenario. This qualitative understanding can be applied to future heating prescriptions as well.

MAD accretion flows produce regions of high magnetization throughout the simulation domain, where truncation errors can lead to negative internal energy which are corrected by imposing numerical floors. The convergence properties of MADs have been explored \citep{white2019mad}, but remain less well understood than for the SANE case. For calculating radiative models of MADs, the treatment of high magnetization regions plays an important role \citep{chael2019}. Here we follow \citet{EHTPaperV} and exclude emission from regions where $\sigma > 1$. That choice is arbitrary and for MAD models effects the NIR flux density (see appendix \ref{app:sigmacut}).

We have also assumed a thermal electron distribution function, while non-thermal emission can broaden the submm spectrum \citep{ozel2000,yuan2003,broderick2009} and increase the image size \citep{ozel2000,mao2017,chael2017,davelaar2018}. In particular, it may be possible to find viable SANE/reconnection models when non-thermal electrons are included. Their inclusion is also promising for comparing theoretical models with X-ray flare data \citep[e.g.,][]{ball2016}. We have also assumed an accretion flow angular momentum axis aligned with that of the black hole spin. This may be a poor assumption in the Galactic center. The orientation of the stellar winds providing the extremely low accretion rate \citep{quataert2004,cuadra2006,ressler2018} is unlikely to align with those of earlier accretion episodes. Disk tilt can change both the image morphology and spectrum \citep{dexter2013,white2020,chatterjee2020} as a result of shock heating \citep{fragile2008,white2019tilt}. We also neglect radiative cooling, which has found to be unimportant for the low accretion rate of Sgr A* \citep{dibi2012,ryan2018}.

With the above caveats, our results favor a strongly magnetized MAD accretion flow in Sgr A*. The resulting magnetic field structure is consistent with that inferred from the combined time-variable polarization and astrometric motions seen in NIR flares \citep{gravityflare}. The MAD limit is associated with the strongest Poynting flux driven jets from black holes \citep{tchekhovskoy2011}, but no powerful jet has conclusively been found from Sgr A*. The kinetic luminosity of our models remains modest compared to the energy associated with $\gtrsim 1$ pc scales in the Galactic center. Our models are also restricted to a limited computational domain. They do not make predictions for non-thermal radio emission on larger scales, as seen in jetted systems. For a MAD to operate in Sgr A*, either the acceleration mechanism is inefficient in the Galactic center, it only accelerates particles relatively close to the black hole where the extended radio emission can be hidden by interstellar scattering along the line of sight \citep[e.g.,][]{markoff2007}, and/or the black hole spin is small in magnitude resulting in a low BZ jet power.

\section{Conclusions}

We have carried out a large parameter survey of GRMHD models of Sgr A* with self-consistent electron heating. We have considered a range of (prograde) black hole spin, vertical magnetic field strength (weak/SANE or strong/MAD), and sub-grid electron heating models (relatively uniform heating, ``reconnection" or strongly $\beta$-dependent, ``turbulent" heating). We have studied radiative models of the radio to NIR emission from Sgr A* based on our model survey. The main findings are as follows:

\begin{itemize}
    \item Parameter combinations of magnetic fields and electron heating of SANE/turbulence and MAD/reconnection can explain the mm to NIR SED shape, variability including large-amplitude NIR flares, and mm/submm source sizes;
    \item SANE/turbulence models are heavily depolarized due to Faraday rotation effects, while some of our MAD/reconnection models remain viable for explaining a wide range of Sgr A* observations;
    \item MAD models show azimuthal (``twisty'') EVPA maps due to significant near horizon poloidal fields and this pattern should be apparent especially at $345$ GHz or higher frequency where Faraday rotation becomes negligible;
    \item limitations include uncertainty in the sub-grid electron heating schemes, the use of thermal electron distribution functions, the treatment of highly magnetized regions, particularly for NIR emission, and an accretion flow aligned with the spin axis of the central black hole. 
\end{itemize}

\section*{Acknowledgements}
We thank the anonymous referee for constructive suggestions, which led to an improved paper. J.D. and A.J.-R. were supported in part by a Sofja Kovalevskaja award from the Alexander von Humboldt foundation, by a CONACyT/DAAD grant (57265507), and by NASA Astrophysics Theory Program Grant 80NSSC20K0527. S.M.R. was supported by the Gordon and Betty Moore Foundation through Grant GBMF7392. The calculations presented here were carried out on the MPG supercomputers Hydra and Cobra hosted at MPCDF.

\bibliographystyle{mnras}


\appendix

\section{Updated Faraday rotation coefficient for a thermal plasma}
\label{app:rhov_appendix}

\citet{dexter2016} presented approximate, analytic forms for synchrotron and Faraday coefficients appropriate for relativistic electrons ($\theta_e \gtrsim 1$), including for the case of purely thermal electrons. Their expression for the Faraday rotation coefficient ($\rho_V$, their equation B14) fails at low $\nu/\nu_c$ and $\theta_e \ll 1$. This can be seen from:

\begin{equation}
    \rho_V = \rho_{V,\rm NR} \, f(\theta_e, \nu/\nu_c),
\end{equation}

\noindent where $\nu_c = (3 e B / 4\pi m c) \, \theta_e^2 \, \cos{\theta_B}$ is the critical synchrotron frequency and $\theta_B$ is the angle between the emission and magnetic field directions in the fluid frame. The coefficient $\rho_{V,\rm NR}$ is the correct non-relativistic limit, 

\begin{equation}
    \rho_{V,\rm NR} = \frac{2 n e^3 B \cos{\theta_B}}{m^2 c^2 \nu^2},
\end{equation}

\noindent related for example to the usual Faraday rotation measure. For $\nu/\nu_c \gg 1$, the function $f(\theta_e, \nu/\nu_c) = K_0(\theta_e^{-1})/K_2(\theta_e^{-1})$ \citep{shcherbakov2008}, while for general $\nu/\nu_c$ and $\theta_e \gg 1$, there is an additive correction term \citep{jones1979} which \citet{dexter2016} included approximately with a fitting function $\Delta J_5(\nu/\nu_c)$:

\begin{equation}
    \rho_V = \rho_{V, NR} \left(\frac{K_0(\theta_e^{-1}) - \Delta J_5 (\nu/\nu_c)}{K_2(\theta_e^{-1})}\right).
\end{equation}

For cold electrons $\theta_e \ll 1$, the modified Bessel function $K_n (x)$ reaches an asymptotic limit independent of $n$. Then the ratio $K_0/K_2 \rightarrow 1$, correctly reproducing the non-relativistic limit. Adding the $\Delta J_5$ term violates this limit. As a simple fix, in \textsc{grtrans} we now multiply the $\Delta J_5$ term by a narrow step function at $\theta_e = 1$, so that it is suppressed in the non-relativistic limit. We additionally set $f(\theta_e, \nu/\nu_c) = 1$ whenever $\theta_e < 10^{-2}$ to avoid numerical errors in the ratio of modified Bessel functions of large argument. These changes are necessary for accurate calculations of Faraday rotation at larger radii where the electrons can be non-relativistic.

\section{Electron temperature evolution and convergence}
\label{app:te_appendix}

\begin{figure*}
\begin{tabular}{cc}
\includegraphics[width=0.48\textwidth]{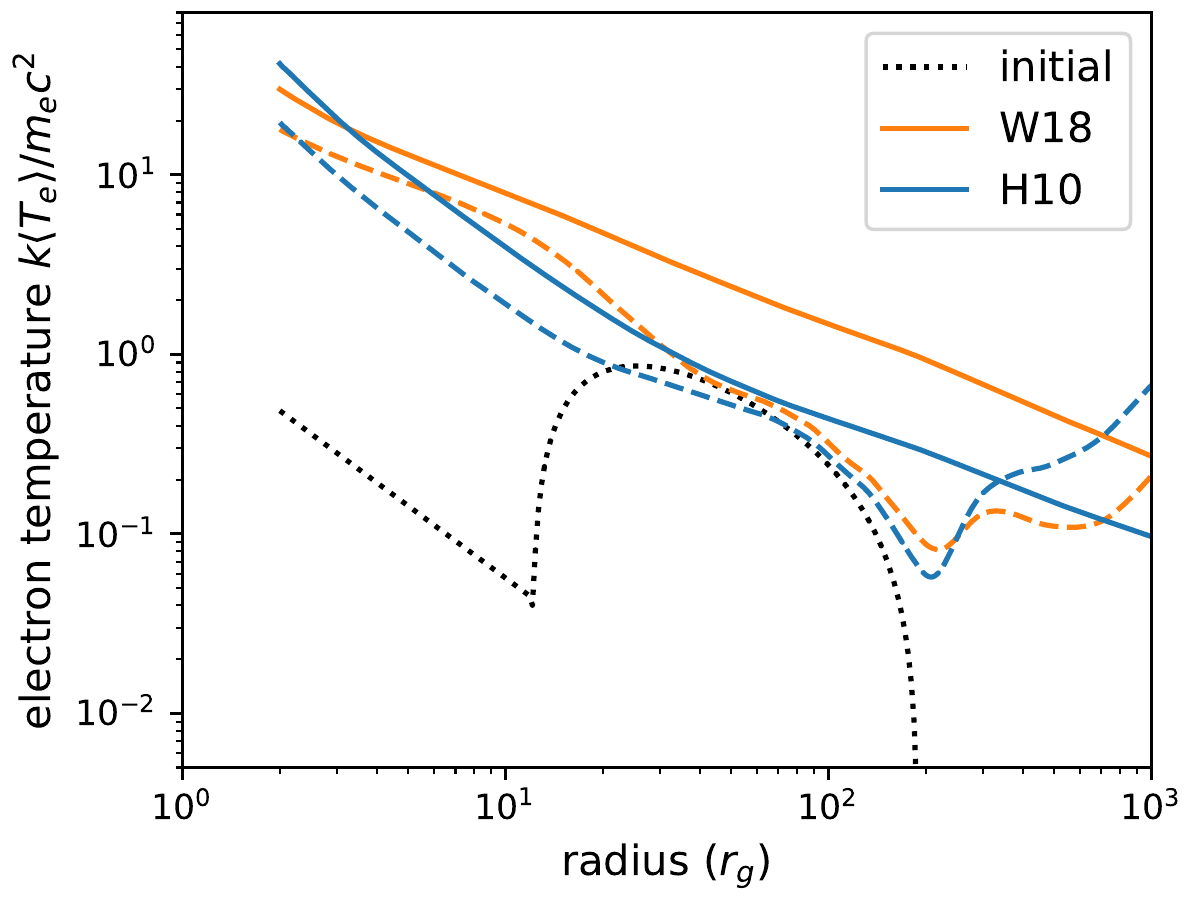} &
\includegraphics[width=0.48\textwidth]{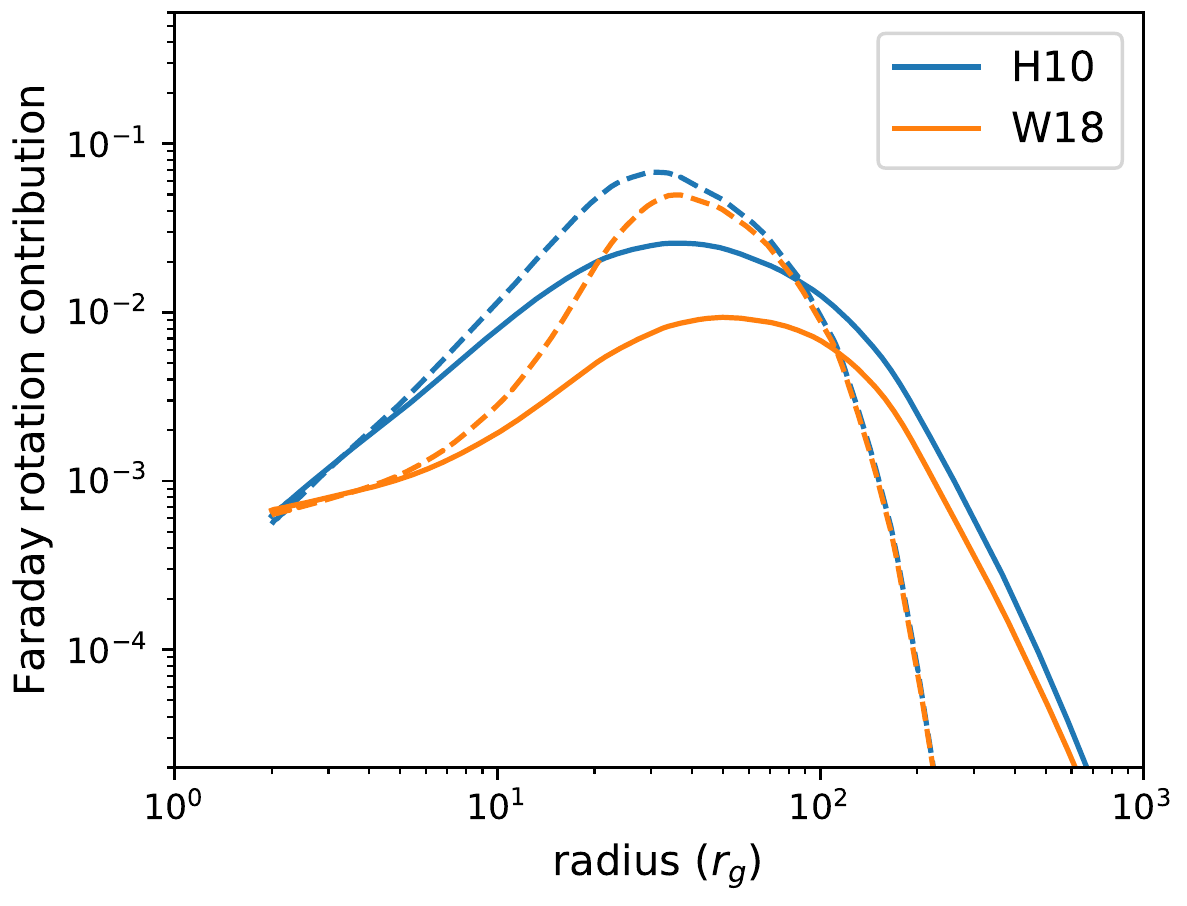}
\end{tabular}
\caption{\label{fig:te_evolution}Shell-averaged electron temperature (left) and relative contribution to the Faraday rotation measure (right) for our long duration SANE $a=0$ simulation. We compare the assumed initial conditions in our simulations (black dotted line) with the converged final state (solid) and time interval used for calculating radiative models (dashed) for the H10 and W18 electron heating models. The electron temperature profile used at early times is systematically colder the converged profile by a factor $\simeq 1.5$ at all radii. For $r \gtrsim 20 r_g$ at early times the solution does not seem to have heated much beyond its initial state. As a result, the W18 model electrons (orange dashed curve) are far too cold at large radius. The effect is much weaker in the H10 case, where the equilibrium electron temperature at large radius happens to be close to that assumed in the initial condition. For both electron models, at earlier times the low temperatures causes a stronger Faraday rotation peak at smaller radii. As described in the main text, we exclude material with $r > 20 r_g$ in calculating radiative models.}
\end{figure*}

Our simulations evolve multiple electron energies starting from an initial condition with $T_p/T_e=10$. The long duration simulations achieve equilibrium electron temperature profiles for roughly the last half of their duration ($1-2 \times 10^5 r_g/c$ for SANE and $4-6\times10^4 r_g/c$ for MAD). \autoref{fig:te_evolution} shows shell-averaged radial profiles of the equilibrium electron temperature for the H10 and W18 models (solid lines) for the SANE simulation, compared to the same profiles for the time interval used for calculating the models in our parameter survey (dashed) and the initial condition (dotted). The electrons heat gradually and are too cold by a factor $\simeq 1.5$ everywhere at early times. For the SANE case at early times, the electron temperature has apparently not relaxed from its initial condition for $r \gtrsim 20 r_g$.

We mitigate this effect in our analysis by excluding material outside of $20 r_g$ in calculating the SANE models, since otherwise the cold electrons Faraday depolarize the emission region. We have also checked that polarization properties excluding $r \gtrsim 20 r_g$ are similar to the full radiative transfer results from the long duration SANE model at late times, once the electron temperature distribution has converged. The effect of excluding emission (or running to very long durations) is large for the W18 model, where the electrons heat efficiently. It is modest for the H10 model, where the equilibrium temperature profile at those radii turns out to be similar to our assumed initial condition. In particular, Faraday rotation through an accretion flow is generally thought to be dominated by the location where $\theta_e = 1$ \citep[e.g.,][]{marrone2006}. That location is at $r \simeq 30 r_g$ and $300 r_g$ for the H10 and W18 models. We also measure this directly in the simulation data (\autoref{fig:te_evolution}), where we find contributions to the RM peaking at $\simeq 30 r_g$ and $\simeq 80 r_g$. As a result of cold electrons near the midplane, our H10 models show large Faraday rotation measures and significant depolarization from cold electrons at relatively small radii.

\section{Internal and external Faraday rotation}
\label{app:rm_appendix}

\begin{figure*}
\begin{tabular}{ccc}
\includegraphics[width=0.29\textwidth]{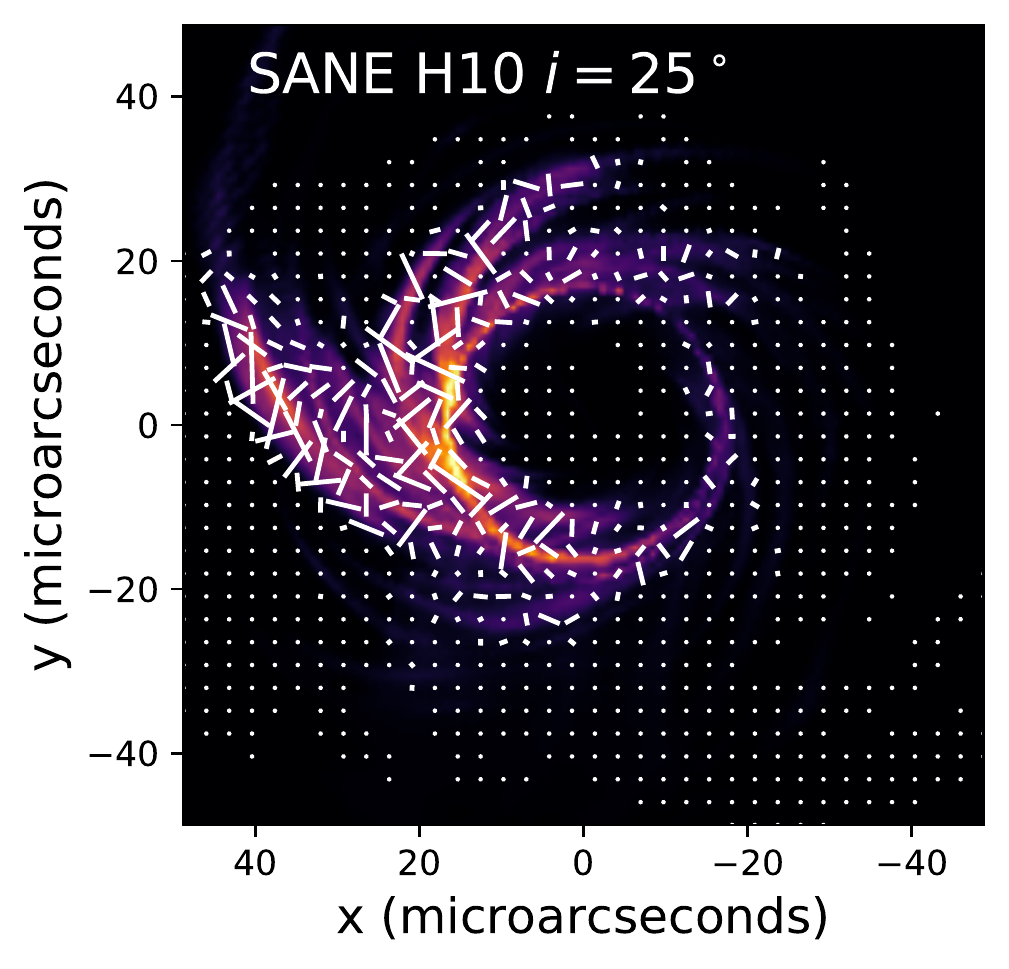} &
\includegraphics[width=0.29\textwidth]{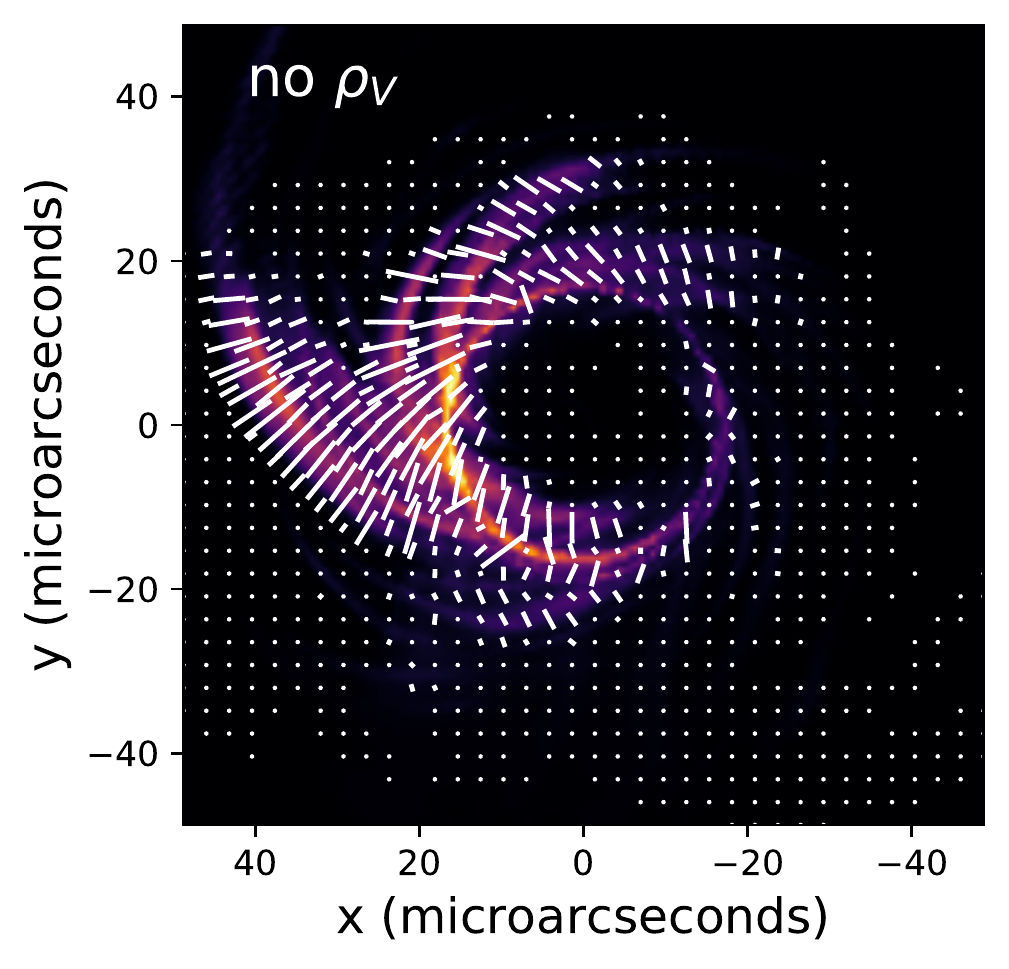}&\includegraphics[width=0.36\textwidth]{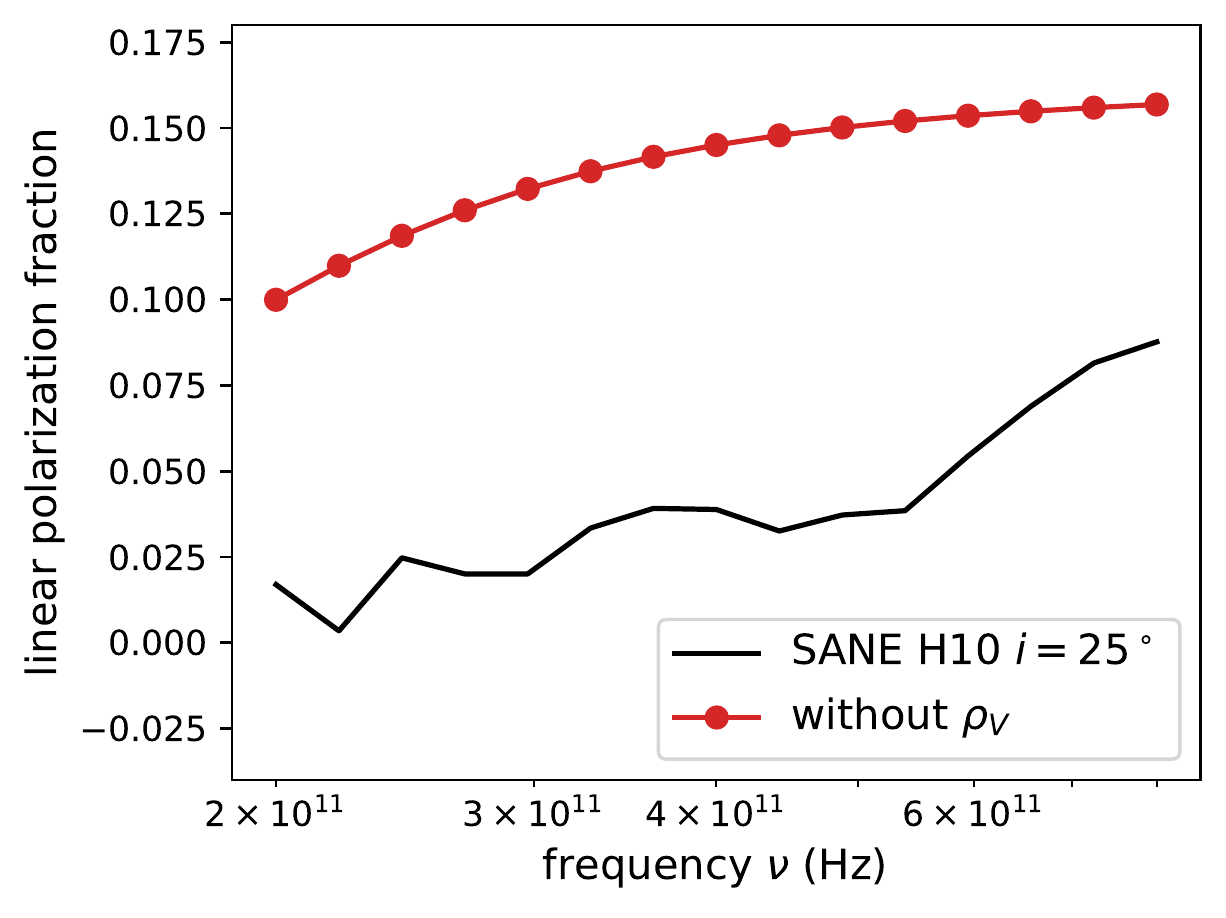}
\end{tabular}
\caption{\label{fig:taufr_compare}Total intensity images (false color) and polarization maps (white ticks, length proportional to polarized flux) for a snapshot from our long duration SANE $a=0$ simulation using the H10 electron heating model and viewed at an inclination of $i=25^\circ$. The full calculation (left) shows substantial disorder in the polarization pattern due to internal Faraday rotation. Setting $\rho_V=0$ (middle) leads to an ordered polarization pattern. The right panel shows polarization fraction as a function of frequency  for the same models. The net linear polarization is much higher when $\rho_V=0$.}
\end{figure*}

The Faraday rotation measure (RM) is related to the Faraday optical depth $\tau_{\rho_V} = \int dl \rho_V$ by $\tau_{\rho_V} = 2 RM \lambda^2$. For Sgr A*, the observed RM magnitude of $6x10^5 \, \, \rm rad \, \, \rm m^{-2}$ at 230 GHz corresponds to $\tau_{\rho_V} \lesssim 1$, depending on the frequency bands used. The total EVPA rotation is still $< \pi$, insufficient to produce strong Faraday depolarization. We report much larger values of  $\tau_{\rho_V}$ in \autoref{tab:phys_props}, despite the fact that the models can show EVPA $\propto \lambda^2$ with realistic values of the RM magnitude \autoref{fig:external_rm}. 

The RM is measured from the change in EVPA, which comes from the observed polarized flux. Strongly depolarized regions contribute little polarized flux. As a result, Faraday thick regions internal to the source do not necessarily lead to large RMs. \citet{moscibrodzka2017} found that in submm models of M87 from SANE GRMHD simulations, the RM measured from the change in EVPA could be roughly constant, even as $\tau_{\rho_V}$ varied by several orders of magnitude. 

\autoref{fig:taufr_compare} shows sample polarization maps for a snapshot of our long duration SANE $a=0$ simulation at late times using the H10 electron model and viewed at $i=25^\circ$. The full calculation shows a scrambled polarization map due to Faraday rotation internal to the emission region. The RM inferred for this snapshot is only $\simeq -3\times 10^5 \, \, \rm rad \, \, m^{-2}$. When we neglect Faraday rotation by setting $\rho_V=0$, the polarization map appears ordered and the inferred RM drops to $\simeq 0$. The net linear polarization is also much higher when $\rho_V=0$, $\simeq 12\%$ at 230 GHz compared to $\simeq 2\%$ in the full calculation. Evidently the depolarization in SANE H10 models is due to Faraday rotation, even when viewed at low inclination. Internal Faraday rotation can also be strong enough to substantially depolarize the image without showing up as a large RM as inferred by the change of EVPA with frequency.

\section{Effect of emission from highly magnetized regions}
\label{app:sigmacut}

\begin{figure*}
\begin{tabular}{cc}
\includegraphics[width=0.48\textwidth]{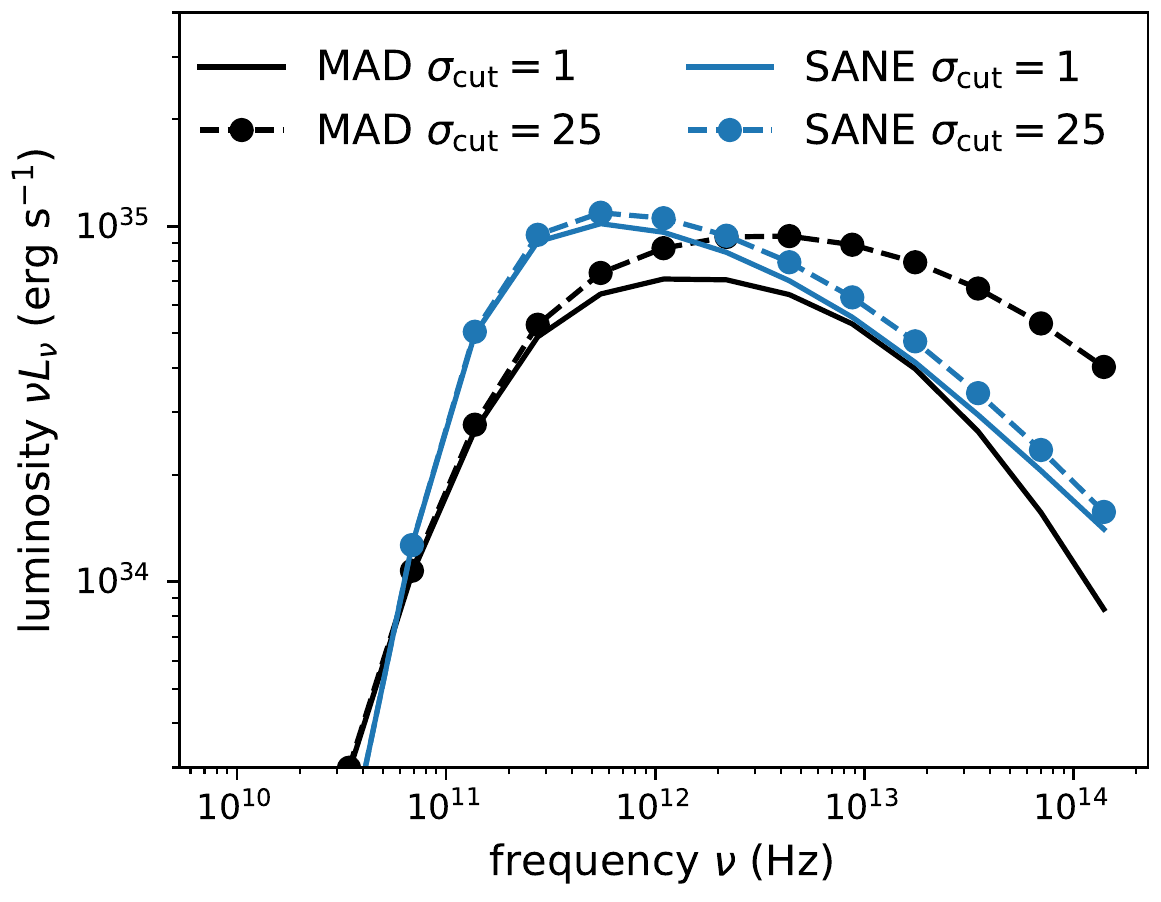} &
\includegraphics[width=0.48\textwidth]{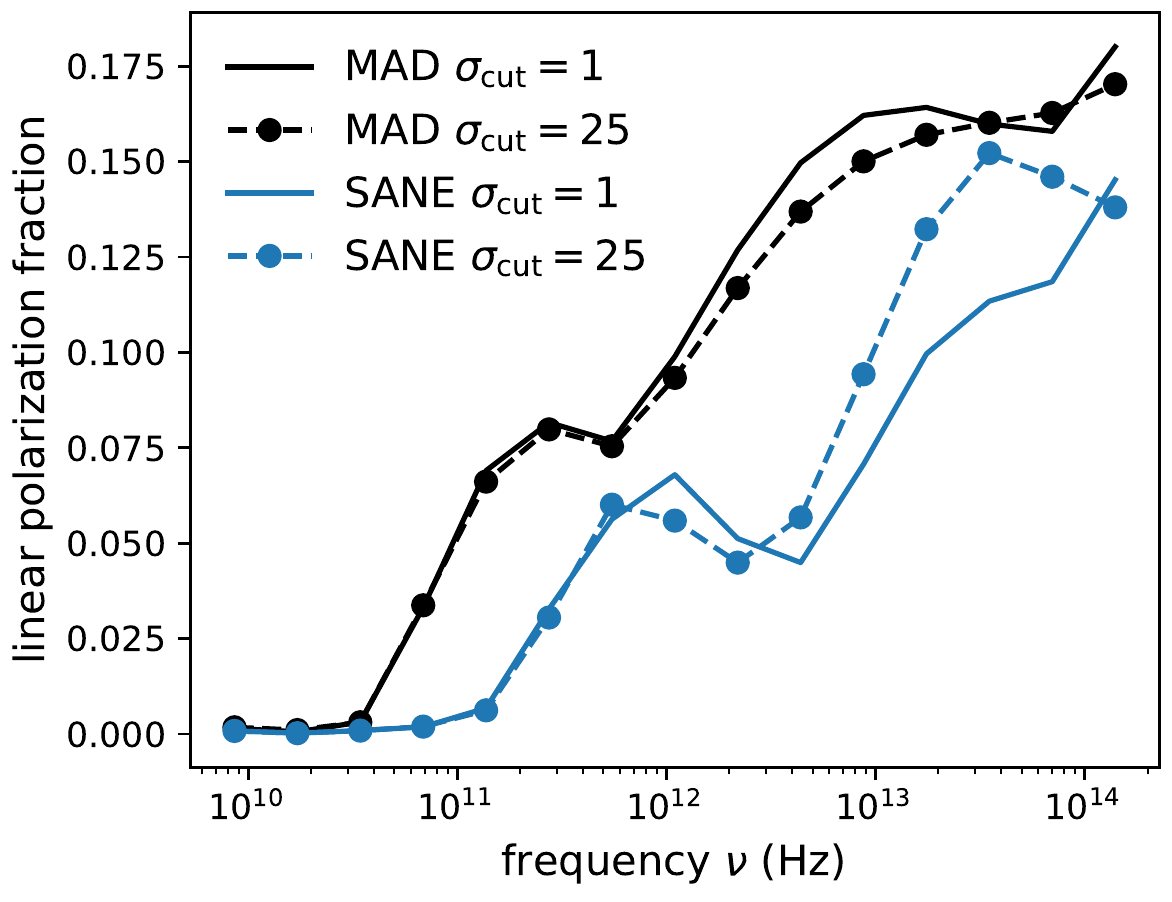}
\end{tabular}
\caption{\label{fig:sigma_cut}Sample spectra (left) and linear polarization fractions (right) for snapshots from near the end of our long duration SANE $a=0$ and MAD $a=0.9375$ simulations, using the H10 and W18 electron heating models respectively. Each model is calculated with magnetization cutoffs of $\sigma_{\rm cut}=1$ and $25$. High $\sigma > 1$ plasma contributes little of the total emission at any frequency in the SANE case. In the MAD case, we see little difference in the polarization fraction but the submm spectral slope and particularly the NIR flux density can change by factors of several depending on the choice of $\sigma_{\rm cut}$.}
\end{figure*}

In this work we follow \citet{EHTPaperV} and neglect emission from all regions where the magnetization $\sigma > 1$. Highly magnetized regions are difficult to evolve accurately in ideal MHD and may have mixed with artificially injected mass and energy (due to ``floors"). \citet{ressler2017} show that this choice makes little difference for SANE models, where most of the fluid is weakly magnetized. Highly magnetized regions are more prevalent in MAD models, and \citet{chael2019} explored the effects of various cuts on $\sigma$ in their images and spectra of M87.

\autoref{fig:sigma_cut} shows Sgr A* spectra and linear polarization fractions for two sample snapshots, one each from late times in our long duration SANE $a=0$ and MAD $a=0.9375$ simulations. We adopt the H10 (SANE) and W18 (MAD) electron heating models since those best describe the Sgr A* spectrum. In the SANE case, we confirm that high $\sigma$ material does not contribute significantly to the radio to NIR emission.

In the MAD case, $\sigma > 1$ plasma produces an increasing fraction of the emission at higher frequencies beyond the THz spectral peak and dominates the radiation in the NIR. Adopting a higher $\sigma$ cutoff value would therefore lead to higher NIR flux densities and slight changes to the submm spectral index. The choice of $\sigma$ cutoff value remains interesting to explore further in future work, but seems unlikely to be a major source of uncertainty in the analysis presented here.

\bsp	\label{lastpage}
\end{document}